\RequirePackage{fix-cm}
\documentclass[iicol,sn-basic]{sn-jnl}

\catcode`\|=12\relax 

\usepackage{hyperref}       
\usepackage{url}            
\usepackage{booktabs}       
\usepackage{amsfonts}       
\usepackage{multirow, makecell}

\usepackage{amsmath}
\usepackage{amsthm}
\usepackage{amssymb}

\usepackage{siunitx}
\usepackage{array}
\newcolumntype{P}[1]{>{\centering\arraybackslash}p{#1}}
\newcolumntype{M}[1]{>{\centering\arraybackslash}m{#1}}
\usepackage{tabularx}
\usepackage{diagbox}

\usepackage{mathtools}
\usepackage{graphicx}
\usepackage{color,soul}
\usepackage{subcaption}

\usepackage{algorithm}
\usepackage{listings}
\usepackage{xcolor}

\definecolor{ao(english)}{rgb}{0.0, 0.5, 0.0}
\definecolor{electricpurple}{rgb}{0.75, 0.0, 1.0}

\newcommand{\rone}[1]{#1}
\newcommand{\rtwo}[1]{#1}
\newcommand{\rme}[1]{#1}

\definecolor{codegreen}{rgb}{0,0.6,0}
\definecolor{codegray}{rgb}{0.5,0.5,0.5}
\definecolor{codepurple}{rgb}{0.58,0,0.82}
\definecolor{backcolour}{rgb}{0.95,0.95,0.92}

\lstdefinestyle{mystyle}{
    backgroundcolor=\color{backcolour},   
    commentstyle=\color{codegreen},
    keywordstyle=\color{magenta},
    numberstyle=\tiny\color{codegray},
    stringstyle=\color{codepurple},
    basicstyle=\ttfamily\footnotesize,
    breakatwhitespace=false,         
    breaklines=true,                 
    captionpos=b,                    
    keepspaces=true,                 
    numbers=left,                    
    numbersep=5pt,                  
    showspaces=false,                
    showstringspaces=false,
    showtabs=false,                  
    tabsize=2
}

\usepackage{multicol}
\usepackage{bm}
\usepackage{tikz}
\usetikzlibrary{patterns, patterns.meta}
\usetikzlibrary{fadings}
\usetikzlibrary{calc}

\makeatletter
\DeclareRobustCommand{\pder}[1]{%
  \@ifnextchar\bgroup{\@pder{#1}}{\@pder{}{#1}}}
\newcommand{\@pder}[2]{\frac{\partial#1}{\partial#2}}
\makeatother

\makeatletter
\DeclareRobustCommand{\der}[1]{%
  \@ifnextchar\bgroup{\@der{#1}}{\@der{}{#1}}}
\newcommand{\@der}[2]{\frac{d#1}{d#2}}
\makeatother

\newcommand{\mb}[1]{\mathbf{#1}}

\newcommand*\diff{\mathop{}\!\mathrm{d}}

\newcommand{\gfp}{\tilde\gamma}

\jyear{2021}%
\begin{document}
\title[Optimization]{Topology optimization for the design of porous electrodes}

\author[1]{\fnm{Thomas} \sur{Roy}}\email{roy27@llnl.gov}
\equalcont{These authors contributed equally to this work.}

\author[1]{\fnm{Miguel A.} \sur{Salazar de Troya}}\email{salazardetro1@llnl.gov}
\equalcont{These authors contributed equally to this work.}

\author[2]{\fnm{Marcus A.} \sur{Worsley}}\email{worsley1@llnl.gov}

\author*[1]{\fnm{Victor A.} \sur{Beck}}\email{beck33@llnl.gov}

\affil[1]{\orgdiv{Computational Engineering Division}, \orgname{Lawrence Livermore National Laboratory}, \orgaddress{\street{7000 East Ave}, \city{Livermore}, \postcode{94550}, \state{California}, \country{USA}}}
\affil[2]{\orgdiv{Materials Science Division}, \orgname{Lawrence Livermore National Laboratory}, \orgaddress{\street{7000 East Ave}, \city{Livermore}, \postcode{94550}, \state{California}, \country{USA}}}

\raggedbottom

\abstract{
Porous electrodes are an integral part of many electrochemical devices since they have high porosity to maximize electrochemical transport and high surface area to maximize activity. Traditional porous electrode materials are typically homogeneous, stochastic collections of small-scale particles and offer few opportunities to engineer higher performance. Fortunately, recent breakthroughs in advanced and additive manufacturing are yielding new methods to structure and pattern porous electrodes across length scales. These architected electrodes are emerging as a promising new technology to continue to drive improvement; however, it is still unclear which structures to employ and few tools are available to guide their design. In this work we address this gap by applying topology optimization to the design of porous electrodes. We demonstrate our framework on two applications: a porous electrode driving a steady Faradaic reaction and a transiently operated electrode in a supercapacitor. We present computationally designed electrodes that minimize energy losses in a half-cell. For low-conductivity materials, the optimization algorithm creates electrode designs with a hierarchy of length scales. Further, the designed electrodes are found to outperform undesigned, homogeneous electrodes. Finally, we present three-dimensional porous electrode designs. We thus establish a topology optimization framework for designing porous electrodes.
}


\maketitle

\section{Introduction}

Electrochemical devices are ubiquitous across society and play an increasingly critical role in addressing our global energy storage and conversion challenges. Beyond the more familiar applications like primary and secondary (rechargeable) batteries, these devices are enabling many of the novel, large-scale electrical energy storage (EES) technologies critical for driving the adoption of renewable electricity \citep{Gur_2018}. Large-scale integration is crucial, as renewable energies are the key for a sustainable and carbon-free future \citep{Chu_2012, Chu_2016}.

Secondary batteries, flow batteries, and supercapacitors can be employed to directly store and dispatch electrical energy and provide a buffer between the intermittency of electricity supply and demand \citep{Gur_2018, sawant_harnessing_2021}.
Alternatively, electrochemical reactors and electrolyzers provide a pathway for converting electrical energy into chemical potential energy in the form of carbon and hydrogen fuels \citep{Ager_2018, stockl_optimal_2021}.
In addition to providing a useful conduit for otherwise wasted excess electricity, for many of these reactors the feedstock is carbon dioxide, leading additionally to a reduction in greenhouse gases \citep{obrien_single_2021, lamaison_designing_2021}.
More broadly, electrochemical devices employed as reactors are driving the electrification of industrial chemical manufacturing, further reducing or eliminating emissions \citep{schiffer_electrification_2017, yan_synthetic_2017, shatskiy_organic_2019, stankiewicz_beyond_2020, barton_electrification_2020}.
Nevertheless, significant research, engineering and design effort is required to realize the benefits that these technologies offer, ensuring economic viability and guaranteeing widespread deployment \citep{Chen_2018,Gur_2018}.

\begin{figure}[!htbp]
    \centering
    \begin{subfigure}{7cm}
        \includegraphics[]{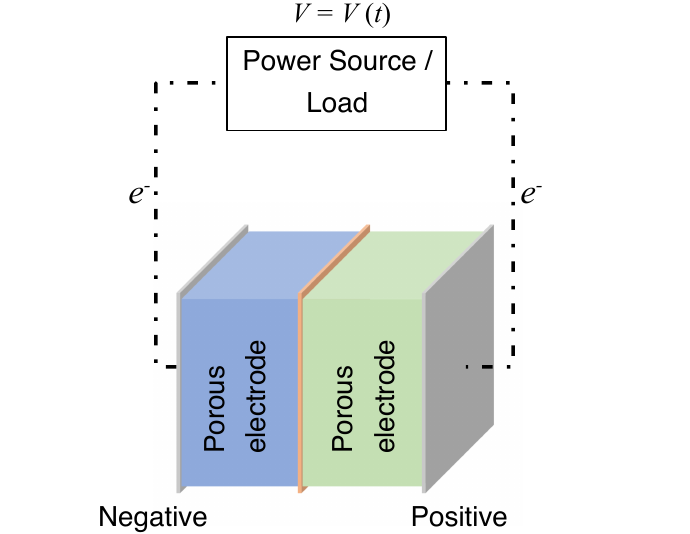}
        \caption{Transient}
    \end{subfigure}
    \begin{subfigure}{7cm}
        \includegraphics[]{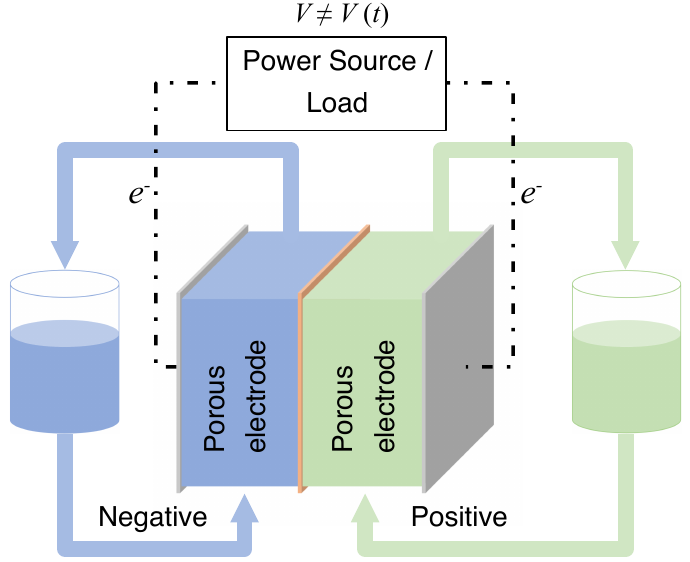}
        \caption{Steady-state}\label{fig:echem_diagram_ss}
    \end{subfigure}
    \caption{Model electrochemical devices composed of two porous electrodes separated by a membrane or separator. The porous electrodes contact current collectors, which are electrically connected to a power source or load. (a) Typical devices operate transiently and transform material \emph{in situ}. (b) Steady-state operation requires the continuous introduction and removal of material using flow. }
    \label{fig:echem_diagram}
\end{figure}

Electrochemical devices include galvanic cells, which output energy, electrolytic cells, which require energy to perform a chemical transformation, and supercapacitors, which can dispatch energy through stored charge \citep{fuller2018electrochemical}. Examples of galvanic cells include fuel cells and discharging batteries. Electrolytic cells include charging batteries, electrochemical reactors and electrolyzers. All of these devices share a common, core architecture typically composed of two porous electrodes immersed in an electrolyte and separated by a membrane or separator, as shown schematically in Figure \ref{fig:echem_diagram}. 

The electrochemical processes driving performance occur at the electrode-electrolyte interface, and generally the higher the surface area, the higher the currents, power outputs, and energy storage. The porous electrode is often a coherent, stochastic collection of conductive micron-scale particles, and is principally engineered to yield the highest surface area per volume possible while minimizing electrical, fluid flow, and diffusive resistances as required by the specific application \citep{wang20083d, weber_redox_2011}. The result is a monolith that can be characterized by a single porosity and single length scale for the constituent particles, leaving few opportunities for further engineering of the structure.

An emerging alternative to further increase electrochemical device performance is to employ architected electrodes \citep{forner-cuenca_engineering_2019, park_perspective_2020, zhang_design_2021}. Instead of a simple, single-porosity monolith, the electrode can be composed of multiple materials, and shaped and patterned at multiple length scales. The concepts have been applied to generate high-performance, dual-scale lithium-ion batteries composed of energy-dense regions accessed by tailored, high-mobility channels \citep{bae_design_2013,Cobb_2014,nemani2015design,zhang_tunable_2021}. Similar ideas have also been used to improve redox flow battery performance \citep{Zhou_2016}. Spurred by advances in additive manufacturing and 3D printing \citep{ambrosi_3d-printing_2020}, these efforts have been further extended to create patterned supercapacitors with superior energy density \citep{zhu2016supercapacitors} and electrochemical flow reactors with improved productivity and mass transfer \citep{beck_inertially_2021}. These new manufacturing techniques are especially exciting, as they offer the promise of near arbitrary control over the porous electrode structure.   

Advanced design tools are required to fully exploit these new manufacturing techniques and further improve electrochemical device performance. Topology optimization offers a novel opportunity to automatically design the multiscale, porous electrodes. The first application of these techniques focused on addressing design challenges in structural mechanics. In the pioneering work of
\cite{bendsoe}, the optimal design geometry was first formulated as a material distribution problem wherein a volume fraction field models the solid, void and intermediate material phases and incorporated the Solid Isotropic Material with Penalization (SIMP) method to penalize the intermediate phase and recover a discrete design. An alternative approach based on the level-set method \citep{wang2003level,allaire,SETHIAN2000489} uses the zero isocontour of a level-set function to define the solid-void interface, thereby avoiding the intermediate phase. 

Topology optimization techniques continue to be extended to more physically complex system including thermal, fluid, and coupled systems \citep{alexandersen_review_2020}, but there have been limited applications of the ideas to electrochemical devices. \rtwo{Early work used level-set approaches to design electrodes for solid oxide fuel cells} \citep{iwai2011power} \rtwo{and lithium-ion batteries} \citep{zadin2013designing}. \rtwo{Topology optimization was also used to design a porous ionic-conducting scaffold for solid oxide fuel cell cathodes} \citep{song20132d}. More recent efforts have generally focused on the design of flow fields, which deliver fluids
to porous electrodes, in the context of redox flow batteries \citep{yaji2018topology, chen2019computational, lin2022topology} and fuel cells \citep{behrou2019topology}. While porous electrodes are part of these models, flow field architectures are the goal of the design problems. Other work on optimizing electrode structure has been limited to considering a smooth variation of the porosity field applied to the design of lithium-ion batteries \citep{ramadesigan2010optimal,golmon2012multiscale,Golmon_2014,xue2016lithium} and redox flow batteries \citep{Beck_2021}. In both cases, computational design and optimization lead to significant performance improvements over conventional, homogeneous-porosity electrodes. Performance improvements using graded porosity have also recently been experimentally demonstrated \citep{Lu_2020}. \rtwo{To the authors' knowledge density-based topology optimization has not yet been applied to the design of the porous electrode itself, nor have these techniques been extended to transient and three-dimensional systems}.

\begin{figure}[!htbp]
    \centering
    \includegraphics[]{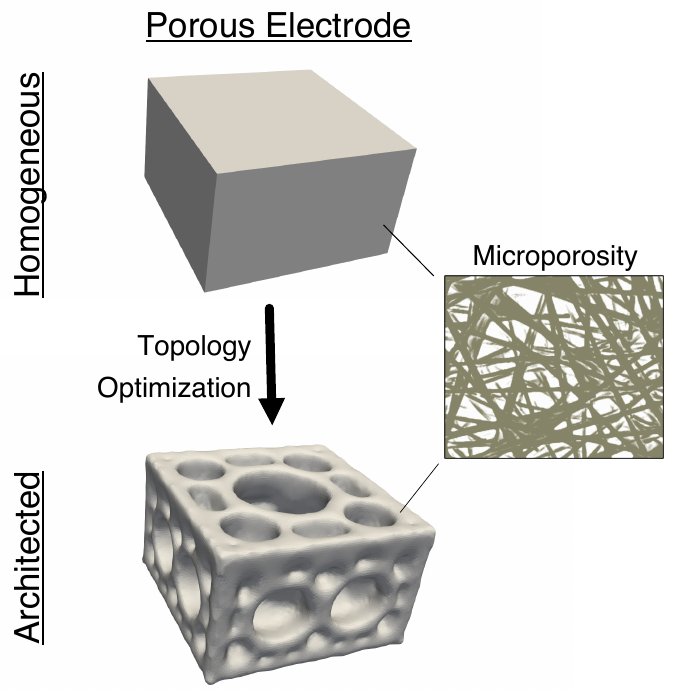}
    \caption{A typical porous electrode is a homogeneous collection of small-scale particles. The porous homogeneous electrode can be sculpted using topology optimization to create an architected porous electrode. Though represented as solids, the material composing the electrodes is in fact porous, and is labeled `microporous' to distinguish the length scale from the pores generated by the optimization procedure.  }
    \label{fig:design_diagram}
\end{figure}

In this manuscript, we introduce the application of topology optimization to design porous electrodes for use in electrochemical devices.
As shown in Figure \ref{fig:echem_diagram}, we separately consider the design of porous electrodes for steady-state and transient operation. Flow batteries, fuel cells and electrochemical reactors are ideally operated at steady-state and thus require constant fluid flow to supply and remove material, while batteries and supercapacitors transform material \emph{in situ} and thus inherently operate transiently. We focus on simulation and optimization of the secondary current distribution, which provides a description of electrode performance when concentration polarization effects are unimportant \citep{haverkort_theoretical_2019}. In brief, the optimization algorithm is used to sculpt a porous material, here labeled `microporous' to distinguish it from the larger pores generated by the optimization process and to emphasize the multiscale nature of the design, into an optimal architecture as shown in Figure \ref{fig:design_diagram}.

In Section \ref{sec:topopt}, we give a brief introduction of topology optimization applied to the design of porous electrodes.
Then, in Section \ref{sec:problems}, we define the systems of partial differential equations (PDEs) for the ionic and electronic potentials describing the secondary current distribution in the porous electrodes. We then specialize these equations to describe model steady and transient applications, a porous electrode driving a reduction-oxidation Faradaic reaction (i.e., a `redox electrode') and a charging supercapacitor, respectively. The equations are nondimensionalized in the context of seminal work on porous electrode theory to yield familiar dimensionless groups and provide generality \citep{newman1975porous}. Implementation details and solution techniques are given in Section
\ref{sec:implementation}. Specifically, we include preconditioning techniques to
accelerate the solution of the PDEs.
In Section \ref{sec:results}, we show various two-dimensional designs optimized for power efficiency, in the case of the redox electrode and for ohmic losses and stored energy in the supercapacitor electrode.
We discuss the impact of the different dimensionless parameters and compare our designs with monolithic, homogeneous electrodes.
Finally, we present examples of designed three-dimensional electrodes for each system.

\section{Topology Optimization}\label{sec:topopt}
The porous electrode design problem is to find the optimal distribution of material $\chi$ in a design domain $\Omega$, i.e.
\begin{equation}
    \begin{aligned}
         & \min _{\chi \in\{0,1\}} \theta_{0}(\chi) = \int_{\Omega} \pi(\chi, \Phi_1, \Phi_2) \diff V                \\
         & \text{s.t.} ~\Phi_1, \Phi_2  ~\text{satisfies } F(\chi, \Phi_1, \Phi_2)=0                   \\
         & \theta_{i}(\chi) = \int_{\Omega} g_{i}(\chi, \Phi_1, \Phi_2) \diff V \leq 0~ i=1,2 \ldots \mathrm{n}_{i},
        \label{eq:topopt}
    \end{aligned}
\end{equation}
where $\theta_0$ is the cost function to be minimized and $\theta_i$ are the $\text{n}_i$ design constraints.
The design variable $\chi$ defines the presence of a two porosity electrode ($\chi=0$ for material $M$ with porosity $\epsilon_M$ and $\chi=1$ for material $N$ with porosity $\epsilon_N$). Note that this definition includes the extreme cases of the materials being pure electrolyte, $\epsilon_i  = 1$, or pure solid, $\epsilon_i  = 0$. The response functions $\Phi_1$ and $\Phi_2$ model the electronic and ionic potentials, respectively, according to the system of equations  $F(\chi, \Phi_1, \Phi_2)=0$, whose details we explain in the next section.

\rtwo{In the context of density-based optimization \cite{bendsoe}}, the discrete nature of the design variable $\chi$ prevents using gradient-based algorithms.
We therefore make the \rtwo{set of optimization variables} convex by replacing $\chi \in \{0, 1\}$ with the
continuous volume fraction variable $\gamma \in [0,1]$.
As explained in Section \ref{sec:problems}, the constitutive equations modeling the conductivities inherently penalize intermediate values of $\gamma$.
We therefore do not use a penalization scheme such as SIMP \citep{bendsoe}.

Design problems in topology optimization are commonly ill-posed.
Optimal designs consist of a non-converging sequence of highly oscillatory
structures that maximize the surface area ad-infinitum.
To obtain a well-posed problem, several techniques impose a minimum
length scale in the design.
Chief among them is the diffusion-reaction PDE filter \citep{lazarov2011filters}
\begin{equation}
    \begin{aligned}\label{eq:filter}
        -r^2 \nabla^2 \tilde \gamma + \tilde\gamma & = \gamma, \quad & \text{in}\quad\Omega\,,   \\
        r^2 \nabla \tilde{\gamma} \cdot \mathbf{n}   & =0 \quad        & \text{on} \quad \partial\Omega\,,
    \end{aligned}
\end{equation}
where we solve for the filtered volume fraction $\tilde\gamma$ given the design volume fraction $\gamma$.
The filter radius $r$ controls the minimum length scale of the designs, $\partial \Omega$ denotes the boundary of $\Omega$, and $\mathbf{n}$, the outward pointing unit normal.

Since filtering inherently produces gray transition regions of intermediate material, projection of the filtered volume fraction $\tilde\gamma$ to 0--1 values is often used to obtain sharper designs \citep{guest2004achieving,wang2011projection}.
However, continuation strategies are required to avoid convergence to low quality local minima.
\rone{In this work, we obtain designs that are mostly discrete without projecting $\tilde\gamma$, so we leave out projection for simplicity.
Future work could apply a projection method to obtain a sharper geometry.}

\section{Governing equations}\label{sec:problems}
In this section, we first describe a generic model that is applicable to most porous electrodes. We then apply it to the specific cases of a porous electrode driving a Faradaic reduction-oxidation reaction (i.e., a porous redox electrode) and a porous electrode used as an electrical double layer capacitor (EDLC) or supercapacitor.

Porous electrodes consist of a porous solid matrix of an electrically conductive material immersed in an electrolyte solution (c-f. Figure \ref{fig:design_diagram}).
Their modeling is determined by the conservation of the ionic and electronic current densities, on the assumption of constant concentration of chemical species and no charge separation \citep{newman1975porous}.
The current density transfer between the ionic and electronic current densities occurs at the interface, and it can be caused by either Faradaic, electrochemical reactions or charge storage at the Electrical Double Layer.

High-fidelity models of porous electrodes are computationally expensive due to their complex geometry and length scale disparity.
Reduced-order models that capture essential quantities of interest without the costly modeling of the geometry are therefore desired.
Most notably, the large-scale separation between the small pore structure (i.e., microporosity) and the electrode's dimension permits the use of inexpensive models based on averaging techniques.
These models resolve the potential fields at the electrode scale using effective material properties that capture the pore-scale effects \citep{newman1975porous,newman2012electrochemical}.
Indeed, these effective properties are averages over representative elementary volumes containing both the solid matrix and the electrolyte phases.
The Bruggeman correlation \citep{bruggeman1935berechnung} is one such model in which the porosity $\epsilon$ and the tortuosity factor $\eta=\frac{3}{2}$ characterize the effective ionic conductivity
\begin{equation}
    \label{eq:kappaeff}
    \kappa = \epsilon^{\eta} \kappa_0,
\end{equation}
where $\kappa_0$ is the conductivity of the electrolyte phase
and the effective electronic conductivity
\begin{equation}
    \label{eq:sigmaeff}
    \sigma = (1-\epsilon)^{\eta} \sigma_0,
\end{equation}
where $\sigma_0$ is the conductivity of the solid phase.
We remark here that the exponent $\eta=\frac{3}{2}$ penalizes the conductivity of intermediate densities as in SIMP \citep{bendsoe}.
The current density transfer at the electrode-electrolyte interface is interpreted as an effective current per unit volume $ai_n(\Phi_1, \Phi_2)$, where $a$ is the surface area of the interface between the microporous structure and the electrolyte per unit volume of the total electrode.

\begin{figure}[!htbp]
    \centering
    \begin{subfigure}{\linewidth}
    \centering
    \begin{tikzpicture}[>=latex,scale=0.82]
        \draw[preaction={fill, red!15}, pattern=dots, pattern color=red!40] (0,0) -- (8,0) -- (8,4) -- (0,4) -- (0,0);
        \draw[fill=blue!15] (1,4) -- plot [smooth] coordinates {(1, 4) (1.5,2.9) (2,1) (3,2.2) (3.4, 1.2) (4,0.8) (4.8,1.6) (5.5,0.7) (6.2,1.9) (7,4)};
        \draw[line width=1.2pt] (1,0) -- (7,0);
        \node (a) at (4.,0.25) {$\Gamma_1$};
        \draw[line width=1.2pt] (1,4) -- (7,4);
        \node (b) at (4.,3.75) {$\Gamma_2$};
        \draw[thin,|<->|] (0, -0.33) -- node[fill=white,pos=0.5]{$W$}(8, -0.33);
        \draw[thin,|<->|] (1, 4.33) -- node[fill=white,pos=0.5]{$w$}(7, 4.33);
        \draw[thin,|<->|] (8.33, 0) -- node[fill=white,pos=0.5]{$L$}(8.33,4);
        \node at (4,2.7) {{\small Electrolyte} ($\gfp=0$)};
        \node at (1.9,0.5) {\small
            \addtolength{\tabcolsep}{-5pt}  
            \begin{tabular}{cl}
                Microporous & \multirow{2}{*}{($\gfp=1$)} \\
                material & 
            \end{tabular}
            \addtolength{\tabcolsep}{3pt}  
            };
    \end{tikzpicture}
    \caption{Two-dimensional design space.}\label{fig:diagramelectrode}
    \end{subfigure}
    \begin{subfigure}{\linewidth}
    \centering
    \begin{tikzpicture}[>=latex,scale=2]
        \pgfmathsetmacro{\La}{0.4}
        \pgfmathsetmacro{\L}{1.2}
        \pgfmathsetmacro{\h}{1}
        \pgfmathsetmacro{\hz}{2}
        \pgfmathsetmacro{\x}{2*\La+\L}
        \pgfmathsetmacro{\y}{\h}
        \pgfmathsetmacro{\z}{\hz}
        \coordinate (A1) at (\La,0,\z-\La);
        \coordinate (A2) at (\La+\L,0,\z-\La);
        \coordinate (A3) at (\La+\L,0,\La);
        \coordinate (A4) at (\La,0,\La);
        \coordinate (B1) at (\La,\y,\z-\La);
        \coordinate (B2) at (\La+\L,\y,\z-\La);
        \coordinate (B3) at (\La+\L,\y,\La);
        \coordinate (B4) at (\La,\y,\La);
        \draw[thin,|<->|] ($(A1)+(-45:-2pt)$) -- node[fill=white,pos=0.5,sloped]{\small $w$}($(A4)+(-45:-2pt)$);
        \draw[thin,|<->|] ($(A1)+(0,-2pt)$) -- node[fill=white,pos=0.5]{\small $w$}($(A2)+(0,-2pt)$);
        \draw[fill=gray!30] (A1)--(A2)--(A3)--(A4)--cycle;
        \path (0,0,\z) coordinate (A) (\x,0,\z) coordinate (B) (\x,0,0) coordinate (C) (0,0,0)
        coordinate (D) (0,\y,\z) coordinate (E) (\x,\y,\z) coordinate (F) (\x,\y,0) coordinate (G)
        (0,\y,0) coordinate (H);
        \draw (A)--(B)--(C)--(G)--(F)--(B) (A)--(E)--(F)--(G)--(H)--(E);
        \draw (A)--(D)--(C) (D)--(H);
        \draw[thin,|<->|] ($(A)+(0,-4pt)$) -- node[fill=white,pos=0.5]{$W$}($(B)+(0,-4pt)$);
        \draw[thin,|<->|] ($(C)+(4pt,0)$) -- node[fill=white,pos=0.5,sloped]{$L$}($(G)+(4pt,0)$);
        \draw[thin,|<->|] ($(B)+(-45:4pt)$) -- node[fill=white,pos=0.5,sloped]{$W$}($(C)+(-45:4pt)$);
        \draw[fill=gray!30] (B1)--(B2)--(B3)--(B4)--cycle;
        \node[] at ($(A1)!0.5!(A3)$) {$\Gamma_1$};
        \node[] at ($(B1)!0.5!(B3)$) {$\Gamma_2$};
    \end{tikzpicture}
    \caption{Three-dimensional design space.}
    \label{fig:diagramelectrode3D}
    \end{subfigure}
    \caption{Diagrams of the design spaces for porous electrodes. Centered on their respective boundaries, $\Gamma_1$ represents a current collector and $\Gamma_2$, a membrane.}
    \label{fig:diagrams}
\end{figure}
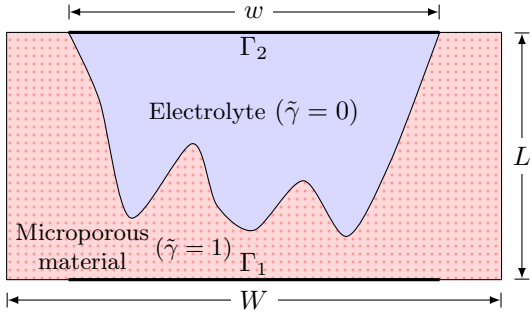
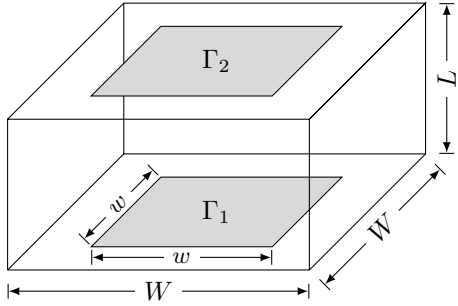

With the effective properties, we can model the porous electrode as a continuum, and we therefore solve for the electronic potential $\Phi_1$ and ionic potential $\Phi_2$ in the entire domain.
\begin{subequations}
    \label{eq:porous_electrode_problem}
    \begin{align}
        \label{eq:solid}
        -\nabla \cdot \left( \sigma \nabla \Phi_1 \right) & = -a i_n(\Phi_1, \Phi_2) & \text{ in } & \Omega,                             \\
        \label{eq:liquid}
        -\nabla \cdot \left( \kappa \nabla \Phi_2 \right) & = a i_n(\Phi_1, \Phi_2)  & \text{ in } & \Omega,                             \\
        \label{eq:dirichlet}
        \Phi_1                                            & = 0                      & \text{on }  & \Gamma_{1},                       \\
        \label{eq:noflux1}
        \sigma \nabla \Phi_1 \cdot \mathbf{n}             & = 0                      & \text{on }  & \partial\Omega\setminus\Gamma_{1} , \\
        \label{eq:robinbc}
        B_p\Phi_2 + B_c\kappa \nabla \Phi_2 \cdot \mathbf{n}  & = g                      & \text{on }  & \Gamma_{2},\\
        \label{eq:noflux2}
        \kappa \nabla \Phi_2 \cdot \mathbf{n}             & = 0                      & \text{on }  & \partial\Omega\setminus\Gamma_{2}.
    \end{align}
\end{subequations}
The domain boundary $\partial\Omega$ consists of two complementary regions for $\Phi_1$: $\Gamma_{1}$, and $\partial\Omega\setminus\Gamma_{1} $ over which, Dirichlet and homogeneous Neumann boundary conditions are applied, respectively.
Similarly, $\partial\Omega$ is split for $\Phi_2$ in $\Gamma_{2}$ and $\partial\Omega\setminus\Gamma_{2} $ over which Robin and homogeneous Neumann boundary conditions are applied, respectively.
The Robin boundary condition coefficients allows us to choose between applying only a potential, i.e. $B_p=1$ and $B_c=0$, or only a current density $B_p=0$ and $B_c=1$.
We do not consider other options in this manuscript.
These boundary conditions are for designing a porous electrode in a half-cell of an electrochemical device, where the membrane splitting the cell is the boundary $\Gamma_2$ (c-f. Figure \ref{fig:design_diagram}). The domain is explicitly depicted in 2D and 3D in Figure \ref{fig:diagrams}. Design of the entire cell, i.e. with two electrodes, is left for future work. Finally, $\mathbf{n}$ denotes the outward pointing unit normal.

We use the filtered design variable $\gfp$ to parametrize the porosity
\begin{equation}
    \epsilon = (1-\gfp)\epsilon_M + \gfp \epsilon_N,
    \label{eq:porosity}
\end{equation}
and the effective specific area per volume
\begin{equation}
    a = (1 - \gfp)a_M + \gfp a_N,
\end{equation}
where $a_M$ and $a_N$ are their specific area per volume for materials $M$ and $N$, respectively.
In this article, for simplicity, we consider $\epsilon_M=1$ and $a_M=0$, i.e. a pure electrolyte phase for $\gfp=0$.

The Bruggeman correlation has been shown to overestimate the ionic conductivity for porous electrodes with different particle arrangement \citep{tjaden2018tortuosity,TJADEN201644}.
Several researchers have proposed correction factors but there is still no consensus model \citep{koresh, thorat}. 
In addition to the original Bruggeman correlation in \eqref{eq:kappaeff}, we thus consider a modification that results in lower effective ionic conductivity of the form $\kappa = f_\kappa \epsilon^\eta \kappa_0$, where $f_\kappa$ takes values in the range $[0.02, 0.79]$ depending on the electrode material \citep{Madabattula_2020}.
We penalize the electrode porosity $\epsilon_N$ in \eqref{eq:porosity}, so we do not affect the conductivity in the pure electrolyte phase, i.e. $\kappa=\kappa_0$ for $\gfp=0$:
\begin{equation}
    \hat\epsilon = (1-\gfp) + f_{p}\gfp \epsilon_N\,.
    \label{eq:porosity_two}
\end{equation}
where $f_{p} = f_\kappa^{1/\eta}$, and use it in \eqref{eq:kappaeff} instead of \eqref{eq:porosity}, i.e.
\begin{equation}
    \label{eq:kappaeff_correction}
    \kappa = \hat\epsilon^{\eta} \kappa_0.
\end{equation}

\subsection{Porous redox electrode}
We consider the case of a porous electrode where a reduction-oxidation
(redox) reaction occurs
inside the porous electrode \citep{newman1975porous,newman2012electrochemical}.
This is common in most porous electrodes operated at steady-state and constant concentration and often requires flowing electrolyte as shown in Figure \ref{fig:echem_diagram_ss}. To distinguish this example we specify the model name as a ``porous redox electrode.''
Depending on the direction of the current and reaction
term, this model could be used for devices such as discharging flow batteries, where a chemical reaction
spontaneously generates current, or electrolyzers and charging batteries, where energy is inputted to activate the reaction.

The current generated due to the simple one electron transfer redox reaction $Ox + e^- \rightarrow Red$, is described by the Butler-Volmer relationship \citep{fuller2018electrochemical}
\begin{multline}\label{eq:in}
    i_n(\Phi_1, \Phi_2) = \frac{i_0}{C_\mathrm{ref}} \left[C_R \exp \left(\frac{\alpha_A F}{RT} \Delta \Phi \right)\right. \\
        \left.- C_O \exp \left(\frac{-\alpha_C F}{RT} \Delta \Phi \right)\right],
\end{multline}
where $i_0$ is the exchange current density, or just ``exchange current,'' corresponding to the reference concentration
$C_\mathrm{ref}$ \citep{newman1975porous}, and $C_R$ and $C_O$ are the concentration of reductant, $Red$, and oxidant, $Ox$, respectively.
The anodic and cathodic charge transfer (dimensionless) coefficients are $\alpha_A$ and $\alpha_C$, respectively.
For simplicity, we assume $C = C_R = C_0$ and $\alpha = \alpha_A = \alpha_C$.
$T$ is the absolute temperature, $F$ is Faraday's constant, and $R$ is the universal gas constant.
Finally, $\Delta \Phi = \Phi_1 - \Phi_2 - U_0$ is defined as the potential difference and $U_0$ is the standard potential for this reaction.

We supply the system with only an ionic current density
\begin{equation}
    \label{eq:IA}
    g = I/A,
\end{equation}
on $\Gamma_{2}$, i.e. $B_p=0$ and $B_c=1$ in \eqref{eq:robinbc}, and ground the electronic potential at $\Gamma_{1}$, cf. \eqref{eq:dirichlet}.
The rest of the boundary is electrically insulated, cf. \eqref{eq:noflux1} and \eqref{eq:noflux2}.
The total current supplied to the system, $I$, is specified over the membrane area, $A$.
A positive current $I$ drives a chemical reaction and induces the porous redox electrode to act as an electrolyzer or charging battery.

Let $L$ be the thickness of the electrode, as illustrated in Figure \ref{fig:diagrams}.
We replace the expressions \eqref{eq:in}, \eqref{eq:IA}, and the nondimensional variables:
\begin{align}
    \hat{\Phi}_1 & = \frac{\kappa_0 A}{L I} \Phi_1 , \label{eq:Phi1hat}                   \\
    \hat{\Phi}_2 & = \frac{\kappa_0 A}{L I} \left(\Phi_2 + U_0\right), \label{eq:Phi2hat} \\
    \hat{\mb x}  & = \frac{1}{L}\mb x, \label{eq:xhat}
\end{align}
in the porous electrode system \eqref{eq:porous_electrode_problem} to obtain the nondimensional equations of the porous redox  electrode:
\begin{subequations}
\label{eq:redox_eq}
    \begin{align}
        \label{eq:redox_Phi1}
        -\hat\nabla \cdot \left( (1-\epsilon)^\frac{3}{2}\hat\nabla \hat\Phi_1 \right) & =
        -\frac{\delta}{\mu} \frac{\tau}{1+\tau} \hat i                                             & \text{ in } & \hat\Omega,                                              \\
        \label{eq:redox_Phi2}
        -\hat\nabla \cdot \left( \epsilon^\frac{3}{2} \hat\nabla \hat\Phi_2 \right)    & =
        \frac{\delta}{\mu} \frac{1}{1+\tau}  \hat i                                                & \text{ in } & \hat\Omega,                                              \\
        \hat\Phi_1                                                                     & = 0         & \text{on }  & \hat\Gamma_{1}, \\
        (1-\epsilon)^\frac{3}{2}\hat\nabla \hat\Phi_1 \cdot \mathbf{n}                 & = 0         & \text{on }  & \partial\hat\Omega\setminus\hat\Gamma_{1}, \\
        \epsilon^\frac{3}{2} \nabla \hat\Phi_2 \cdot \mathbf{n}                        & = 1       & \text{on }  & \hat\Gamma_{2},  \\
        \epsilon^\frac{3}{2} \hat\nabla \hat\Phi_2 \cdot \mathbf{n}                    & = 0         & \text{on }  & \partial\hat\Omega\setminus\hat\Gamma_{2},
    \end{align}
\end{subequations}
where
\begin{align}
    \label{eq:delta}
    \delta = \frac{\alpha F L^2 a_N i_0 C}{R T C_\mathrm{ref}}\left(\frac{1}{\sigma_0} + \frac{1}{\kappa_0}\right),
\end{align}
is the ratio of ohmic and kinetic resistances (it is an inverse Wagner number \citep{fuller2018electrochemical}),
\begin{align}
    \tau & = \frac{\kappa_0}{\sigma_0},\label{eq:tau} 
\end{align}
is the ratio of liquid and solid conductivities and
\begin{align}
    \label{eq:mu}
    \mu & = \frac{\alpha F L}{R T \kappa_0 } \frac{I}{A},
\end{align}
is the dimensionless applied current density.
Finally, the nondimensionalized Butler-Volmer term is
\begin{equation}
    \label{eq:ihat}
    \hat i = \gfp\left[\exp\left(\mu\Delta\hat\Phi\right)-\exp\left(-\mu\Delta\hat\Phi\right)\right],
\end{equation}
with $\Delta\hat\Phi = \hat{\Phi}_1 - \hat{\Phi}_2$.
Note that in the linear regime of Butler-Volmer, $\mu$ cancels in the right-hand sides of \eqref{eq:redox_Phi1} and \eqref{eq:redox_Phi2}.

\subsection{Supercapacitor electrode}
\newcommand{\phimax}{\Phi_{\text{max}}}
We also model a porous electrode operating under transient conditions. We again focus on a half-cell and specifically consider a single porous electrode of an EDLC, also commonly referred to as a supercapacitor \citep{fuller2018electrochemical}. This serves as a model, transiently operated electrochemical device and can be extended to battery simulation. The latter involves added complexity in the governing equations, but the methodology for applying topology optimization to a transient electrochemical system is nevertheless well illustrated using an EDLC as a model. The supercapacitor behavior is governed by \eqref{eq:porous_electrode_problem} and a time-dependent current transfer at the electrode-electrolyte interface
\begin{equation}
    i_n = C_d \frac{\partial(\Phi_{1}-\Phi_{2})}{\partial t},
    \label{eq:current_cap}
\end{equation}
due to charge accumulation at the electrical double layer, where $C_d$ is the double layer capacitance \citep{newman1975porous}. 

We apply a charging current to the electrode by specifying a time-dependent ionic potential
\begin{align}
    g = \nu t ~\text{for}~ t\in[0, \phimax/\nu],
    \label{eq:charging_pot}
\end{align}
on $\Gamma_{2}$, i.e. $B_p=1$ and $B_c=0$ in \eqref{eq:robinbc}, and ground the electronic potential at $\Gamma_{1}$, cf. \eqref{eq:dirichlet}.
The rest of the boundary is electrically insulated, cf. \eqref{eq:noflux1} and \eqref{eq:noflux2}.
The charging rate $\nu$ and the maximum potential $\phimax$ modulate the input power and energy into the system.
The timescales in the system are the characteristic charging time $\frac{a_N C_{d} L^2}{\kappa_0}$ and the input charging time $\frac{\phimax}{\nu}$.

Replacing the expressions \eqref{eq:current_cap}, \eqref{eq:charging_pot} and the nondimensional variables
\newcommand{\hphi}{\hat{\Phi}}
\begin{align}
    \hphi_1          & = \frac{\Phi_1}{\phimax}\,,                    \\
    \hphi_2          & = \frac{\Phi_2}{\phimax} \,,                   \\
    \hat{t}          & = \frac{t}{\frac{a_N C_{d} L^2}{\kappa_0}} \,, \\
    \hat{\mathbf{x}} & = \frac{\mathbf{x}}{L}\,,
\end{align}
in \eqref{eq:porous_electrode_problem}, we obtain the problem: Find $\hphi_1$ and $\hphi_2$ such that
\newcommand{\hnabla}{\hat{\nabla}}
\begin{subequations}\label{eq:supercap_eq}
    \begin{alignat}{2}
        -\hnabla \cdot \left( (1 - \epsilon)^{3/2} \hnabla \hphi_{1} \right)    & = - \tau \gamma\frac{\partial(\hphi_{1}-\hphi_{2})}{\partial \hat{t}} &  & \text{ in } \hat\Omega,                             \label{eq:supercap_Phi1} \\
        -\hnabla \cdot \left(\epsilon^{3/2} \hnabla \hphi_{2} \right) & =  \gamma\frac{\partial(\hphi_{1}-\hphi_{2})}{\partial \hat{t}}       &  & \text{ in }\hat\Omega,                              \label{eq:supercap_Phi2} \\
        \hat\Phi_1                                               & = 0                                                                   &  & \text{ on } \hat\Gamma_1,                            \\
        (1-\epsilon)^{3/2}\hat\nabla \hat\Phi_1 \cdot \mathbf{n} & = 0                                                                   &  & \text{ on } \partial\hat\Omega\setminus\hat\Gamma_1, \\
        \hat\Phi_2                                               & = \xi  \hat{t}                                                        &  & \text{ on }\hat\Gamma_2,                             \\
        \epsilon^{3/2} \hat\nabla \hat\Phi_2 \cdot \mathbf{n}    & = 0                                                                   &  & \text{ on } \partial\hat\Omega\setminus\hat\Gamma_2,
    \end{alignat}
\end{subequations}
for $\hat t\in[0, 1/\xi]$ and $\hphi_1=\hphi_2=0$ as initial conditions.
Only two nondimensional parameters determine $\hphi_1$ and $\hphi_2$:
$\tau$ as in \eqref{eq:tau}, the ratio of electrolyte conductivity to electrode conductivities, and
\begin{align}
    \xi & = \frac{a_N C_{d} L^2 / \kappa_0}{\phimax / \nu},
    \label{eq:xi}
\end{align}
the ratio of the characteristic time for charging to the total charging time.

\section{Implementation}\label{sec:implementation}
The two potential equations for both the porous redox electrode and the supercapacitor
are solved using the finite element library \texttt{Firedrake} \citep{rathgeber2016firedrake}, which uses \texttt{PETSc} \citep{petsc-user-ref} as the backend
for the linear algebra. The sensitivities are automatically derived by pyadjoint \citep{mitusch2019dolfin}.
\rme{The results of this paper can be reproduced using TOPE \citep{thomas_roy_2022_6366849}.}
The MMA algorithm \citep{svanberg1987method} solves the optimization problems via the \texttt{Python}
implementation \texttt{pyMMAopt} \citep{miguel_salazar_de_troya_2021_4456055, salzardetroya2021}.
\rtwo{We consider the optimized designs to have converged at three hundred iterations, as shown in Section \ref{sec:convergence}.}

Both problems are discretized using piecewise linear finite elements on triangular (2D) or tetrahedral (3D) meshes.
The porous redox electrode problem is linearized using Newton's method with an $L^2$ norm linesearch.
The supercapacitor equation is integrated in time using a backward Euler scheme,
and the cost function $\theta_0^\mathrm{sp}$ from \eqref{eq:supercap_problem}, with a trapezoidal scheme.
We integrate in time using 200 time steps for all simulations.
An adaptive scheme is more efficient, but we leave it for future work.
\rtwo{The calculation of sensitivities for large-scale transient problems can run into memory bottlenecks when using the adjoint method.
Indeed, the method requires saving the entire state of the forward problem, i.e. the potentials at each time step, to calculate the adjoint variable.
We did not encounter this problem, but future research with larger problems can use checkpointing schemes such as in \cite{revolve} or \cite{zhang2022petsc} to overcome it.}

For both the porous redox electrode and the supercapacitor, the linear system
of equations resulting from the finite element discretization is of the form
\begin{equation}\label{eq:block}
    \begin{bmatrix}
        A_{11} & A_{12} \\
        A_{21} & A_{22} \\
    \end{bmatrix}
    \begin{bmatrix}
        \phi_1 \\
        \phi_2
    \end{bmatrix}
    =
    \begin{bmatrix}
        r_1 \\
        r_2
    \end{bmatrix},
\end{equation}
where $A_{11}$ is the electronic potential coefficients, $A_{22}$, the ionic
potential coefficients, and $A_{12}$ and $A_{21}$, their respective couplings.
Correspondingly, we denote the electronic potential unknowns and residual by
$\phi_1$ and $r_1$, respectively, and the ionic potential unknowns and
residual by $\phi_2$ and $r_2$.

Instead of solving this linear system using a direct method, we consider a preconditioned iterative method \citep{wathen2015preconditioning} to improve scalability and solution time, especially for 3D simulations.
The structure of the system enables preconditioning approaches where iterative methods are used for the different blocks.

First, the system \eqref{eq:block} \rtwo{can easily be made} symmetric for both the redox and supercapacitor problems \rtwo{by multiplying the ionic potential equations \eqref{eq:redox_Phi2} and \eqref{eq:supercap_Phi2} by $\tau$, respectively}.
Second, the diagonal blocks $A_{11}$ and $A_{22}$ result from the discretization
of elliptic operators, making them ideal candidates for multigrid methods \citep{brandt1977multi}. We
thus use the Conjugate Gradient method with a blockwise symmetric Gauss-Seidel
preconditioner of the form
\begin{equation}\label{eq:precon}
    \begin{bmatrix}
        A_{11} & A_{12} \\
        0      & A_{22} \\
    \end{bmatrix}
    \begin{bmatrix}
        A_{11}^{-1} & 0           \\
        0           & A_{22}^{-1} \\
    \end{bmatrix}
    \begin{bmatrix}
        A_{11} & A_{12} \\
        0      & A_{22} \\
    \end{bmatrix}^\top.
\end{equation}
The diagonal blocks $A_{11}$ and $A_{22}$ are approximately inverted using a
single AMG V-cycle \citep{ruge1987algebraic} (\texttt{BoomerAMG}
\citep{henson2002boomeramg} from the \texttt{hypre} library \citep{falgout2002hypre}).

The PDE filter \eqref{eq:filter} is solved using a two-point flux approximation scheme described in Appendix \ref{sec:DG0}.
For the linear system resulting from the discretization of the PDE filter, we
use the Conjugate Gradient method preconditioned with an AMG V-cycle.

All \texttt{PETSc} solver options are given in Appendix \ref{sec:solveroptions}.

\section{Optimal designs}\label{sec:results}
We now present optimal designs for both the porous redox and supercapacitor/EDLC electrodes.
The design domain in 2D and 3D, Figure \ref{fig:diagramelectrode} and Figure \ref{fig:diagramelectrode3D}, respectively, receive the current or are subjected to an applied potential on $\Gamma_2$ and collect the current (ground the electronic potential) on $\Gamma_1$.
The computational domains are chosen such that the electrodes are twice as wide as they are thick and i.e. $W=2L$, while the membrane and current collector have lengths 75\% of the electrode width, i.e. $w=1.5L$.

In 2D, we simulate half of the total domain with \rme{a mesh of approximately $15,000$ triangular elements} and use the usual symmetry boundary conditions over the cut edges.
We use a similar approach in 3D to only simulate a quarter of the domain with \rme{a mesh of approximately $6,000,000$ tetrahedral elements,} and symmetry boundary conditions over the cut surfaces.

The first examples in both following subsections consider the original Bruggeman correlation in the effective ionic conductivity, cf.  \eqref{eq:kappaeff}.
The second set of examples uses the modified Bruggeman correlation in \eqref{eq:porosity_two} with $f_p=0.02^{2/3}$, which corresponds to the lowest factor in \cite{Madabattula_2020}.
\rone{The filtered density $\gfp$ is obtained using the PDE filter \eqref{eq:filter} with a filter radius of $r = 0.01$.}

\subsection{Porous redox electrode}
We first design a porous redox electrode to minimize the ionic potential at the membrane for a fixed current density, i.e.
\begin{equation}\label{eq:chemobj}
    \begin{aligned}
        \min_{\gamma\in [0,1]} \theta^{pe}_0 & = \int_{\Gamma_2} \hat{\Phi}_2 \diff s ,                      \\
        \text{s.t. }                          & \hphi_1,\hphi_2~\text{satisfy \eqref{eq:redox_eq}}~, \\
    \end{aligned}
\end{equation}
which is equivalent to driving the overpotential to zero, defined as the excess necessary potential to drive the Faradaic reaction, $\eta = \Phi_1\vert_{\Gamma_1} - \Phi_2\vert_{\Gamma_2} - U_0$.
Regardless of the application, the most power efficient operation of this half-cell occurs when the cell operates as closely as possible to the thermodynamic potential.
As evident in the expression, as $\eta \rightarrow 0$, the potential difference in the half-cell approaches its thermodynamic limit.
This minimization problem is thus equivalent to maximizing the power efficiency of the electrode.

Optimized designs are shown for various values of the conductivity ratio $\tau$ \eqref{eq:tau}, the inverse Wagner number $\delta$ \eqref{eq:delta}, and the dimensionless current density $\mu$ \eqref{eq:mu}.
\rme{The optimized designs start with an initial uniform $\gamma=0.5$ everywhere in the domain.}

\rtwo{Running on one core of an Intel Xeon E5-2695 v4 CPU, each optimization iteration (forward and adjoint problem) took between 0.9 to 1.5 seconds, depending on required the number of Newton iterations (1 to 5).}

\newcommand{\pecost}[1]{$\theta^{\text{pe}}_0=#1$}
\newcommand{\figurestablepe}[8]{
    \renewcommand\theadset{\def\arraystretch{.85}}%
    \setlength{\extrarowheight}{2pt}
    \centering
    \begin{tabular}{llll}
        \Xhline{2\arrayrulewidth}
                                        & \multicolumn{3}{c}{$\delta$}                                                               \\
        \cline{3-4}
                                        &                                 & \multicolumn{1}{c}{1}           & \multicolumn{1}{c}{25} \\
        \Xhline{2\arrayrulewidth}
        \multirowcell{4}[-7em]{$\mu$}   &
        \multirow{2}{*}[-4em]{0.1}      &
        \multicolumn{1}{c}{\pecost{#5}} & \multicolumn{1}{c}{\pecost{#6}}                                                            \\[-1em]
                                        &                                 &  \begin{minipage}[t]{0.03\linewidth} \vspace{0pt}  (a)  \end{minipage}  \begin{minipage}[t]{0.33\linewidth}\vspace{0pt} #1 \end{minipage}                         & \begin{minipage}[t]{0.03\linewidth} \vspace{0pt}  (b)  \end{minipage}  \begin{minipage}[t]{0.33\linewidth}\vspace{0pt} #2 \end{minipage}                      \\[-1em]
                                        & \multirow{2}{*}[-4em]{5}
                                        & \multicolumn{1}{c}{\pecost{#7}} & \multicolumn{1}{c}{\pecost{#8}}                          \\[-1em]
                                        &                                 & \begin{minipage}[t]{0.03\linewidth} \vspace{0pt}  (c)  \end{minipage}  \begin{minipage}[t]{0.33\linewidth}\vspace{0pt} #3 \end{minipage} & \begin{minipage}[t]{0.03\linewidth} \vspace{0pt}  (d)  \end{minipage}  \begin{minipage}[t]{0.33\linewidth}\vspace{0pt} #4 \end{minipage}                     \\[-1em]
        \Xhline{2\arrayrulewidth}
    \end{tabular}
}

\subsubsection{Original Bruggeman correlation}
\label{sec:redox_original}

\begin{figure*}[!htbp]
    \figurestablepe{
        \includegraphics[width=\linewidth]{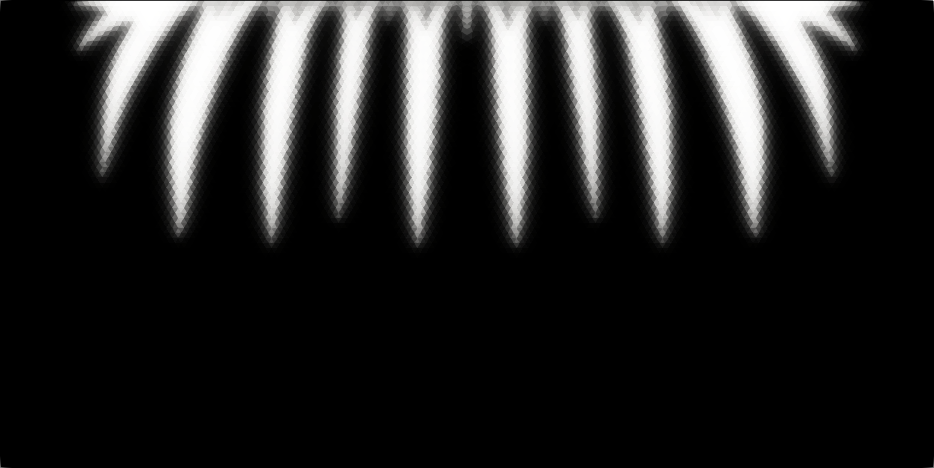}
    }{
        \includegraphics[width=\linewidth]{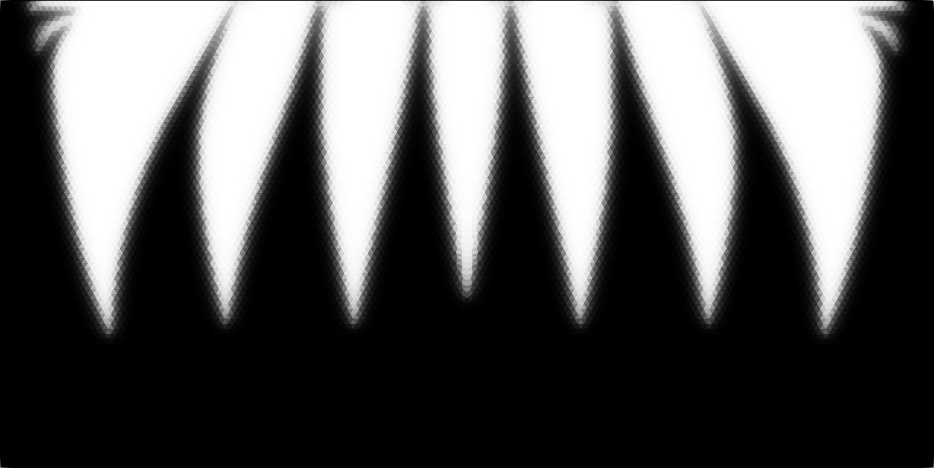}
    }{
        \includegraphics[width=\linewidth]{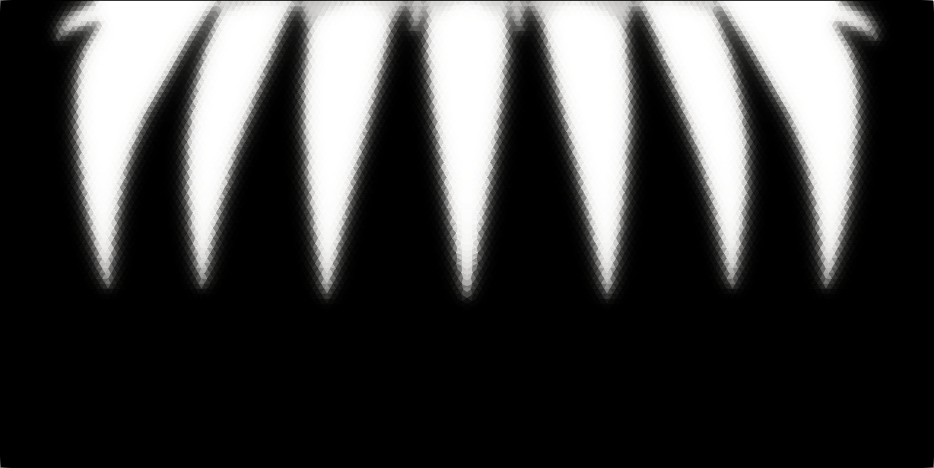}
    }{
        \includegraphics[width=\linewidth]{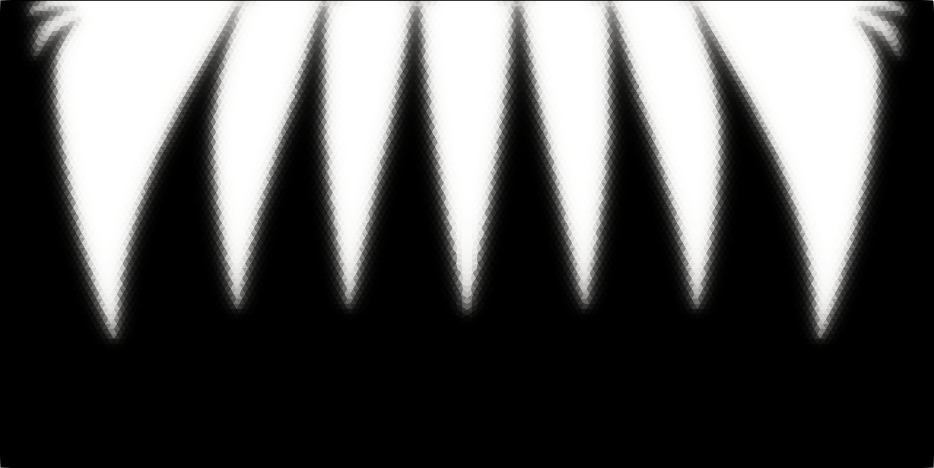}
    }{2.500}{1.3603}{1.9763}{1.3576} %
    {\phantomsubcaption \label{tab:pe_simple1a}}
    {\phantomsubcaption \label{tab:pe_simple1b}}
    {\phantomsubcaption \label{tab:pe_simple1c}}
    {\phantomsubcaption \label{tab:pe_simple1d}}
    \caption{Optimized porous redox electrode designs considering the original Bruggeman correlation and $\tau=0.5$. Black is $\tilde{\gamma} = 1$; white is $\tilde{\gamma} =0$. }
    \label{tab:pe_simple1}
\end{figure*}

\begin{figure*}[!htbp]
    \figurestablepe{
        \includegraphics[width=\linewidth]{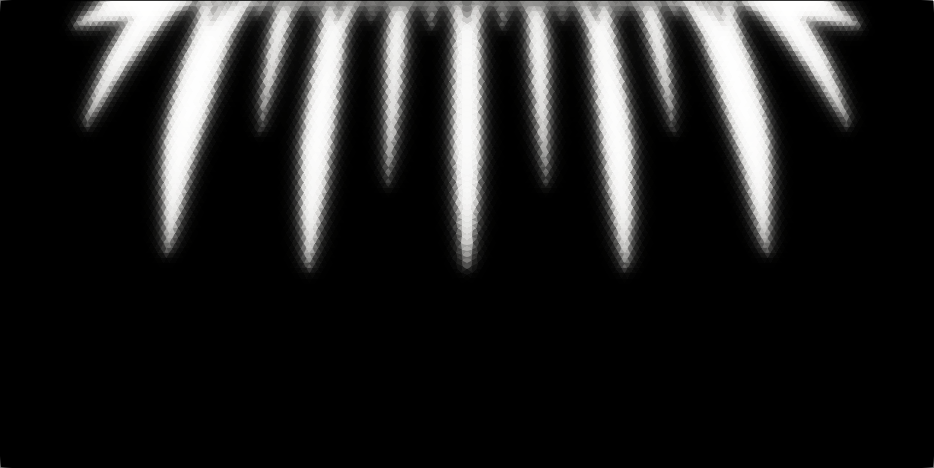}
    }{
        \includegraphics[width=\linewidth]{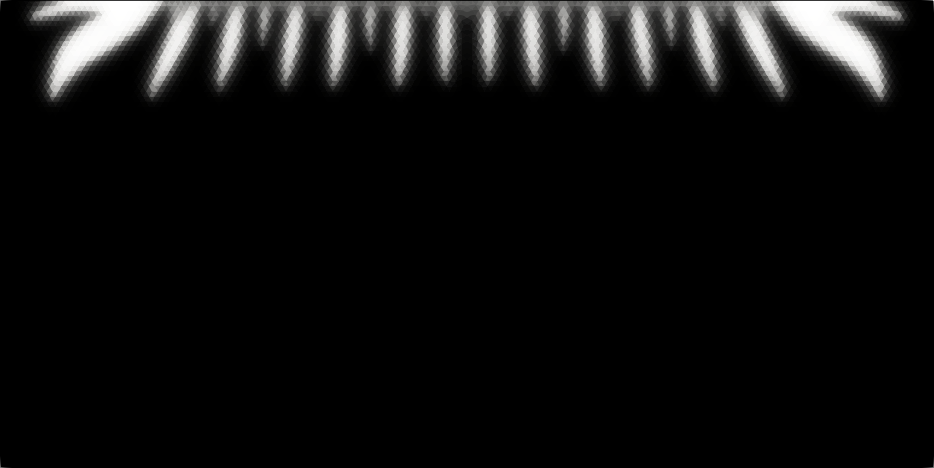}
    }{
        \includegraphics[width=\linewidth]{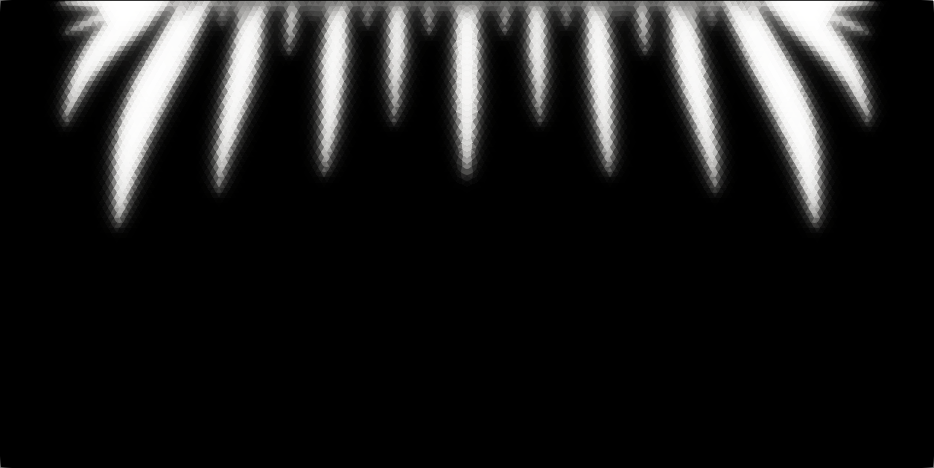}
    }{
        \includegraphics[width=\linewidth]{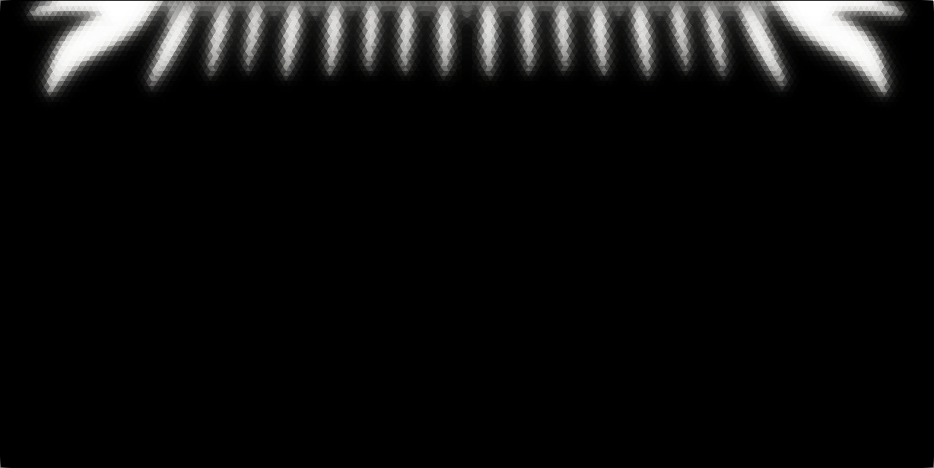}
    }{1.7045}{0.6166}{1.2471}{0.6052}
    {\phantomsubcaption \label{tab:pe_simple2a}}
    {\phantomsubcaption \label{tab:pe_simple2b}}
    {\phantomsubcaption \label{tab:pe_simple2c}}
    {\phantomsubcaption \label{tab:pe_simple2d}}
    \caption{Optimized porous redox electrode designs considering the original Bruggeman correlation and $\tau=0.1$. Black is $\tilde{\gamma} = 1$; white is $\tilde{\gamma} =0$.}
    \label{tab:pe_simple2}
\end{figure*}
The optimized designs with the original Bruggeman correlation \eqref{eq:kappaeff} in Figures \ref{tab:pe_simple1} and \ref{tab:pe_simple2} present a sharp teeth-like pattern.
These teeth facilitate ionic transport since the effective ionic conductivity is higher in the electrolyte-only phase.

We note that for $\epsilon_N = 0.5$, the effective ionic conductivity $\kappa$ \eqref{eq:kappaeff} in the electrolyte-only phase ($\gfp=0$) is greater than the effective electronic conductivity $\sigma$ \eqref{eq:sigmaeff} in the porous phase ($\gfp=1$) when $\tau > 0.5^{1/1.5}$.
This implies that, with respect to reducing ohmic losses, the electrolyte-only phase is preferred over the porous phase when $\tau > 0.5^{1/1.5}$. Manifestly, reducing $\tau$ increases the proportion of the domain occupied by the porous phase. A further reduction of $\tau$ naturally results in a greater dominance of the porous phase.

On a similar note, increasing $\delta$ augments the relative importance of ohmic losses.
Consequently, an increase in $\delta$ corresponds with a preference for the material with less ohmic resistance.
For $\tau=0.5$, this means an increase in electrolyte-only phase, and for $\tau=0.1$, an increase in the porous phase.

The parameter $\delta$ also relates to electrode penetration depth \citep{fuller2018electrochemical}.
A small $\delta$ means that the reaction happens throughout the electrode, while a large $\delta$ means that the reaction happens closer to the membrane (or here closer to the electrolyte-only phase), due to increased ohmic resistance.
In the absence of reactant consumption, a significant portion of the electrode is essentially unused for larger $\delta$ as observed in Figures \ref{tab:pe_simple2b} and \ref{tab:pe_simple2d}.
In that case, a thinner electrode would be more efficient, although here the electrode is forced to fill the domain to retain electrical contact.

The parameter $\mu$ represents the nondimensional current density at the membrane.
In the linear regime of the Butler-Volmer equation \eqref{eq:ihat}, $\mu$ can be cancelled and is therefore assumed to have little effect over the optimal design.
In the nonlinear regime, however, increasing it results in a greater reaction rate at a fixed potential difference in the electrode. Equivalently, at fixed currents the potential drop will be lower.
We can investigate the nonlinearity of the Butler-Volmer equation by looking at the magnitude of $\hat i$ in \eqref{eq:ihat}: a larger value, say greater than 2, indicates that we are in the nonlinear regime.
For the chosen parameters, the cases with $\mu=0.1$ are in the linear regime, while for $\mu=5$, the $\delta=25$ cases are slightly nonlinear and the $\delta=1$ cases are very nonlinear.
It is clear that increasing $\mu$ increases the nonlinearity of $\hat i$.
As for $\delta$, reducing it leads to a decreased ohmic resistance. We thus observe larger potential differences and move further towards the nonlinear regime.
\rtwo{As another indication of nonlinearity, we can look at the number of Newton iterations. The numbers of iterations required for solving the system for the initial design ($\gfp = 0.5$ everywhere) are given in Table \ref{tab:newtons}. We observe that increasing $\mu$ and decreasing $\delta$ increases the number of Newton iterations.}
\begin{table}[htb!]
    \caption{\rtwo {Newton iterations for solving the original Bruggeman system on a uniform design.}}
    \label{tab:newtons}
    \centering
    \begin{subtable}{.3\linewidth}
        \centering
        \subcaption*{$\tau=0.5$}
        \begin{tabular}{ ccc}
            \multirow{2}{*}{$\mu$} & \multicolumn{2}{c}{$\delta$} \\
            \cline{2-3}
            &  1 & 25 \\
            \hline
            0.1 & 2 & 1 \\
            5   & 5 & 3 \\
        \end{tabular}
    \end{subtable}
    \begin{subtable}{.3\linewidth}
        \centering
        \subcaption*{$\tau=0.1$}
        \begin{tabular}{ cc}
            \multicolumn{2}{c}{$\delta$} \\
            \cline{1-2}
                1 & 25 \\
            \hline
            2 & 1 \\
            4 & 2 \\
        \end{tabular}
    \end{subtable}
\end{table}

In the $\tau=0.5$ cases, an increase in $\mu$ results in a slight increase in the electrolyte-only phase, the favored phase in terms of effective ionic conductivity.
This increase is more pronounced for the $\delta=1$ case.
On the other hand, there is little change for $\tau=0.1$, especially for $\delta=25$.
Overall, the effect of $\mu$ is more significant for $\delta=1$, where the Butler-Volmer is more nonlinear.

\subsubsection{Modified Bruggeman correlation}

\begin{figure*}[!htbp]
    \figurestablepe{
        \includegraphics[width=\linewidth]{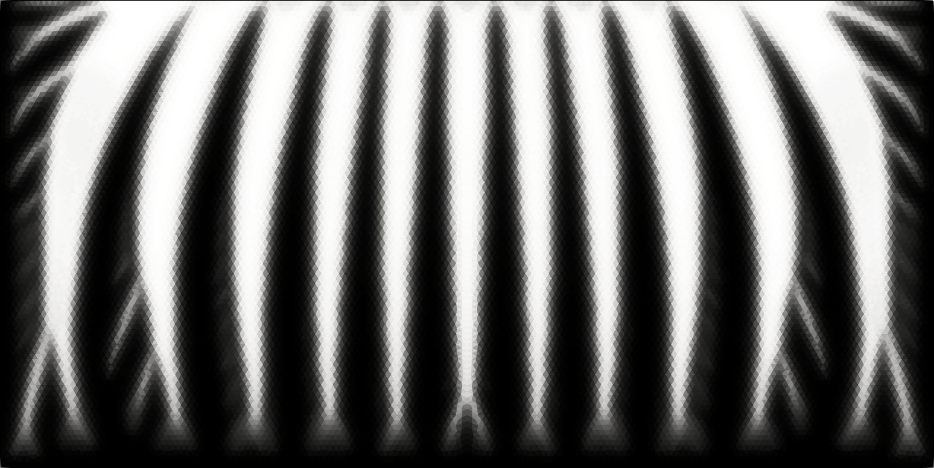}
    }{
        \includegraphics[width=\linewidth]{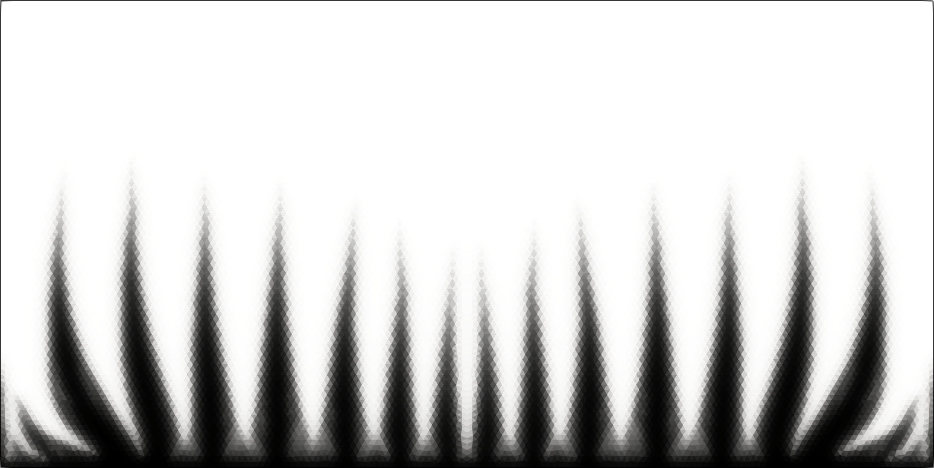}
    }{
        \includegraphics[width=\linewidth]{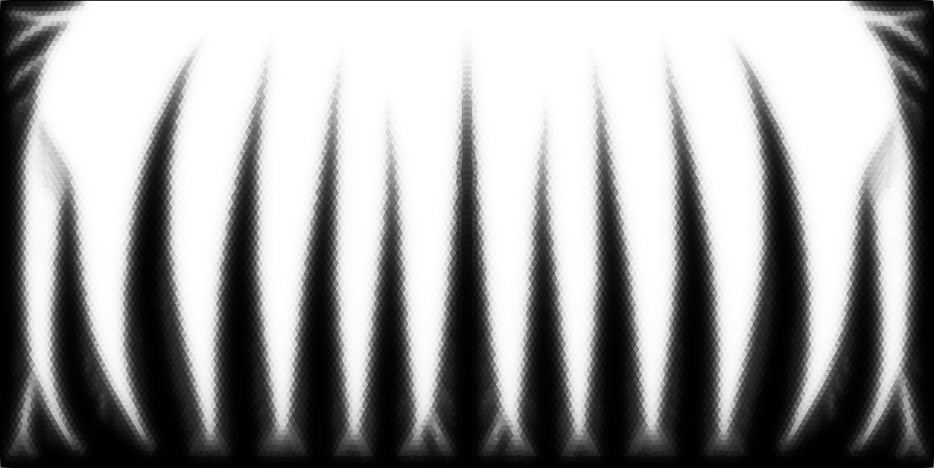}
    }{
        \includegraphics[width=\linewidth]{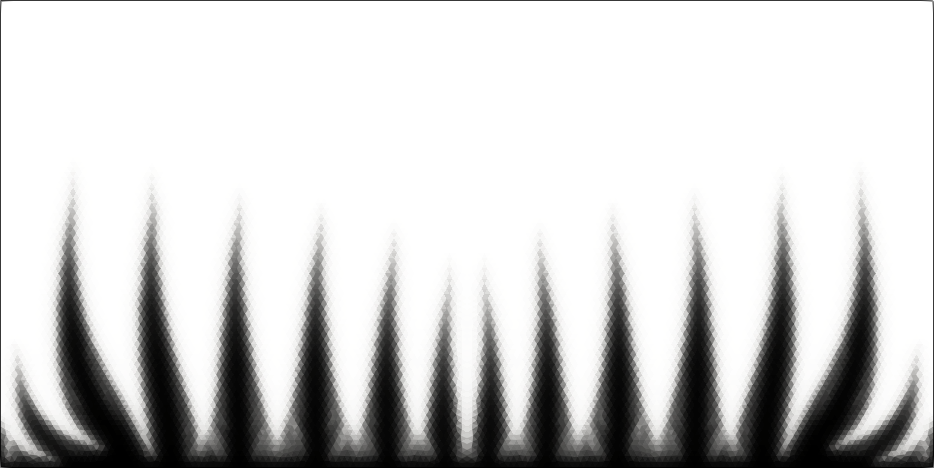}
    }{3.2818}{1.6348}{2.3973}{1.6263}
    {\phantomsubcaption \label{tab:pe_eff1a}}
    {\phantomsubcaption \label{tab:pe_eff1b}}
    {\phantomsubcaption \label{tab:pe_eff1c}}
    {\phantomsubcaption \label{tab:pe_eff1d}}
    \caption{Optimized porous redox electrode designs considering the modified Bruggeman correlation and $\tau=0.5$. Black is $\tilde{\gamma} = 1$; white is $\tilde{\gamma} =0$.}
    \label{tab:pe_eff1}
\end{figure*}

\begin{figure*}[!htbp]
    \figurestablepe{
        \includegraphics[width=\linewidth]{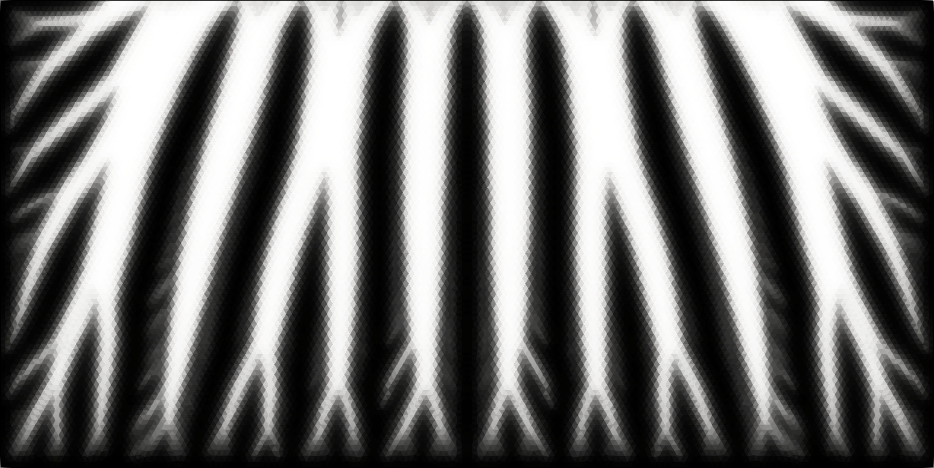}
    }{
        \includegraphics[width=\linewidth]{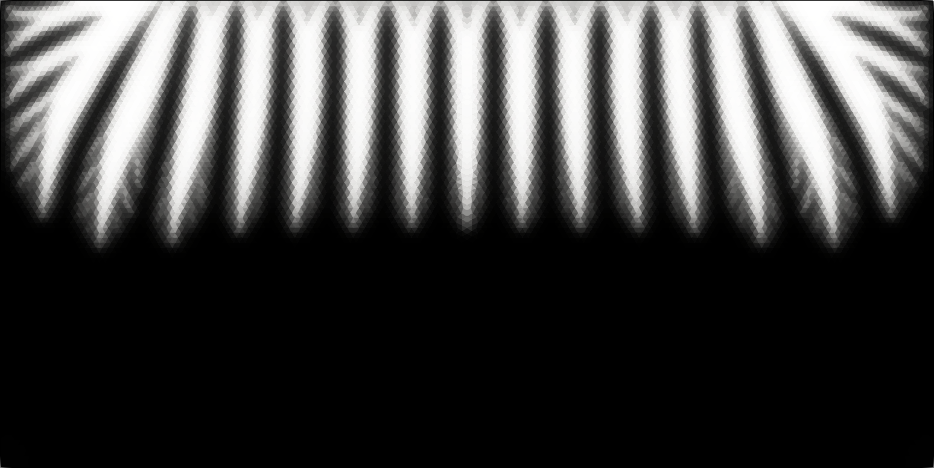}
    }{
        \includegraphics[width=\linewidth]{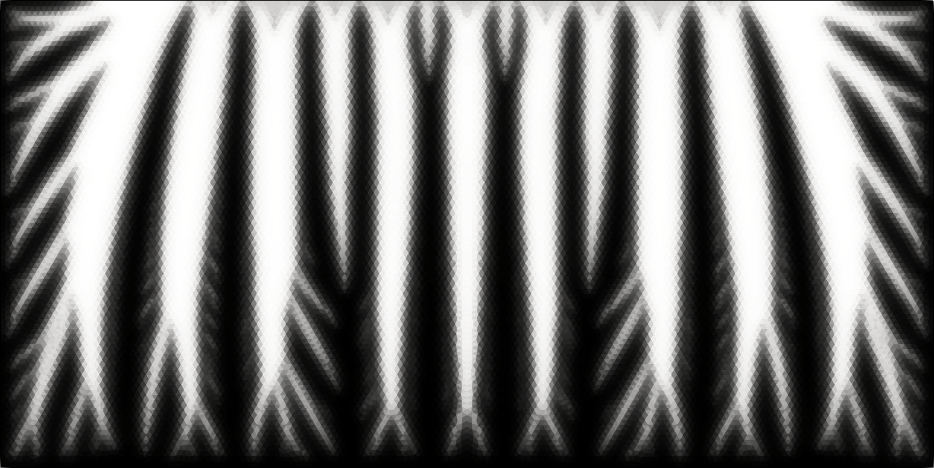}
    }{
        \includegraphics[width=\linewidth]{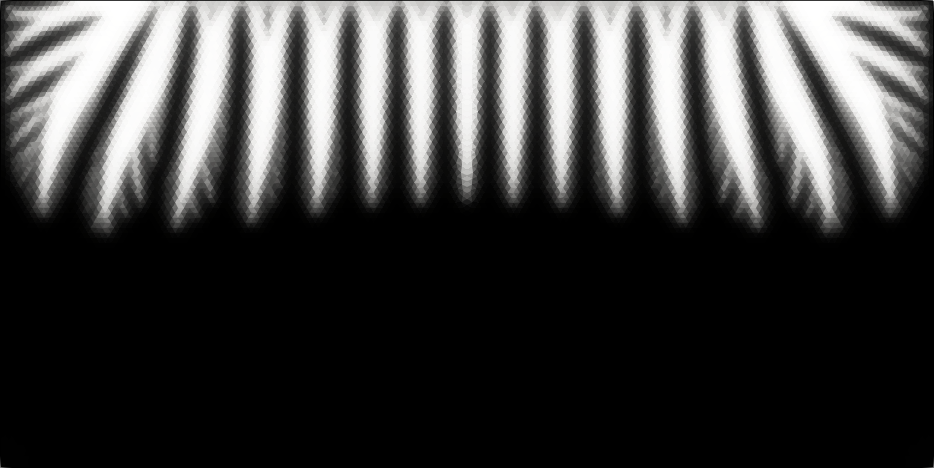}
    }{1.9665}{0.4580}{1.1844}{0.4289}
    {\phantomsubcaption \label{tab:pe_eff2a}}
    {\phantomsubcaption \label{tab:pe_eff2b}}
    {\phantomsubcaption \label{tab:pe_eff2c}}
    {\phantomsubcaption \label{tab:pe_eff2d}}
    \caption{Optimized porous redox electrode designs considering the modified Bruggeman correlation and $\tau=0.005$. Black is $\tilde{\gamma} = 1$; white is $\tilde{\gamma} =0$.}
    \label{tab:pe_eff2}
\end{figure*}

Using the modified Bruggeman correlation \eqref{eq:kappaeff_correction}, the designs in Figures \ref{tab:pe_eff1} and \ref{tab:pe_eff2} exhibit hierarchical root-like patterns, maintaining the teeth-like structure.
The smaller, root-like channels or macropores facilitate ionic transport deeper inside the porous phase to compensate for the lower ionic conductivity. Interestingly, the optimization algorithm converged to a multiscale solution to improve the objective, corroborating experimental efforts suggesting this same strategy \citep{wang20083d}.

Reducing $\tau$ has a different effect in this case.
The effective ionic conductivity in the porous phase is less than $1\%$ of its value in the electrolyte-only phase.
Reducing $\tau$ only favors the porous phase for $\delta=25$, i.e. when ohmic resistance is higher.
Additionally, a reduction of $\tau$ by a factor of 100 appears to increase the prevalence of root-like structures to facilitate ion transport.

As observed for the original Bruggeman correlation, an increase in $\delta$ corresponds with an increase in the phase that reduces ohmic losses, i.e. the electrolyte-only phase for $\tau=0.5$, and the porous phase for $\tau=0.005$.

Increasing $\mu$ also expands the electrolyte-only phase for $\tau=0.5$, especially for $\delta=1$ due to the nonlinearity of Butler-Volmer as discussed in Section \ref{sec:redox_original}.
However, there is little effect for smaller $\tau$.

\subsubsection{Convergence history}
\label{sec:convergence}

\begin{figure*}[!htbp]
    \centering
    \includegraphics[width=0.6\linewidth]{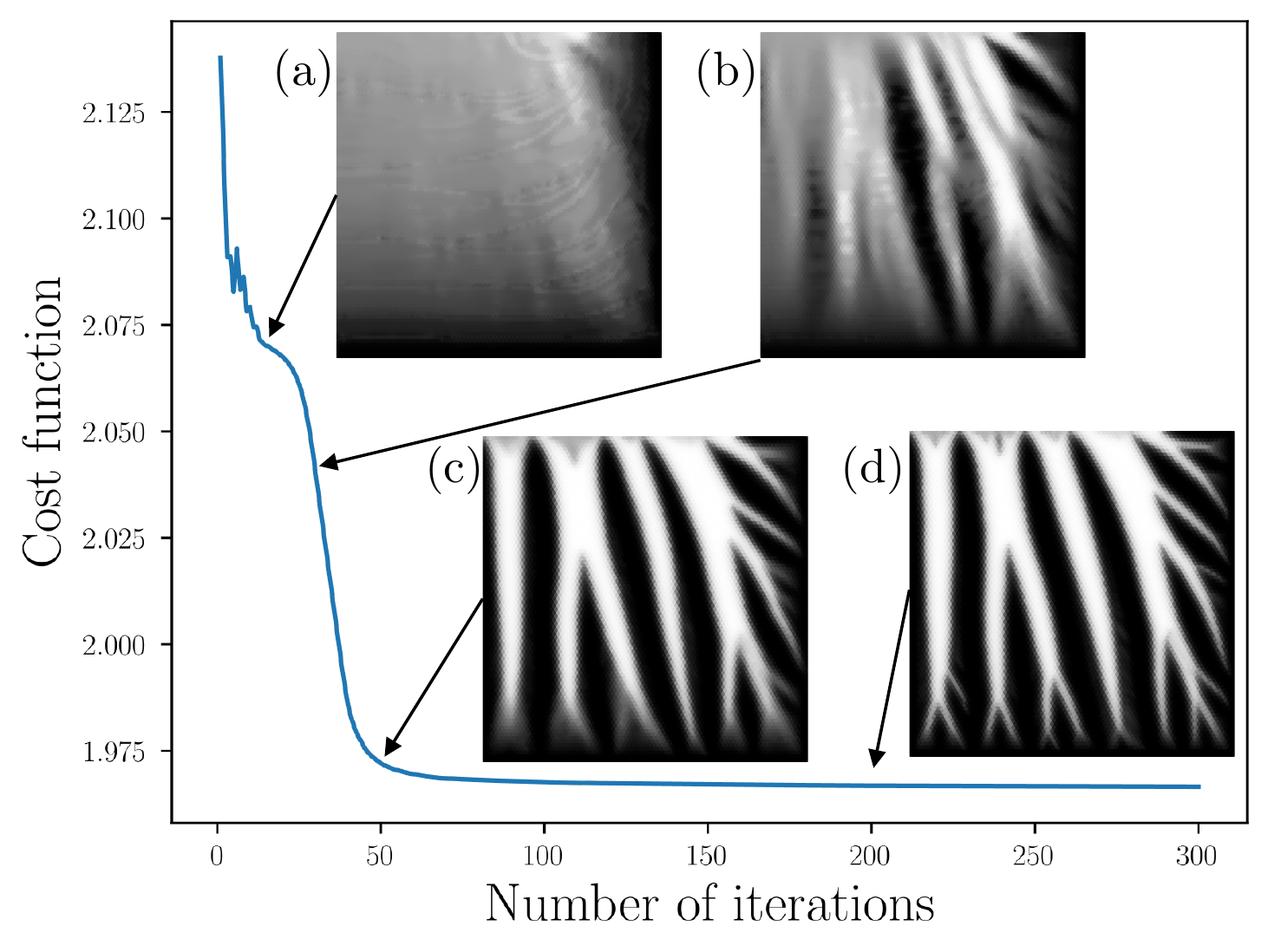}
    \caption{\rone{Cost function history for the redox electrode with $\delta=1, \mu=0.1$, $\tau=0.005$, and the lower effective conductivity. The snapshots are the design geometry at (a) 15, (b) 30, (c) 50, and (d) 200 iterations. Only the symmetric half of the designs are represented to save space in the figure.}}
    \label{fig:redox_history}
\end{figure*}
\rone{We briefly investigate the convergence of the optimization algorithm.}

\rone{
Considering the case with $\delta=1, \mu=0.1$, $\tau=0.005$, and the modified Bruggeman correlation,
we plot the cost function evolution in Figure \ref{fig:redox_history}.
The four insets in Figure \ref{fig:redox_history} represent the evolution of the optimized design.
We observe that most of the cost function reduction happens within the first 50 iterations.
The design at 200 iterations is almost identical to the one obtained after 300 iterations as illustrated in Figure \ref{tab:pe_eff2a}, a sign of convergence.
}

\subsubsection{Comparison to a monolithic electrode}

\begin{figure*}[!htbp]
    \centering
    \begin{multicols}{2}
    \centering
    \begin{subfigure}[b]{\linewidth}
        \centering
        Original Bruggeman correlation \\
        \includegraphics[width=0.49\linewidth]{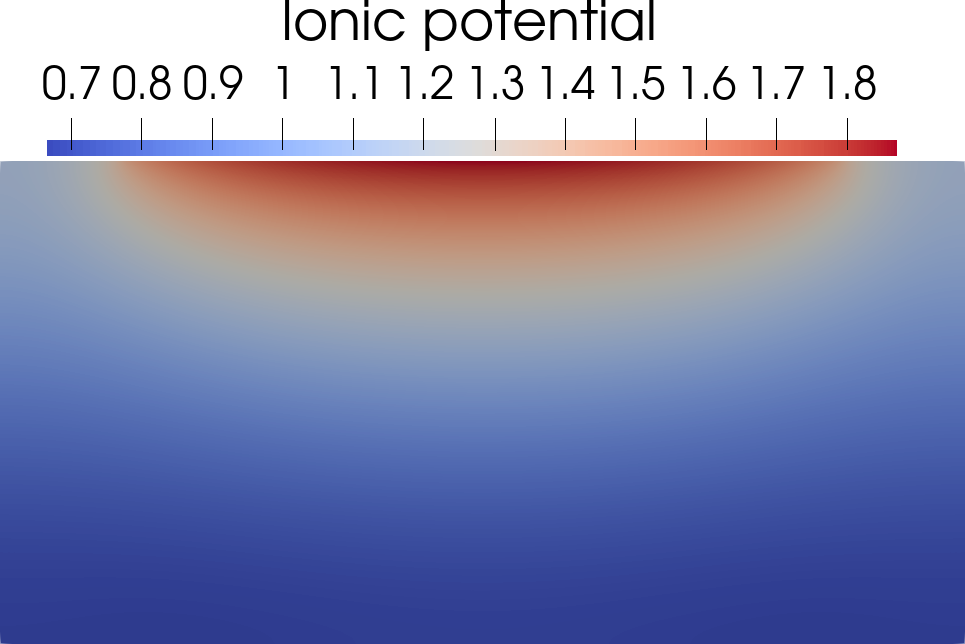}
        \includegraphics[width=0.49\linewidth]{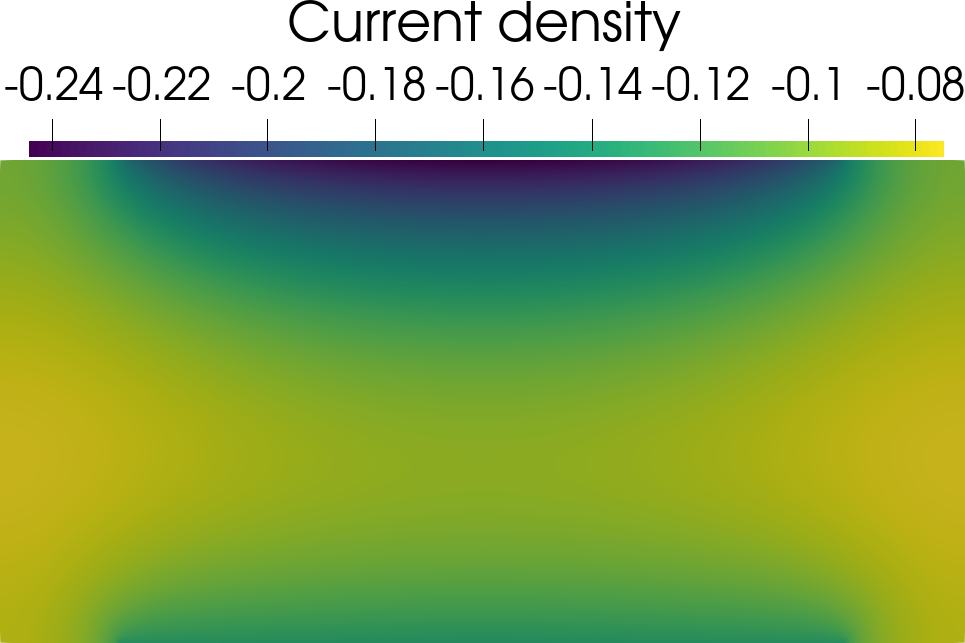}
        \caption{Monolithic electrode \pecost{2.646}}
        \label{fig:mono_simple}
    \end{subfigure}
    \par
    \begin{subfigure}[b]{\linewidth}
        \centering
        \includegraphics[width=0.49\linewidth]{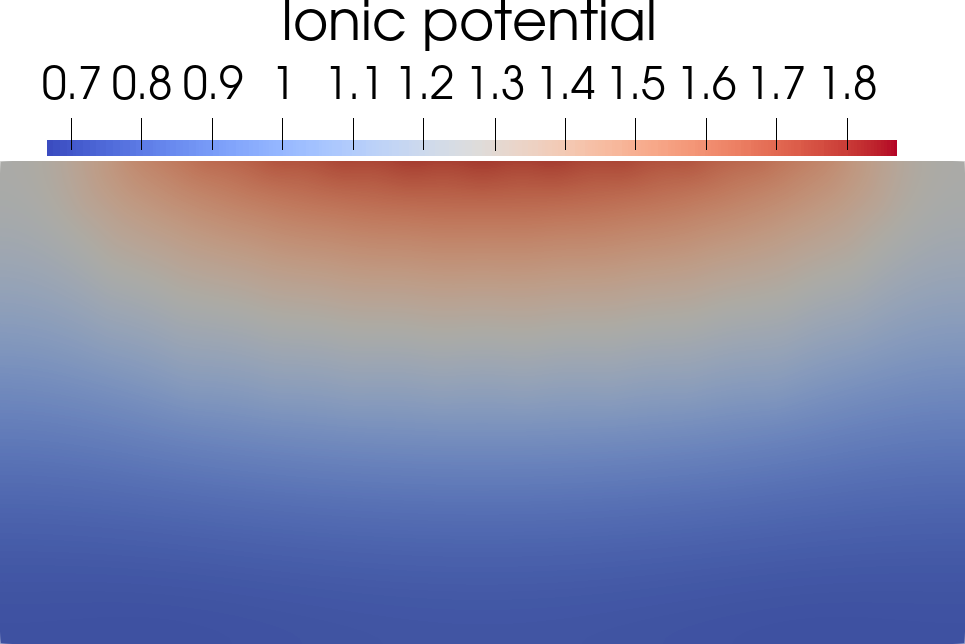}
        \includegraphics[width=0.49\linewidth]{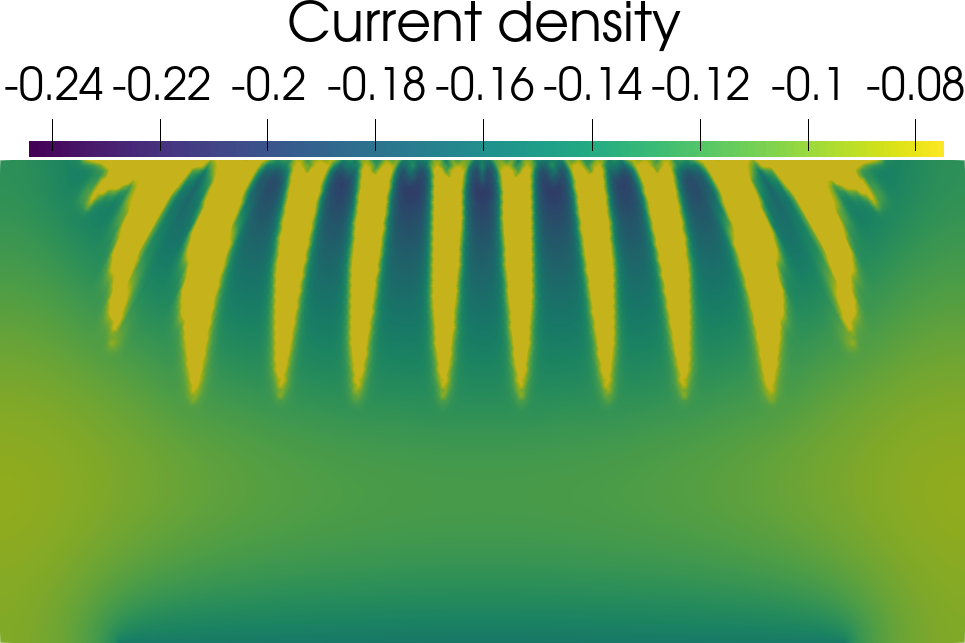}
        \caption{Designed electrode \pecost{2.5}}
        \label{fig:designed_simple}
    \end{subfigure}
    \par
    \begin{subfigure}[b]{\linewidth}
        \centering
        Modified Bruggeman correlation \\
        \includegraphics[width=0.49\linewidth]{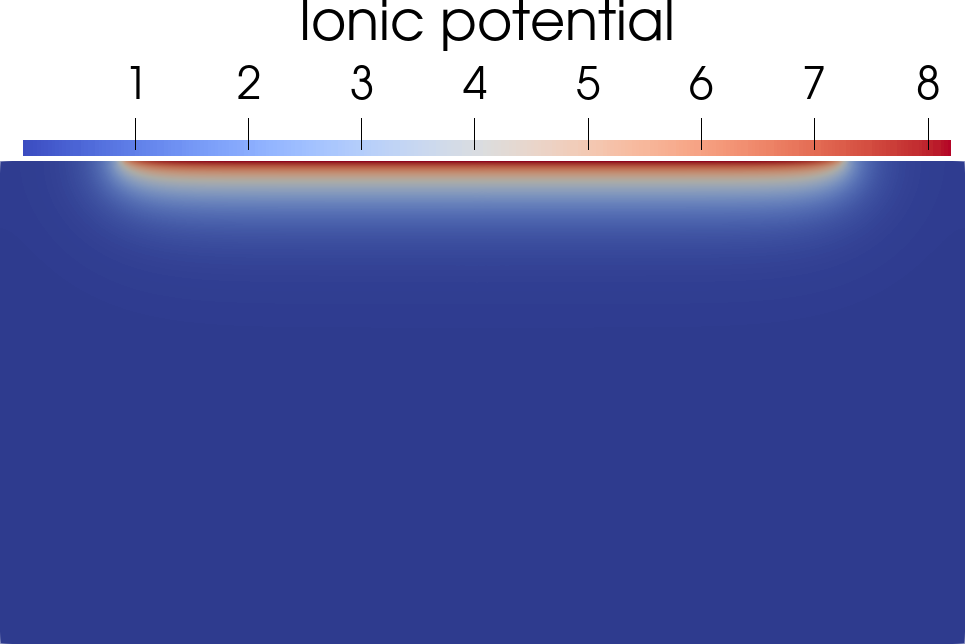}
        \includegraphics[width=0.49\linewidth]{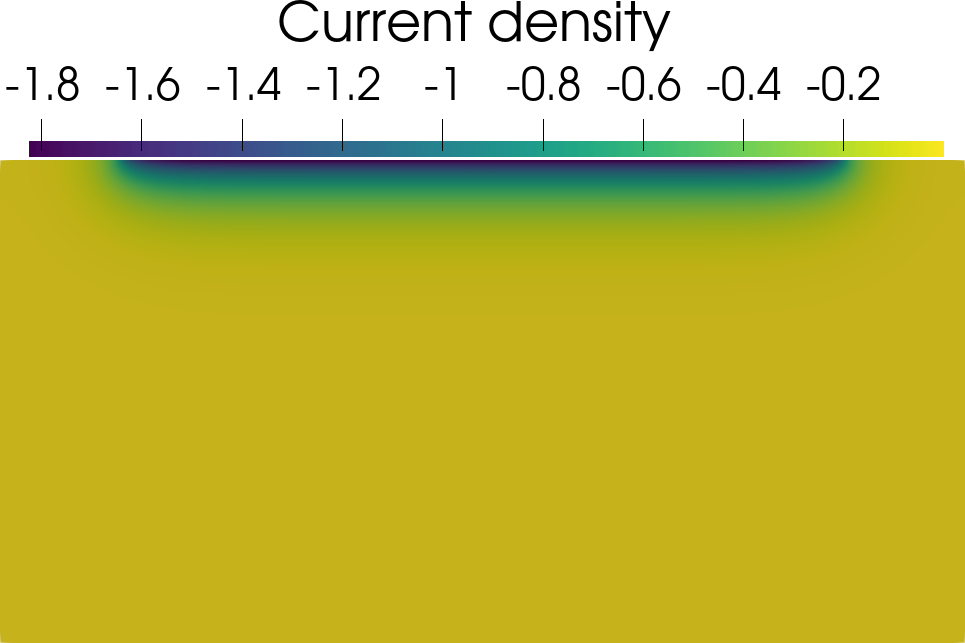}
        \caption{Monolithic electrode \pecost{11.9966}}
        \label{fig:mono_eff}
    \end{subfigure}
    \par
    \begin{subfigure}[b]{\linewidth}
        \centering
        \includegraphics[width=0.49\linewidth]{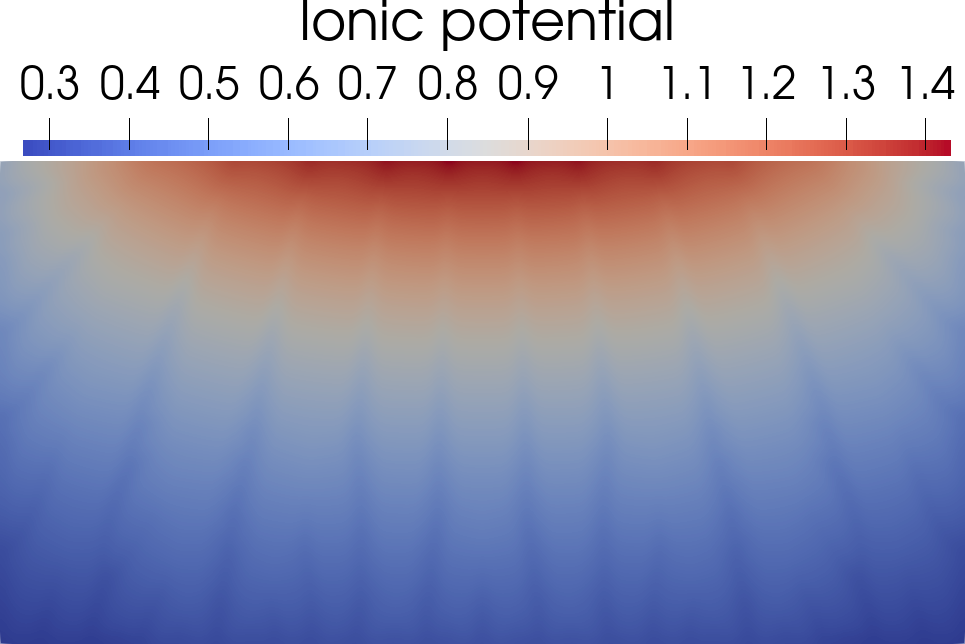}
        \includegraphics[width=0.49\linewidth]{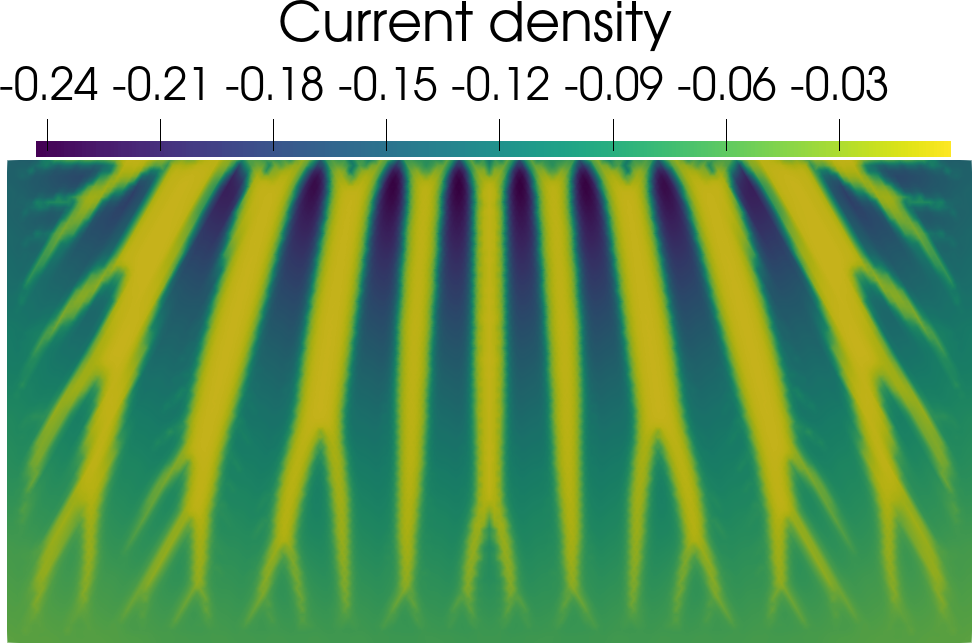}
        \caption{Designed electrode \pecost{1.9596}}
        \label{fig:designed_eff}
    \end{subfigure}
    \end{multicols}
    \caption{Comparison of performance between monolithic and designed porous redox electrodes. Using $\mu=0.1$, $\delta=1$, as well as $\tau=0.5$ and $\tau=0.005$ for the original and modified Bruggeman correlation, respectively.}
    \label{fig:performance}
\end{figure*}

We compare the optimized electrodes to monolithic electrodes, i.e. an electrode consisting solely of the porous phase ($\gfp=1$). For each electrode, we plot the nondimensional ionic potential $\hat\Phi_2$ as well as the nondimensional volumetric charge transfer current density $\hat i$ from \eqref{eq:ihat}, or current density for short.

In Figures \ref{fig:mono_simple}--\ref{fig:designed_simple}, we compare the design for $\delta=1, \mu=0.1$ and $\tau=0.5$ in Figure \ref{tab:pe_simple1a} to a monolithic electrode, both using the original Bruggeman correlation.
We first observe that the distribution of the ionic potential is very similar, with only slightly higher potential values at the membrane for the monolithic electrode, corresponding to a slightly higher cost function $\theta_0^\mathrm{pe}$.
Indeed, the designed electrode leads to a 5.5\% decrease in the average overpotential.
The current density distribution is also similar, but appears more concentrated for the monolithic electrode.
This indicates that spreading the reaction throughout the electrode is more efficient.

Now considering the modified Bruggeman correlation, we compare the designed electrode for $\delta=1, \mu=0.1$ and $\tau=0.005$ in Figure \ref{tab:pe_eff2a} to a monolithic electrode.
Due to the very small effective ionic conductivity in the monolithic electrode, the ionic potential and current density are highly concentrated at the top boundary, cf. Figures \ref{fig:mono_eff}--\ref{fig:designed_eff}.
In contrast, the ionic potential in the designed electrode spreads through the electrolyte-only channels, leading to a better distribution of the current density.
This leads to an 84\% reduction in the average overpotential.
Again there is an apparent benefit to spreading the reaction across the electrode.

In brief, to reduce the overpotential, the optimizer favors designs that spread the current density throughout the electrode.
When using the original Bruggeman correlation, teeth-like designs achieve this goal with a small increase in power efficiency.
On the other hand, the modified correlation with lower effective ionic conductivity leads to root-like designs that are significantly more power efficient.
This suggests that designing electrodes at multiple length scales is especially important in the case of lower effective ionic conductivity.

\subsection{Supercapacitor electrode}

We next pose the EDLC/supercapacitor electrode design problem as
\begin{equation}
    \begin{aligned}
        \min_{\gamma\in [0,1]}          \theta^{\text{sp}}_0(\gamma)= & \int_{0}^{1/\xi}\int_{\hat\Omega} \left(\varepsilon \nabla \hphi_2 \cdot \nabla \hphi_2 \right.                 \\
                                                                      & \left.+  \frac{\hat\varepsilon}{\tau}\nabla \hphi_1 \cdot \nabla \hphi_1 \right) \diff V \diff \hat t                                \\
        \text{s.t.}                                                   & \hphi_1,\hphi_2~\text{satisfy Equation \eqref{eq:supercap_eq}}~                                                   \\
        ~\theta^{\text{sp}}_1(\gamma) =                               & \int_{\hat\Omega} \frac{\tilde\gamma^p}{2}\left(\hphi_1 -\hphi_2\right)^2 \diff V \ge \Sigma E_{\text{max}}.
        \label{eq:supercap_problem}
    \end{aligned}
\end{equation}
This minimizes the ohmic losses in the electrode, i.e. $\theta^{\text{sp}}_0$, subject to a constraint to ensure a minimum amount of stored energy, i.e. $\theta^{\text{sp}}_1$, at the end of the dimensionless charging time, $1/\xi$. Equivalently, this minimizes the energy loss during the charge cycle of the system.

The stored energy is constrained to be greater than a factor $\Sigma$ of the maximum possible stored energy in the system
\begin{equation}
    E_{\text{max}} = \int_{\hat\Omega}  \frac{1}{2} \diff V\,.
\end{equation}
This corresponds to the domain being filled with porous material (i.e., $\tilde{\gamma} = 1$ and the domain is filled with only Material $N$) and the electrode being completely charged such that, $\hphi_1 - \hphi_2 = 1$ at all points in the domain.

The penalization schemes for the nondimensional ionic conductivity
\begin{equation}
    \varepsilon = \tilde\gamma\left((f_p\epsilon_N)^{\frac{3}{2}} - 1\right) + 1,
    \label{eq:penalization_loss}
\end{equation}
and the electronic conductivity
\begin{equation}
    \hat\varepsilon = \tilde\gamma\left(1 - \epsilon_N^{\frac{3}{2}} \right),
    \label{eq:penalization_stored}
\end{equation}
result in intermediate volume fraction values with higher current density than in \eqref{eq:supercap_eq}.
Similarly, these values have a lower contribution to the energy stored calculation due to the penalization \rme{$\tilde\gamma^p$}.
This strategy leaves intermediate values undesirable, and they are thus removed by the optimizer.
\rme{An ultimate value of $p=3$ is desired for proper penalization.
However, starting the optimization process with this value can cause convergence to unfavorable local minima.
To circumvent this, a continuation strategy can be used: starting the optimizer with $p=1$ and changing to $p=3$ after 100 iterations.
This continuation strategy leads to better optimized designs for the higher energy storage cases, but can create undesirable features for the lower energy storage cases.
Therefore, we only use the continuation strategy for $\Sigma = 0.5$.}

\newcommand{\spcost}[1]{$\theta^{\text{sp}}_0=#1$}
\newcommand{\figurestable}[8]{
    \renewcommand\theadset{\def\arraystretch{.85}}%
    \setlength{\extrarowheight}{2pt}
    \centering
    \begin{tabular}{llll}
        \Xhline{2\arrayrulewidth}
                                        & \multicolumn{3}{c}{$\tau$}                                                                    \\
        \cline{3-4}
                                        &                                 & \multicolumn{1}{c}{0.05}        & \multicolumn{1}{c}{0.005} \\
        \Xhline{2\arrayrulewidth}
        \multirowcell{4}[-7em]{$\xi$}   &
        \multirow{2}{*}[-4em]{0.1 }     &
        \multicolumn{1}{c}{\spcost{#5}} & \multicolumn{1}{c}{\spcost{#6}}                                                            \\[-1em]
                                        &                                 &  \begin{minipage}[t]{0.03\linewidth} \vspace{0pt}  (a)  \end{minipage}  \begin{minipage}[t]{0.33\linewidth}\vspace{0pt} #1 \end{minipage}                         & \begin{minipage}[t]{0.03\linewidth} \vspace{0pt}  (b)  \end{minipage}  \begin{minipage}[t]{0.33\linewidth}\vspace{0pt} #2 \end{minipage}                      \\[-1em]
                                        & \multirow{2}{*}[-4em]{0.01}
                                        & \multicolumn{1}{c}{\spcost{#7}} & \multicolumn{1}{c}{\spcost{#8}}                          \\[-1em]
                                        &                                 & \begin{minipage}[t]{0.03\linewidth} \vspace{0pt}  (c)  \end{minipage}  \begin{minipage}[t]{0.33\linewidth}\vspace{0pt} #3 \end{minipage} & \begin{minipage}[t]{0.03\linewidth} \vspace{0pt}  (d)  \end{minipage}  \begin{minipage}[t]{0.33\linewidth}\vspace{0pt} #4 \end{minipage}                     \\[-1em]
        \Xhline{2\arrayrulewidth}
    \end{tabular}
}
For the optimization study, we perform a parameter sweep on the conductivity ratio $\tau$ \eqref{eq:tau},
the timescale ratio $\xi$ \eqref{eq:xi},
and the constraint factor $\Sigma$ \eqref{eq:supercap_problem}.
\rme{The optimized designs start with an initial uniform $\gamma=0.6$ everywhere in the domain, except the design in Figure \ref{tab:bruggeman_effective_02c} for which $\gamma=0.5$ was used because it provided a better-connected design than $\gamma=0.6$.}
\rtwo{Running on one core of an Intel Xeon E5-2695 v4 CPU, each optimization iteration (forward and adjoint problem) took around 180 seconds.}

\subsubsection{Original Bruggeman correlation}
\begin{figure*}[!htbp]
    \figurestable{
        \includegraphics[width=\linewidth]{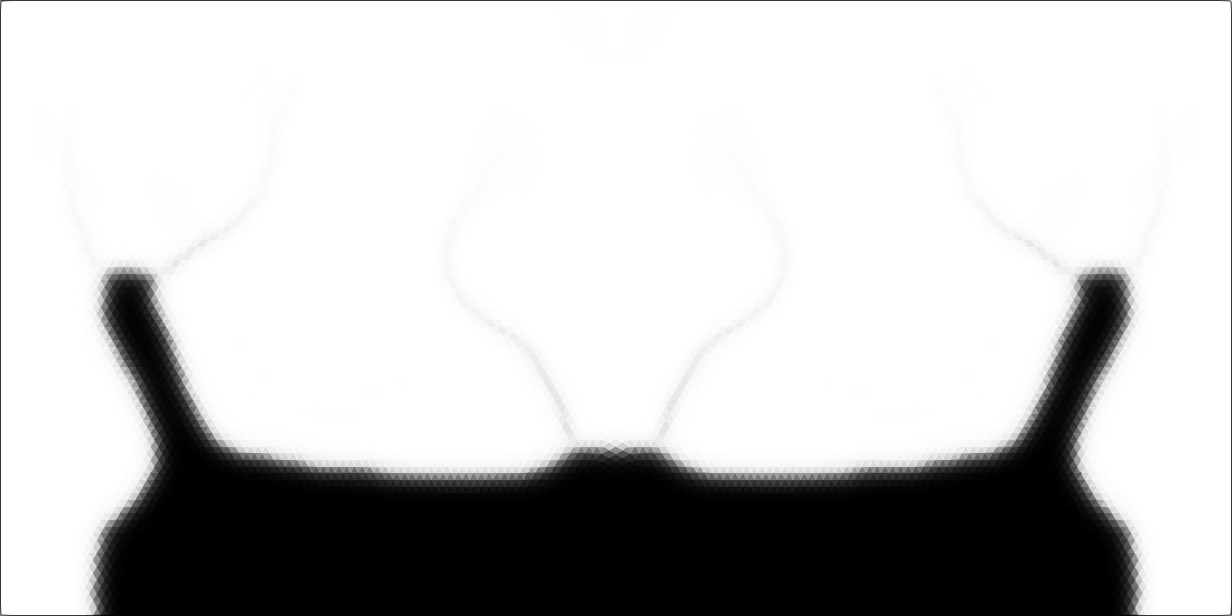}
    }{
        \includegraphics[width=\linewidth]{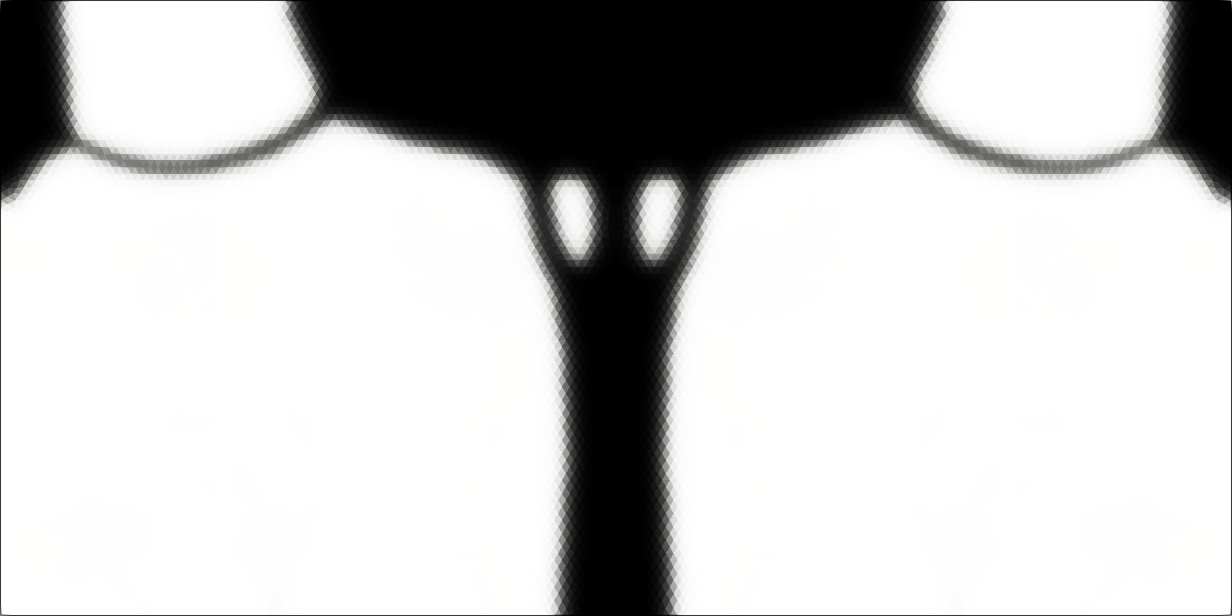}
    }{
        \includegraphics[width=\linewidth]{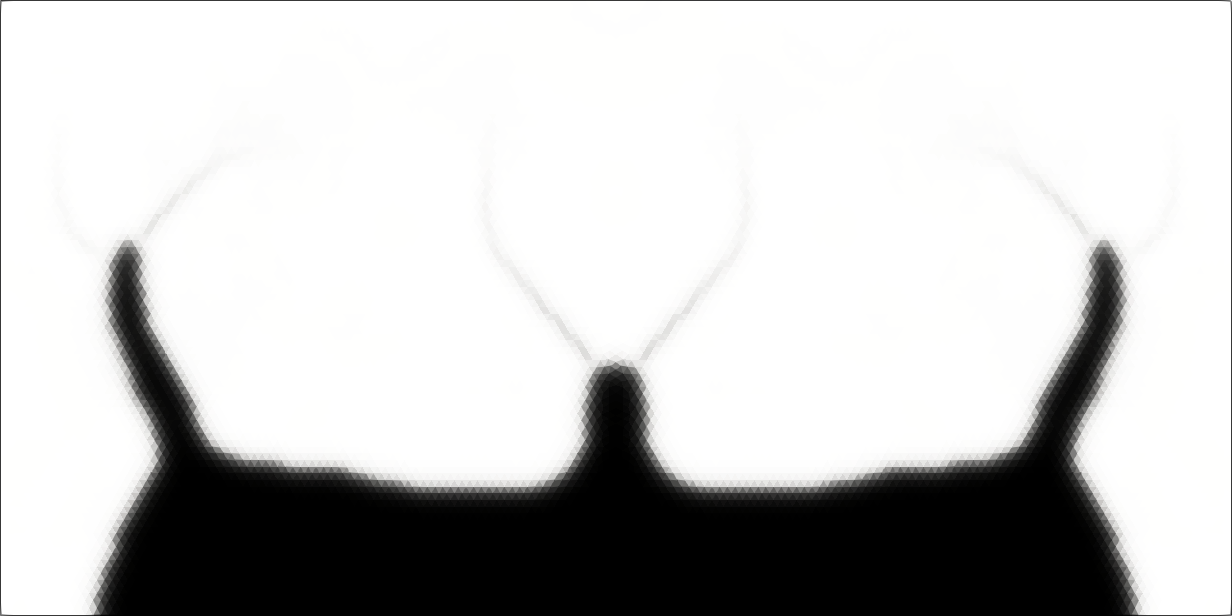}
    }{
        \includegraphics[width=\linewidth]{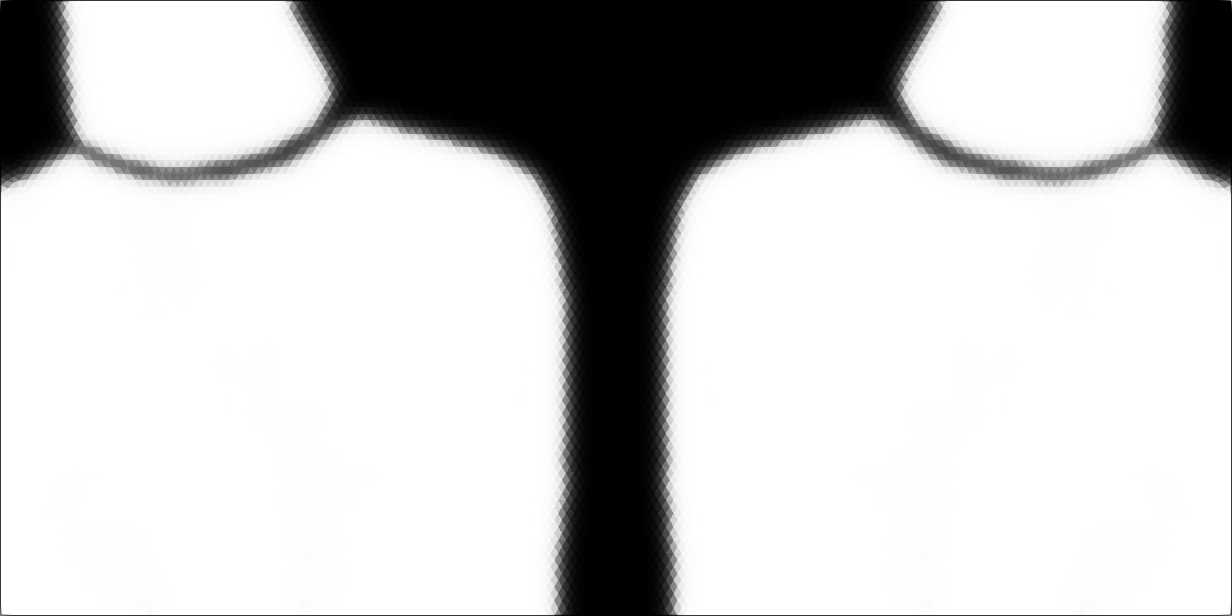}
    }{0.01187}{0.00542}{0.00115}{0.00052}
    {\phantomsubcaption \label{tab:bruggeman_simple_02a}}
    {\phantomsubcaption \label{tab:bruggeman_simple_02b}}
    {\phantomsubcaption \label{tab:bruggeman_simple_02c}}
    {\phantomsubcaption \label{tab:bruggeman_simple_02d}}
    \caption{Optimized supercapacitor designs considering the original Bruggeman correlation and $\Sigma=0.2$ Black is $\tilde{\gamma} = 1$; white is $\tilde{\gamma} =0$.}
    \label{tab:bruggeman_simple_02}
\end{figure*}
\begin{figure*}[!htbp]
    \figurestable{
        \includegraphics[width=\linewidth]{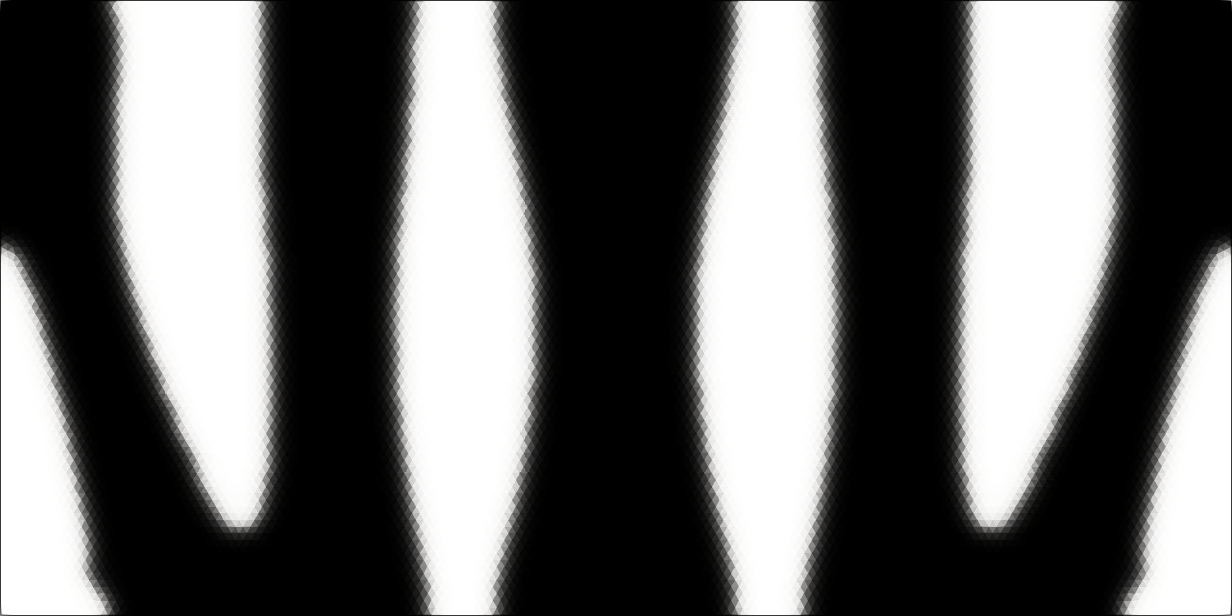}
    }{
        \includegraphics[width=\linewidth]{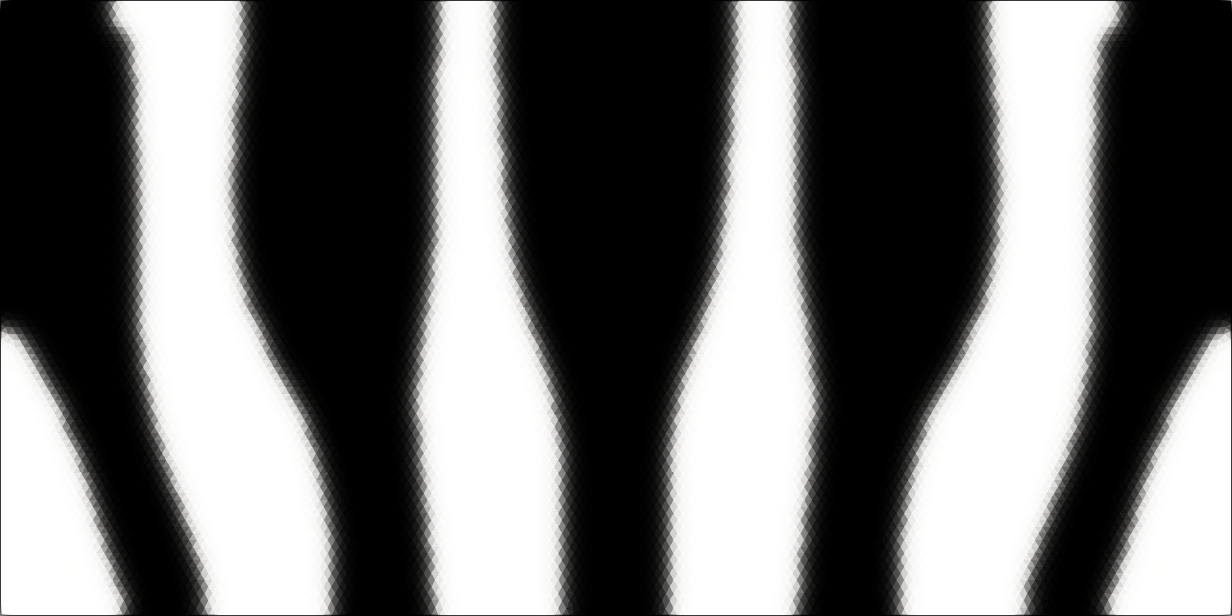}
    }{
        \includegraphics[width=\linewidth]{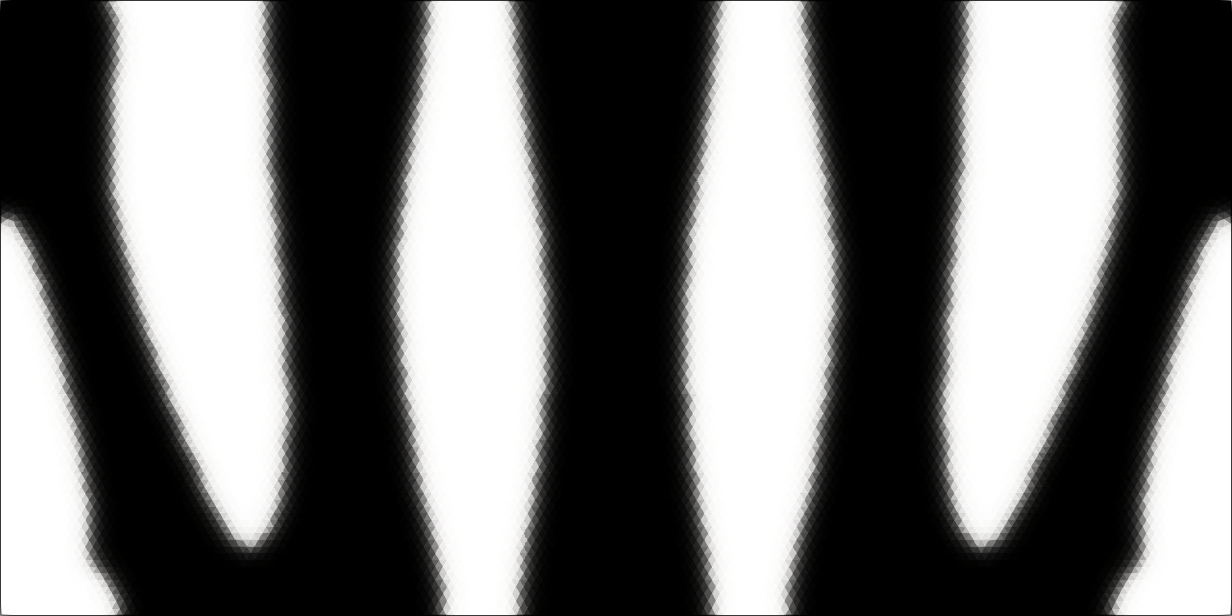}
    }{
        \includegraphics[width=\linewidth]{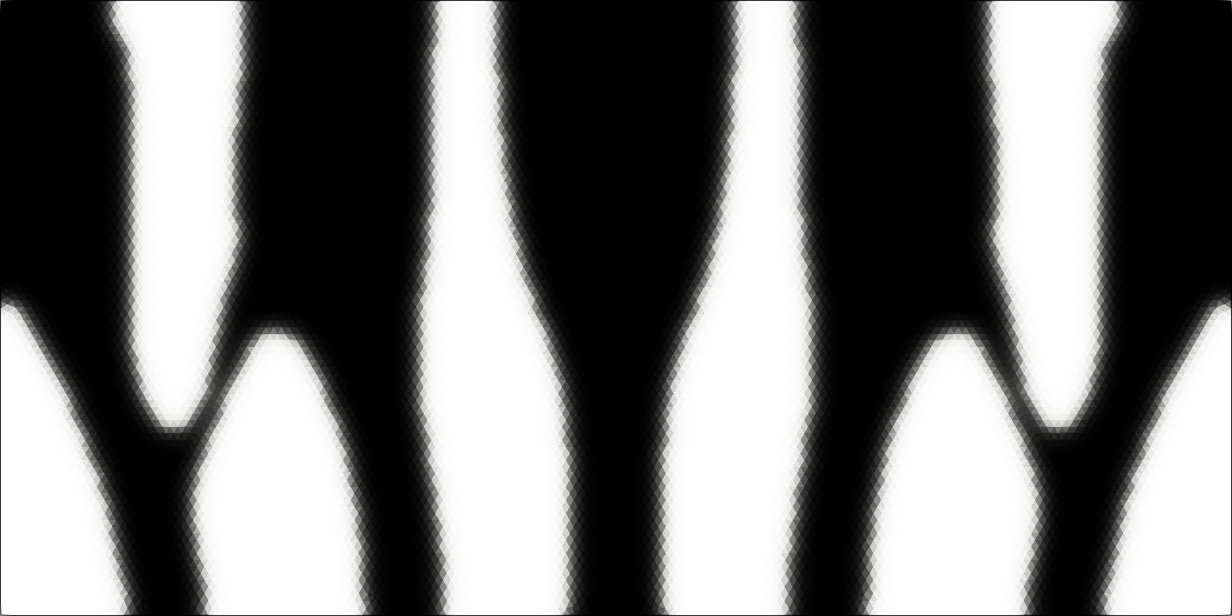}
    }{0.04682}{0.04008}{0.00425}{0.00369}
    {\phantomsubcaption \label{tab:bruggeman_simple_05a}}
    {\phantomsubcaption \label{tab:bruggeman_simple_05b}}
    {\phantomsubcaption \label{tab:bruggeman_simple_05c}}
    {\phantomsubcaption \label{tab:bruggeman_simple_05d}}
    \caption{Optimized supercapacitor designs considering the original Bruggeman correlation and $\Sigma=0.5$ Black is $\tilde{\gamma} = 1$; white is $\tilde{\gamma} =0$.}
    \label{tab:bruggeman_simple_05}
\end{figure*}

The first set of optimized designs in Figures \ref{tab:bruggeman_simple_02} and \ref{tab:bruggeman_simple_05} uses the original Bruggeman correlation \eqref{eq:kappaeff}.
The high ionic conductivity within the porous material ($\tilde{\gamma} = 1$) facilitates ion transport.
As a consequence, the ions can penetrate deeper into the microporous network to access the high surface area within it, and thus the boundaries between the electrolyte and the porous material are generally smooth and have few protuberant, bulbous or jagged features to expose more of the microporous electrode.
Additionally, the ion transport is efficient enough that the design is not substantially affected by lower charging times with respect to the ion transport timescale, i.e. higher $\xi$. However, at lower $\tau$, ionic current transport is less efficient with respect to electronic current transport. As such, the optimized designs have more mass closer to the top boundary, where the current enters the domain. Lastly, designs with higher energy stored requirements, cf. Figures \ref{tab:bruggeman_simple_05}, take more of the design domain to store more energy.

The cost function values of all optimized designs in Figures \ref{tab:bruggeman_simple_02} and \ref{tab:bruggeman_simple_05} show lower energy losses for lower values of $\xi$ (longer charging times).
As expected, for the same maximum potential, longer charging times allow for a more gradual application of the potential field allowing the electrode to absorb the charge while minimizing irreversible, ohmic losses. In the limit of infinite charge times, the losses would further decay as the system approaches the reversible, thermodynamic limit.
A higher energy constraint factor $\Sigma$ yields higher ohmic losses since currents are generally higher and more ions need to travel through the electrode.
On the other hand, decreasing $\tau$ helps to alleviate these losses, especially for lower energy requirements and lower $\xi$.

\subsubsection{Modified Bruggeman correlation}

\begin{figure*}[!htbp]
    \figurestable{
        \includegraphics[width=\linewidth]{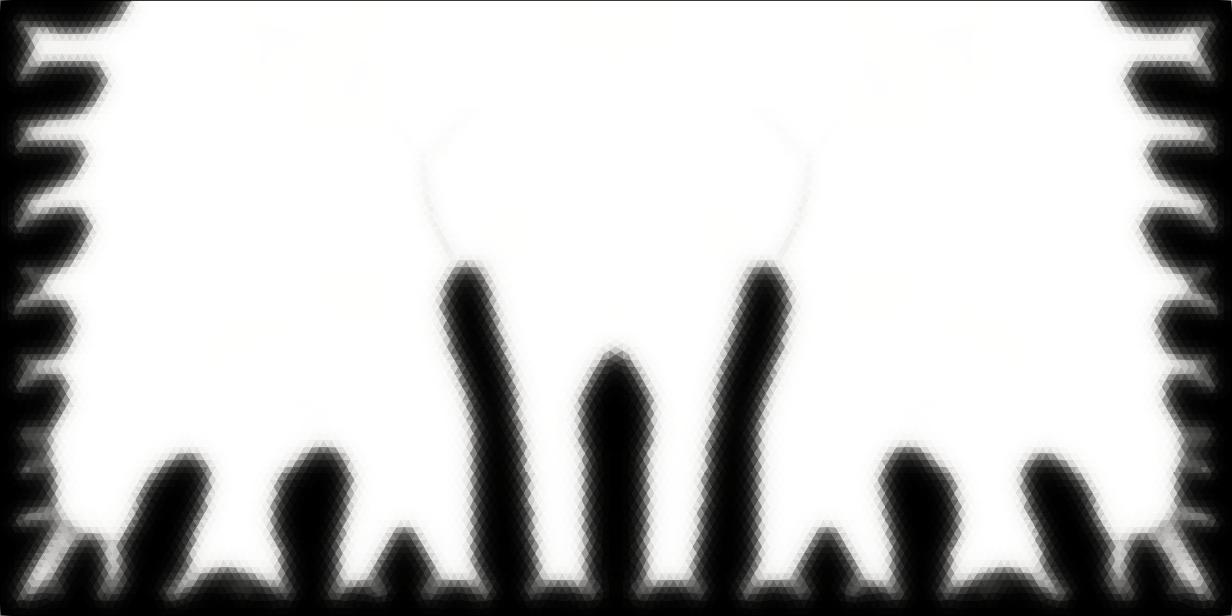}
    }{
        \includegraphics[width=\linewidth]{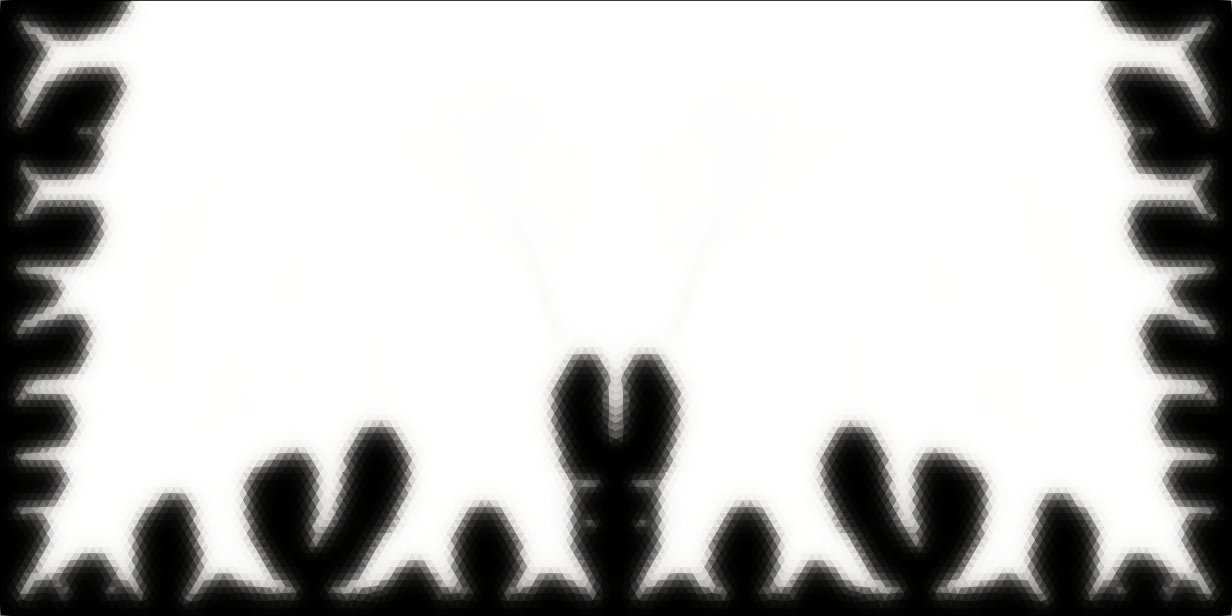}
    }{
        \includegraphics[width=\linewidth]{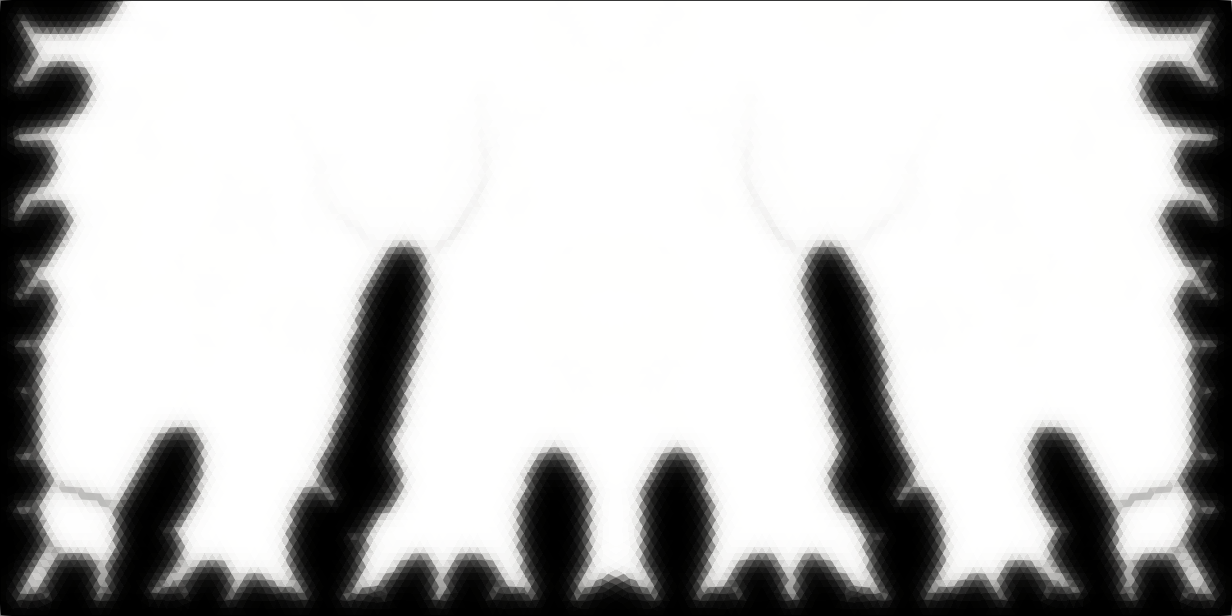}
    }{
        \includegraphics[width=\linewidth]{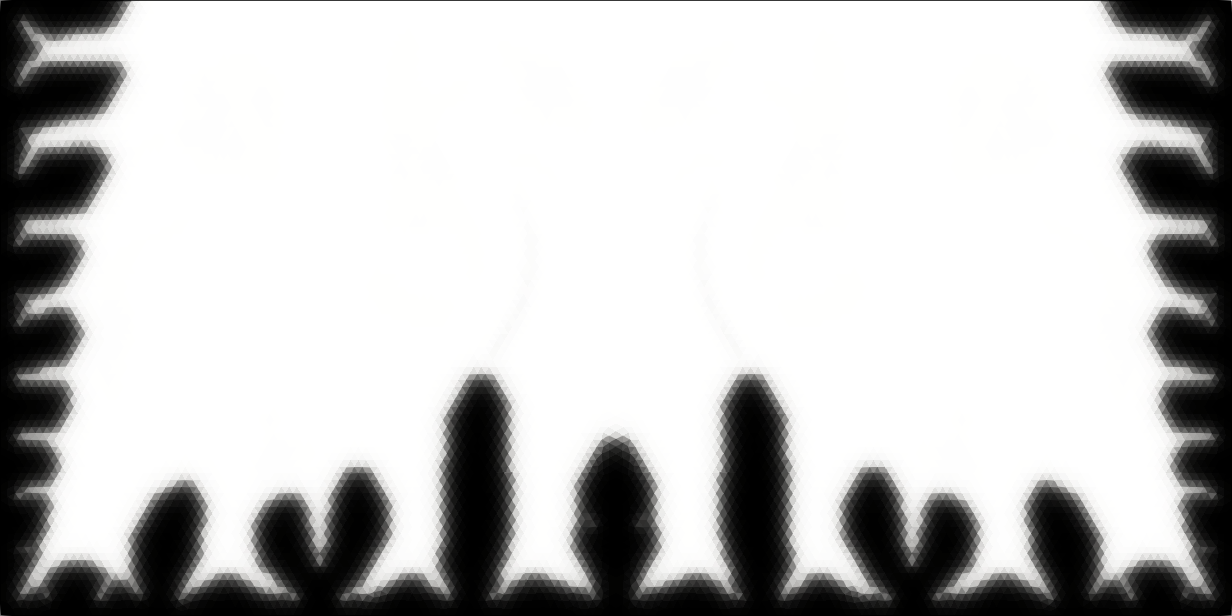}
    }{
        0.01801}{
        0.01486}{
        0.00156}{
        0.00137}
    {\phantomsubcaption \label{tab:bruggeman_effective_02a}}
    {\phantomsubcaption \label{tab:bruggeman_effective_02b}}
    {\phantomsubcaption \label{tab:bruggeman_effective_02c}}
    {\phantomsubcaption \label{tab:bruggeman_effective_02d}}
    \caption{Optimized supercapacitor designs considering the lower effective ionic conductivity and $\Sigma=0.2$ Black is $\tilde{\gamma} = 1$; white is $\tilde{\gamma} =0$.}
    \label{tab:bruggeman_effective_02}
\end{figure*}

\begin{figure*}[!htbp]
    \figurestable{
        \includegraphics[width=\linewidth]{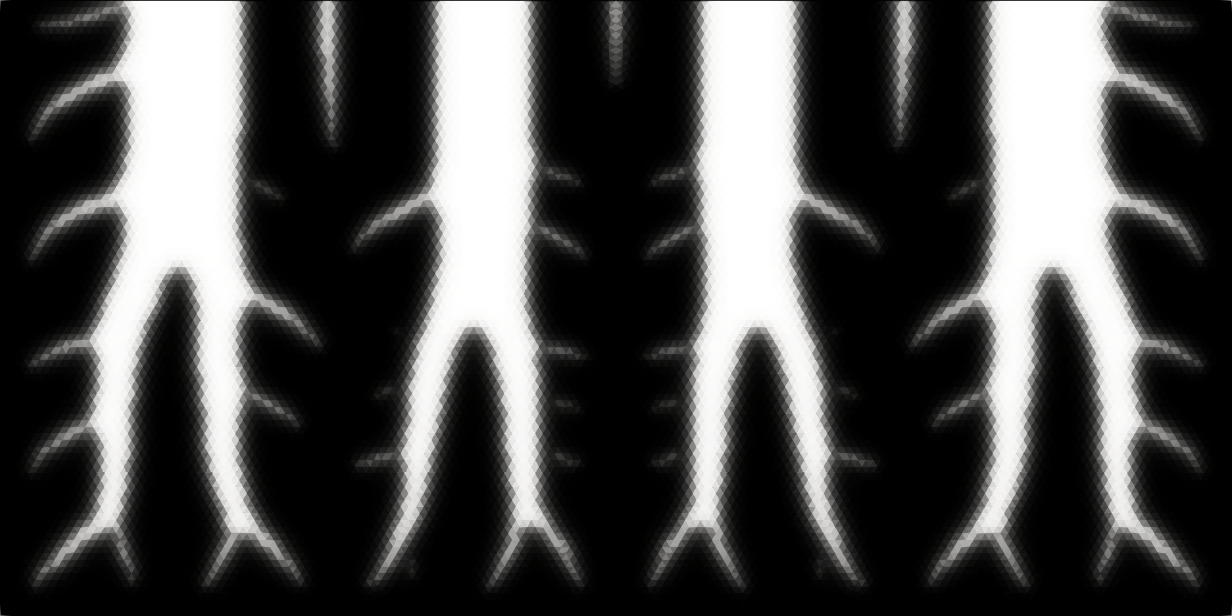}
    }{
        \includegraphics[width=\linewidth]{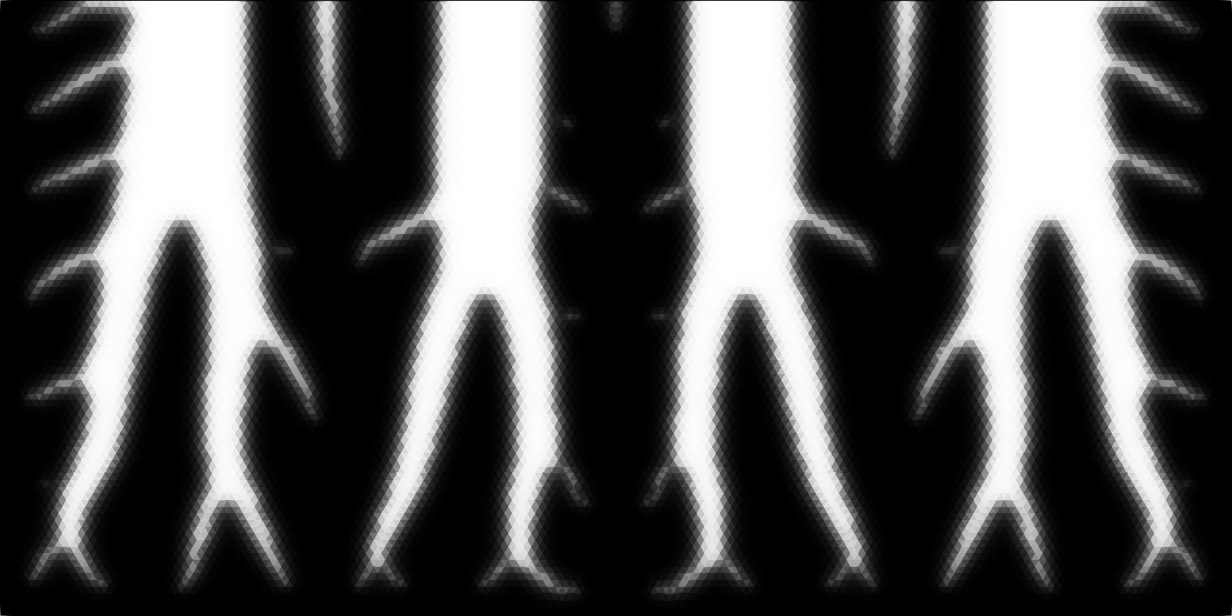}
    }{
        \includegraphics[width=\linewidth]{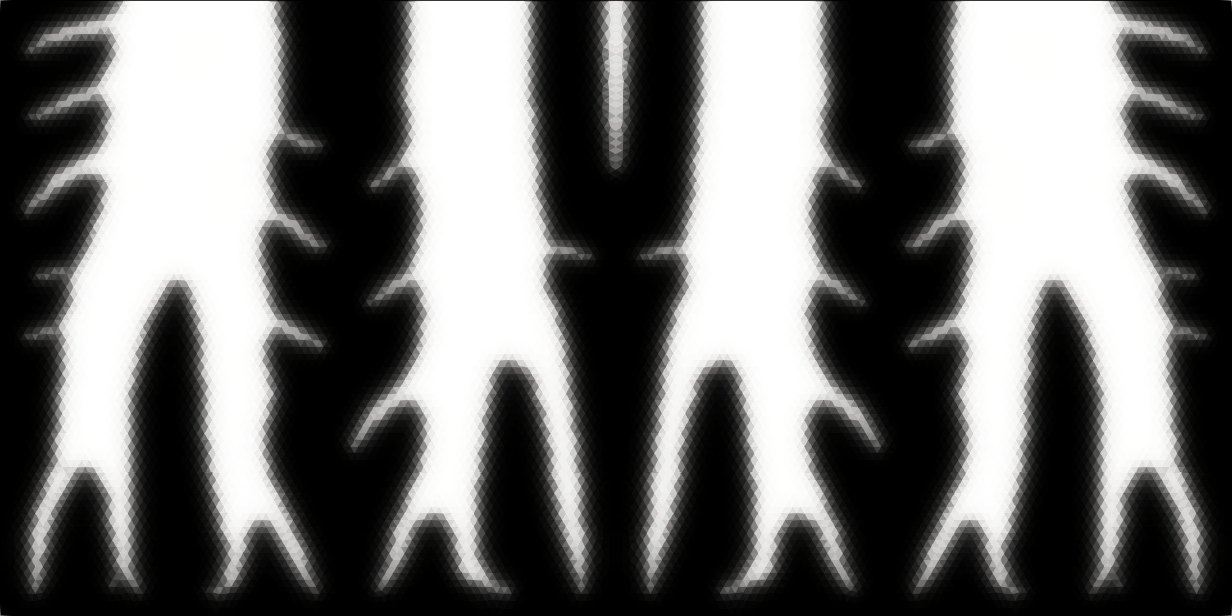}
    }{
        \includegraphics[width=\linewidth]{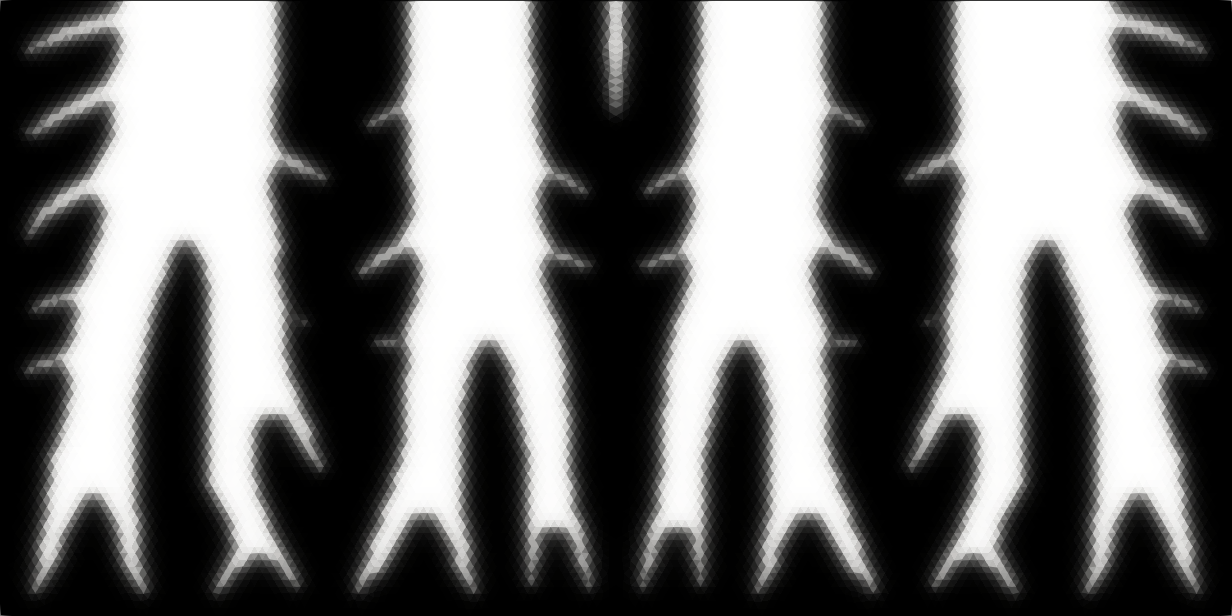}
    }{
        0.16681}{
        0.14420}{
        0.00951}{
        0.00882}
    {\phantomsubcaption \label{tab:bruggeman_effective_05a}}
    {\phantomsubcaption \label{tab:bruggeman_effective_05b}}
    {\phantomsubcaption \label{tab:bruggeman_effective_05c}}
    {\phantomsubcaption \label{tab:bruggeman_effective_05d}}
    \caption{Optimized supercapacitor designs considering the modified Bruggeman correlation and $\Sigma=0.5$ Black is $\tilde{\gamma} = 1$; white is $\tilde{\gamma} =0$.}
    \label{tab:bruggeman_effective_05}
\end{figure*}
The next set of examples uses the modified Bruggeman correlation for the effective ion conductivity, cf. \eqref{eq:porosity_two}.
A lower effective ionic conductivity translates into worse ion transport within the electrode and lower ion penetration, forcing the design to increase the surface area exposed to the electrolyte to maximize the energy stored.
As a result, and in contrast to the smooth designs described in the previous subsection, the optimization algorithm creates a hierarchical porous network with macropores for efficient ion transport, and smaller pores for greater energy storage, cf. the multi-lengthscale structure in Figures \ref{tab:bruggeman_effective_02} and \ref{tab:bruggeman_effective_05}.
This multiscale nature of the designs has been previously highlighted as an important route for attaining improved performance, and we emphasize that here the optimization algorithm automatically converged to a hierarchical design \cite{wang20083d}.

Varying $\xi$ does not noticeably affect the design for $\Sigma=0.2$, cf Figure \ref{tab:bruggeman_effective_02}, but it does for $\Sigma=0.5$, cf. Figure \ref{tab:bruggeman_effective_05}. The greater energy requirement for larger $\xi$, i.e. faster charging times, requires more porous electrode to store more energy, at the cost of greater ohmic losses.
The influence of the conductivity ratio $\tau$ is negligible for the nondimensional values tested here.
Unlike for the original Bruggeman correlation, placing more porous electrode adjacent to the top boundary carries higher ohmic losses due to the lower ionic conductivity.
Increasing the energy stored constraint factor $\Sigma$ did also translate into a growth of the electrode mass within the design domain.

The cost function values in Figures \ref{tab:bruggeman_effective_02} and \ref{tab:bruggeman_effective_05} follow a similar pattern to those in Figures \ref{tab:bruggeman_simple_02} and \ref{tab:bruggeman_simple_05} but with overall higher energy losses due to the reduced ionic conductivity. Importantly, the optimization algorithm again converged to a hierarchical structure to decrease energy loss and improve performance.

\subsubsection{Comparison to a monolithic electrode}

\begin{figure*}[!htbp]
    \centering
    \begin{subfigure}[c]{\linewidth}
        \centering
        \includegraphics[scale=0.3]{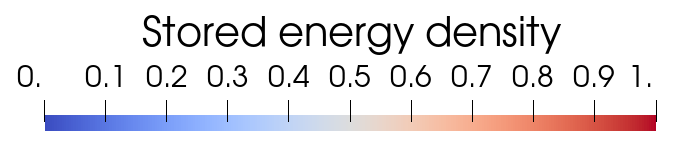}
    \end{subfigure}
    \begin{multicols}{3}
        \centering
        \begin{subfigure}[b]{\linewidth}
            \includegraphics[width=\linewidth]{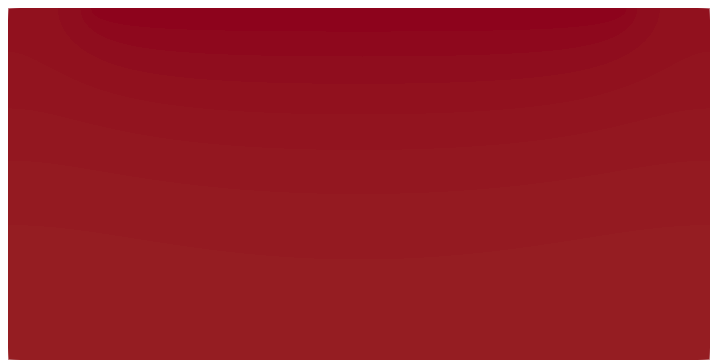}
            \caption{Monolithic electrode at fixed time}
            \label{fig:energy_mono_simple}
        \end{subfigure}
        \begin{subfigure}[b]{\linewidth}
            \includegraphics[width=\linewidth]{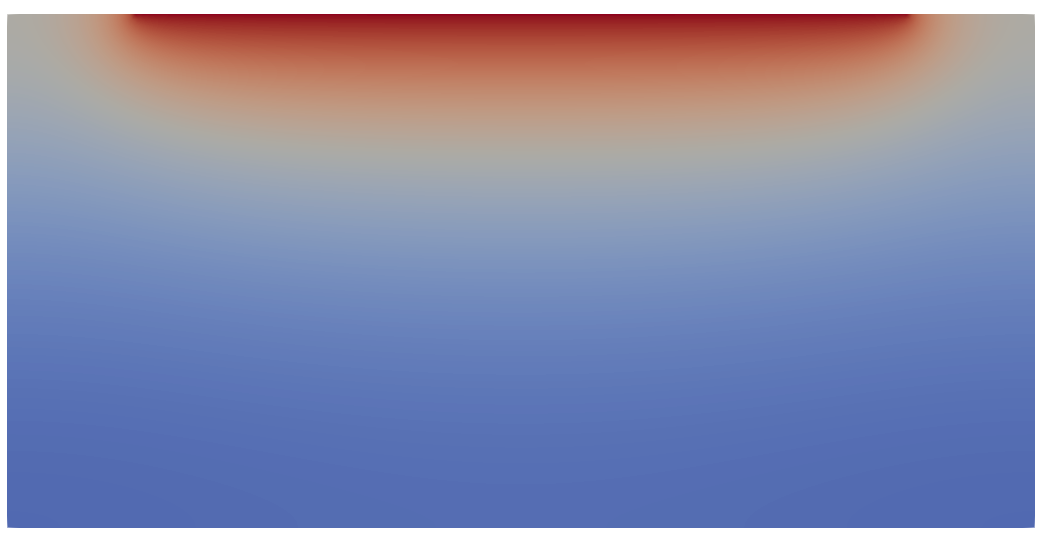}
            \caption{Effective monolithic electrode at fixed time}
            \label{fig:energy_mono_eff}
        \end{subfigure}
        \begin{subfigure}[b]{\linewidth}
            \includegraphics[width=\linewidth]{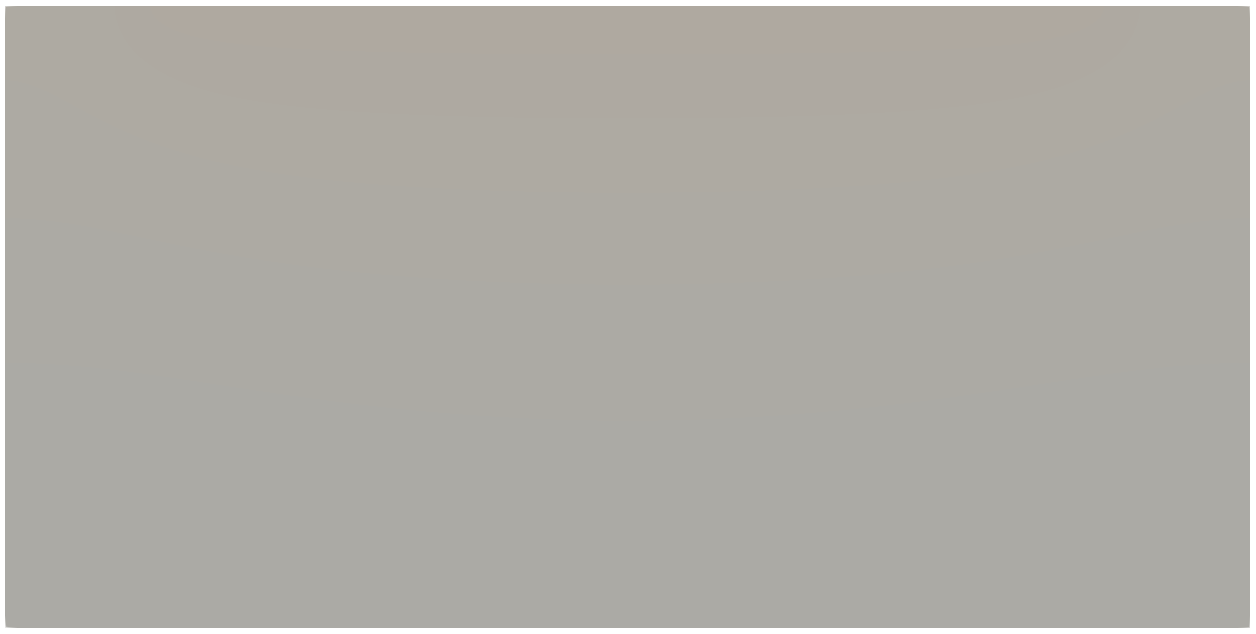}
            \caption{Monolithic electrode at fixed energy stored}
            \label{fig:energy_block_simple_fixed_energy}
        \end{subfigure}
        \begin{subfigure}[b]{\linewidth}
            \includegraphics[width=\linewidth]{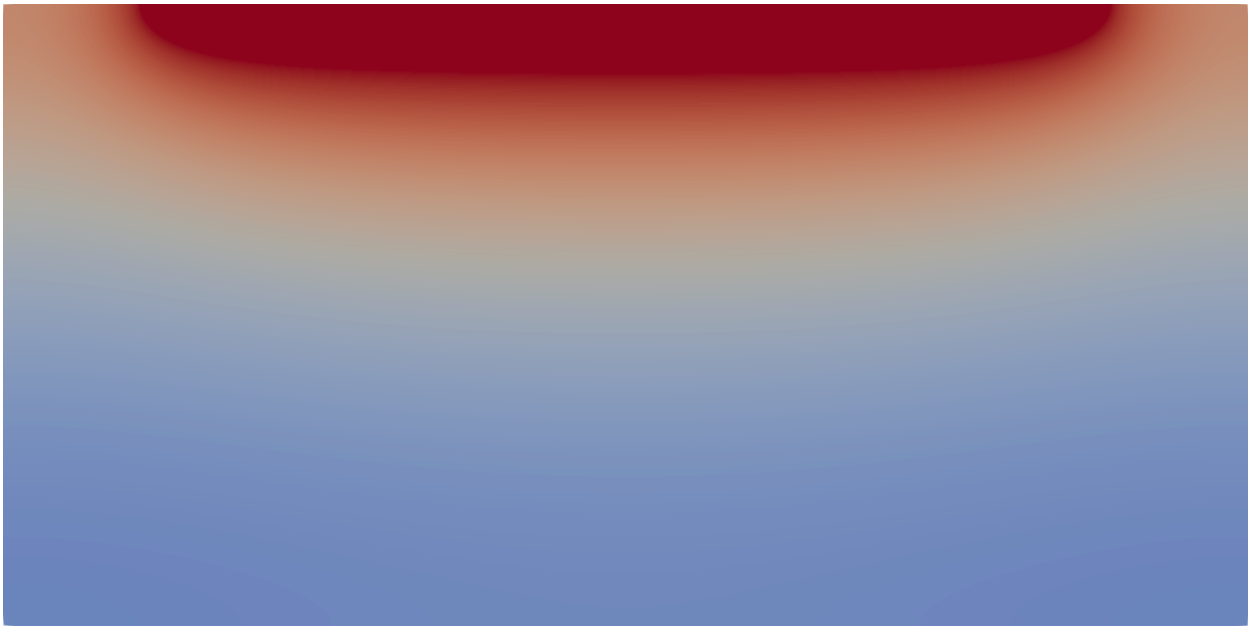}
            \caption{Effective monolithic electrode at fixed energy stored}
            \label{fig:energy_block_eff_fixed_energy}
        \end{subfigure}
        \begin{subfigure}[b]{\linewidth}
            \includegraphics[width=\linewidth]{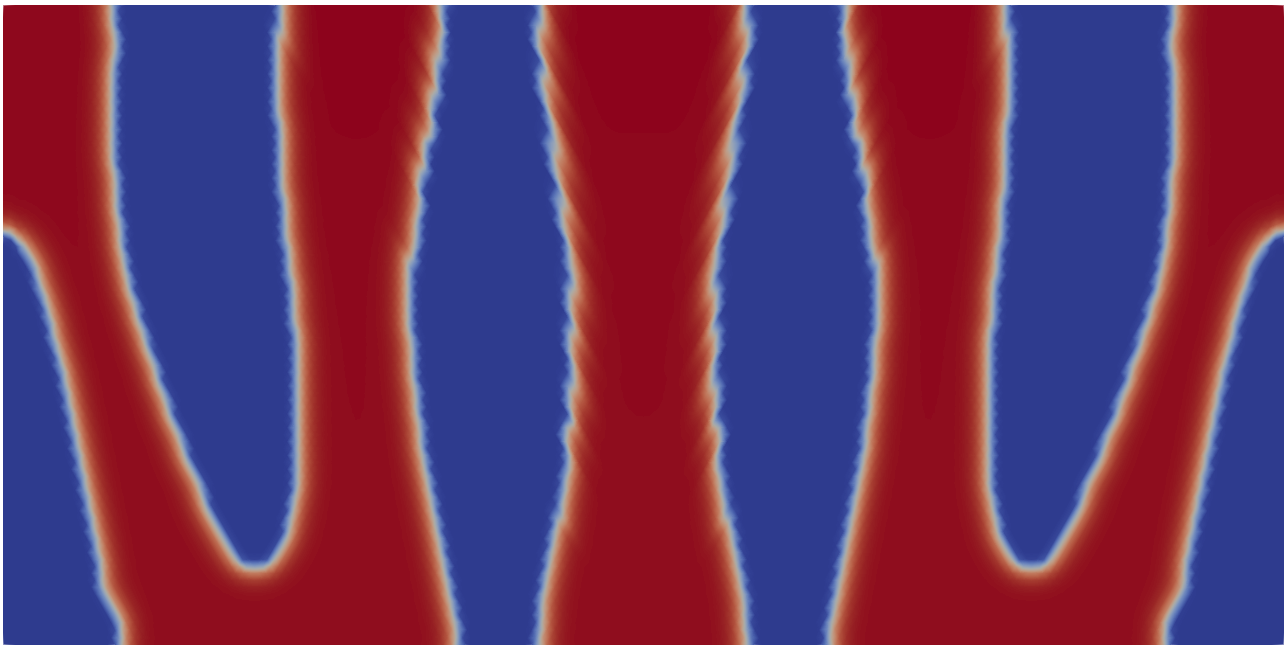}
            \caption{Designed electrode with original Bruggeman}
            \label{fig:energy_designed_simple}
        \end{subfigure}
        \begin{subfigure}[b]{\linewidth}
            \includegraphics[width=\linewidth]{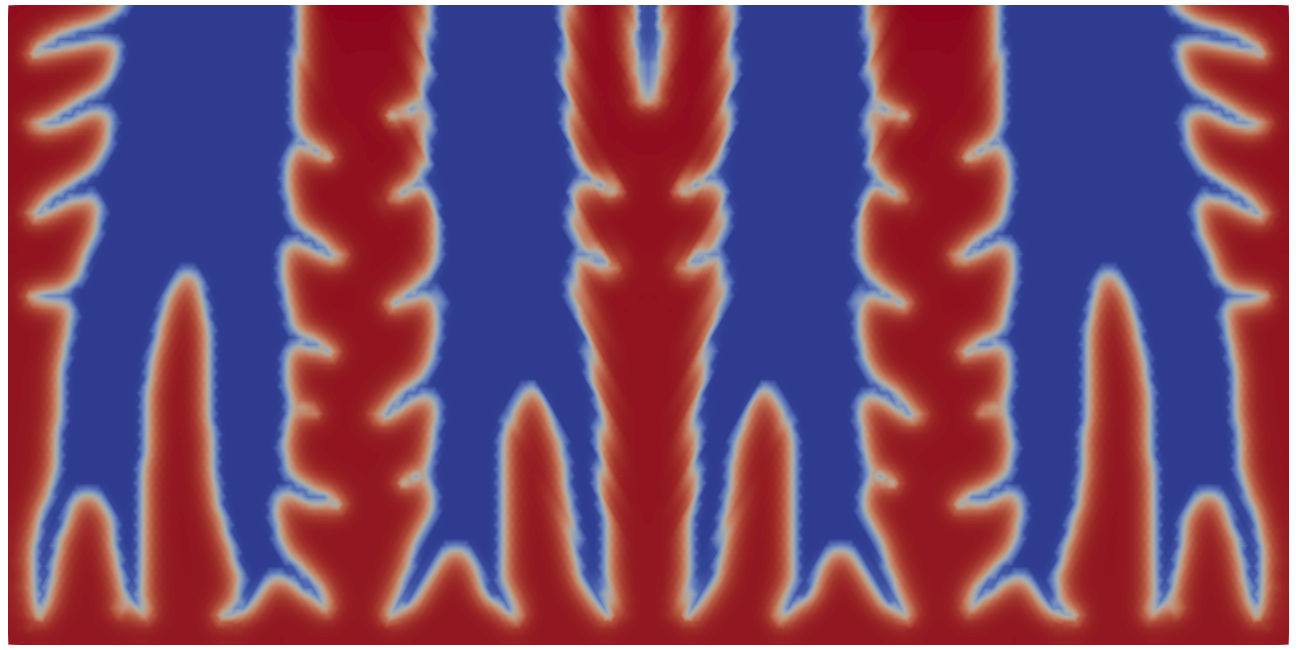}
            \caption{Designed electrode with effective Bruggeman}
            \label{fig:energy_designed_eff}
        \end{subfigure}
    \end{multicols}
    \caption{Energy stored for the monolithic electrode and the optimized design.}
    \label{fig:energy}
\end{figure*}
Our optimized designs for $\xi=0.01$, $\tau=0.05$ and $\Sigma=0.5$ are now compared with a monolithic porous electrode, i.e. $\tilde\gamma=1.0$ everywhere.
We consider both the original and modified Bruggeman correlations and run simulations to compare the energy loss and the energy stored.
We do not use the expressions in \eqref{eq:supercap_problem}, i.e. $\theta^{\text{sp}}_0$ and $\theta^{\text{sp}}_1$ because they penalize intermediate volume fraction values, cf. \eqref{eq:penalization_loss}, \eqref{eq:penalization_stored} in $\theta^{\text{sp}}_0$ and $\gamma^3$ in $\theta^{\text{sp}}_1$, and thus, do not reflect the real energy loss and energy stored.
Instead, we express the energy loss as
\begin{equation}
    \begin{aligned}
    E_{\text{loss}} = &\int_{0}^{1/\xi}\int_{\hat\Omega} \left(\epsilon^{3/2} \nabla \hphi_2 \cdot \nabla \hphi_2 \right. \\
        &+  \left.\frac{(1 - \epsilon)^{3/2}}{\tau}\nabla \hphi_1 \cdot \nabla \hphi_1 \right) \diff V \diff t ,
    \end{aligned}
\end{equation}
and the energy stored as
\begin{equation}
    E_{\text{stored}} = \int_{\hat{\Omega}} \frac{\tilde{\gamma}}{2}\left(\hat{\Phi}_{1}-\hat{\Phi}_{2}\right)^{2} \mathrm{~d} V,
\end{equation}
which are both derived from \eqref{eq:supercap_eq} and together they add up to the total energy inputted to the half-cell.
For the designed electrode with the original Bruggeman correlation, $E_{\text{loss}}=0.004196$ is slightly lower than $\theta^{\text{sp}}_0=0.00425$ in Figure \ref{tab:bruggeman_simple_05} and $E_{\text{stored}}=0.54947$ is slightly larger than 0.5 ($0.5 \times E_{\text{max}}$).
These differences are expected given the residual presence of the intermediate material, which overestimate the energy losses and underestimate the energy stored due to the penalization schemes.

The monolithic electrode simulations were performed for two different scenarios.
First, the monolithic electrode was charged until the energy stored was equivalent to the energy stored in the designed electrode, and, second, the monolith was charged for the same amount of time for which the designed electrode was optimized.

    \begin{table}[htb!]
        \caption{Quantitative comparison of the monolithic and designed electrodes with the original Bruggeman correlation for $\xi=0.01$ and $\tau=0.05$.}
        \label{tab:simple_energies}
        \centering
        \begin{tabular}{ cccc }
            & \thead{Fixed \\Energy\\ Stored} & \thead{Fixed \\Charging\\ Time} & \thead{Optimized\\ Design} \\
            \Xhline{2\arrayrulewidth}
            $E_{\text{stored}}$ & 0.54947& 0.96758& 0.54947\\
            $E_{\text{loss}}$ &0.01655& 0.02207 & 0.004196\\
            \thead{Charging time\\ factor $1/\xi$} & 0.755  & 1.0 & 1.0     \\
        \end{tabular}
    \end{table}

    \begin{table}[htb!]
        \caption{Quantitative comparison of the monolithic and designed electrodes with the effective Bruggeman correlation for $\xi=0.01$ and $\tau=0.05$.}
        \label{tab:effective_energies}
        \centering
        \begin{tabular}{ cccc }
                                   & \thead{Fixed \\Energy\\ Stored} & \thead{Fixed \\Charging\\ Time} & \thead{Optimized\\ Design} \\
            \Xhline{2\arrayrulewidth}
            $E_{\text{stored}}$ & 0.58426& 0.34786& 0.58426\\
            $E_{\text{loss}}$ & 0.47853& 0.32528 & 0.00419\\
            \thead{Charging time\\ factor $1/\xi$} & 1.21  & 1.0 & 1.0     \\
        \end{tabular}
    \end{table}

When employing the original Bruggeman correlation, the designed electrode incurs 0.76\% of the total energy input as ohmic losses, whereas the monolith incurs 2.9\%, cf. Table \ref{tab:simple_energies}. 
The designed electrode, however, is charged more slowly.
Alternatively, when the monolithic electrode is charged for the same amount of time as the designed electrode, a larger amount of energy is stored, but 2.2\% of the inputted energy is lost to ohmic heating, rendering the designed electrode more energy efficient. 
Figures \ref{fig:energy_mono_simple}, \ref{fig:energy_block_simple_fixed_energy} and \ref{fig:energy_designed_simple} compare the energy density field $\frac{\tilde{\gamma} }{2}\left(\hat{\Phi}_{1}-\hat{\Phi}_{2}\right)^{2}$ for the monolithic electrode operated at fixed charging time, fixed energy stored, and the optimized electrode. For these parameters the monolithic electrode charges evenly and shows a uniform stored energy density. For the designed electrode, the introduction of structure leads to non-uniform energy storage distribution but nevertheless leads to improved performance. The open channels lower the ohmic losses.

The impact of optimized structure is even more dramatic when using the modified Bruggeman correlation.
Indeed, as seen in Table \ref{tab:effective_energies}, the energy stored in the monolithic electrode charged for the same amount of time is only 60\% of the optimized design while the ohmic losses are 48\% of the input energy in the monolith and only 0.71\% in the designed electrode.
To attain the same amount of stored energy, the monolithic electrode incurs ohmic losses that are 114 times higher with a charging time 21\% greater.
As seen in Figure \ref{fig:energy_mono_eff} and \ref{fig:energy_block_eff_fixed_energy} the energy is accumulated mostly near the charging boundary because the ions cannot penetrate further into the monolithic electrode.
On the other hand, the hierarchical structure of the optimized design facilitates ion transport to permit an efficient distribution of the energy density in the porous electrode, as observed in Figure \ref{fig:energy_designed_eff}.

\subsection{Three-dimensional optimized designs}

\begin{figure*}[!htbp]
    \centering
    \begin{multicols}{2}
    \centering
    \begin{subfigure}[b]{\linewidth}
        \centering
        \includegraphics[height=7cm]{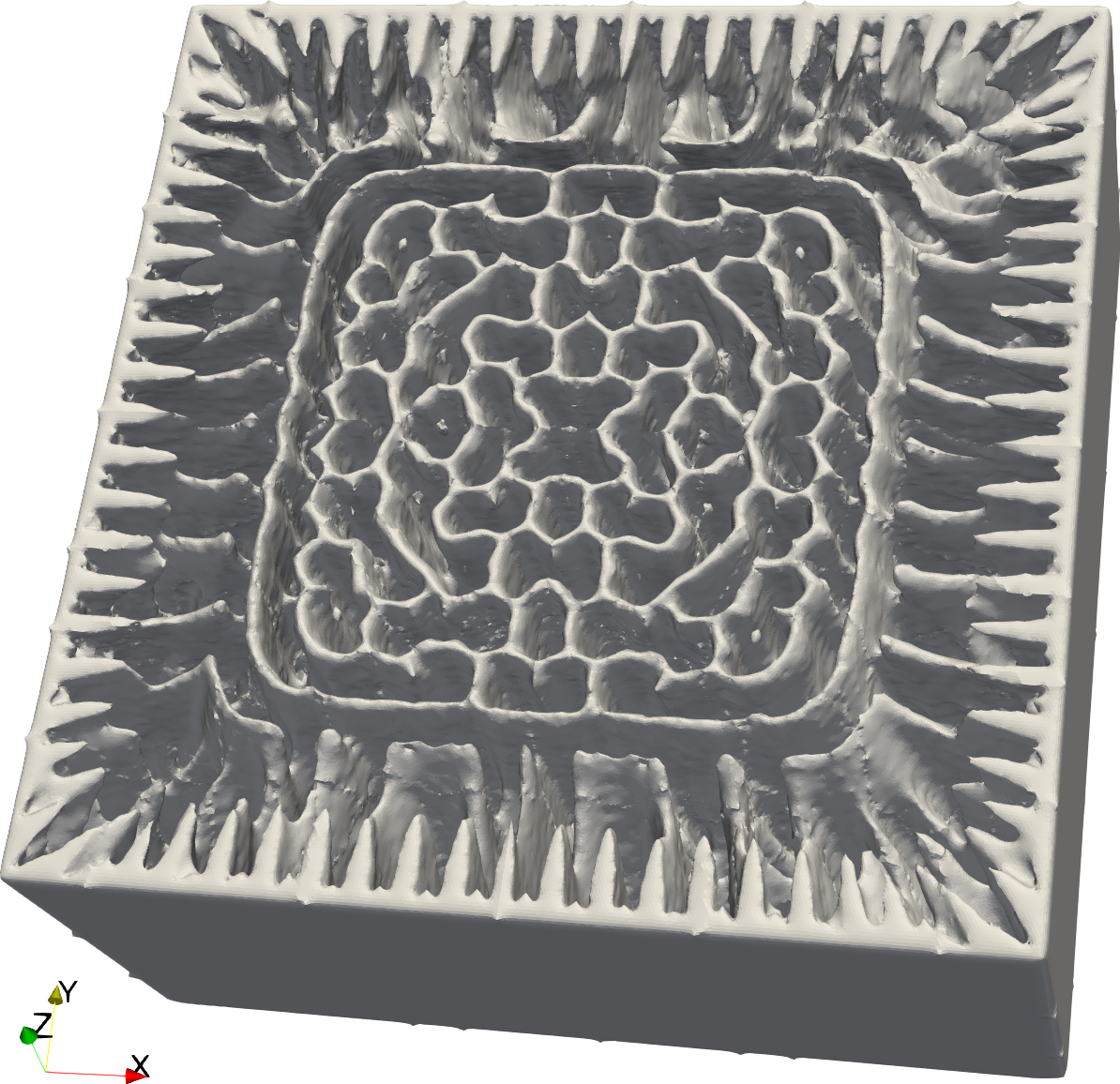}
        \caption{Porous redox electrode optimized design.}
        \label{fig:redoxeff_3D}
    \end{subfigure}
    \par
    \begin{subfigure}[b]{\linewidth}
        \centering
        \includegraphics[width=0.9\linewidth]{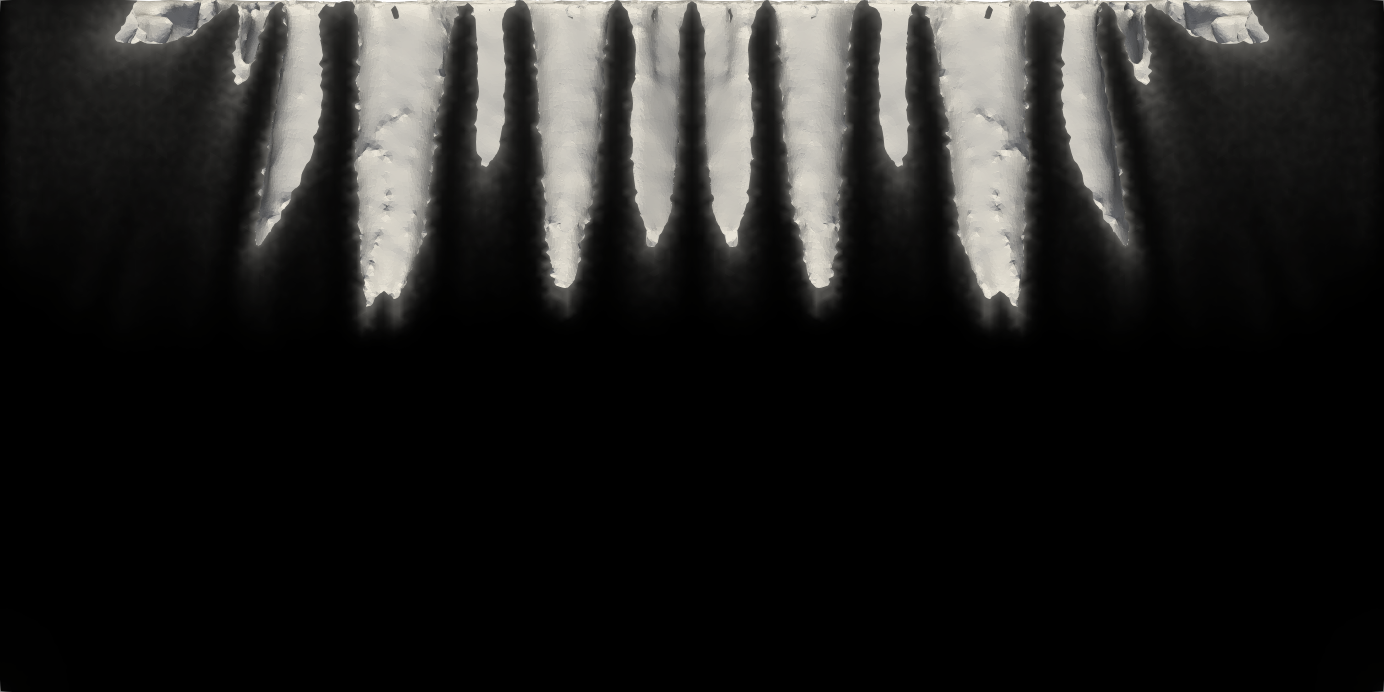}
        \caption{Cross-section of the porous redox electrode optimized design.}
        \label{fig:redoxeff_cross}
    \end{subfigure}
    \par
    \begin{subfigure}[b]{\linewidth}
        \centering
        \includegraphics[height=7cm]{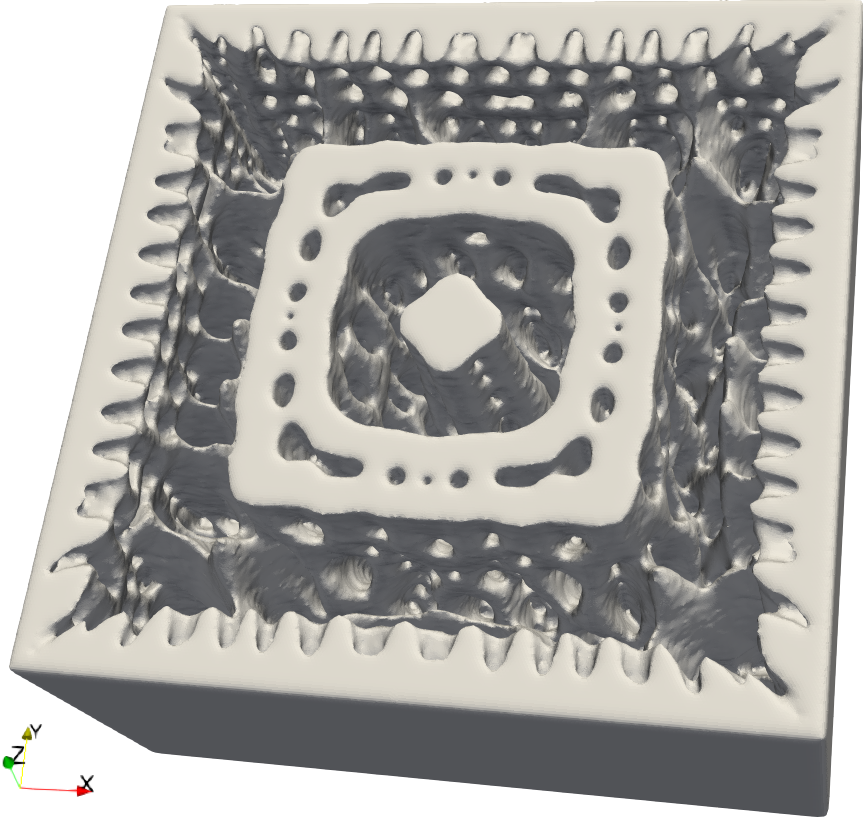}
        \caption{EDLC optimized design.}
        \label{fig:supercap_3D}
    \end{subfigure}
    \par
    \begin{subfigure}[b]{\linewidth}
        \centering
        \includegraphics[width=0.9\linewidth]{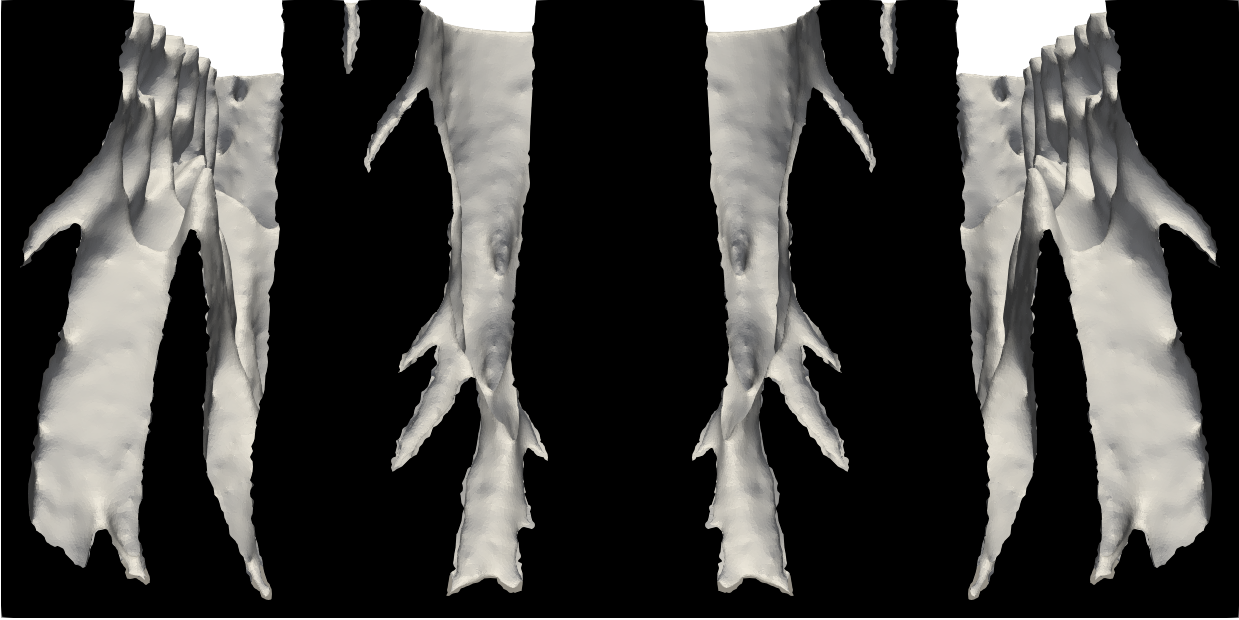}
        \caption{Cross-section of the EDLC optimized design.}
        \label{fig:supercap_3D_cross_section}
    \end{subfigure}
    \end{multicols}
    \caption{Optimized designs in three dimensions.}
    \label{fig:3D}
\end{figure*}

We further demonstrate the design \rtwo{of porous electrodes} for the three-dimensional domain in Figure \ref{fig:diagramelectrode3D}.
Due to the computational cost of each three-dimensional simulation, we present a single optimized design for each problem.

First, we present a 3D optimized porous redox electrode in Figure \ref{fig:redoxeff_3D}, and its cross-section in Figure \ref{fig:redoxeff_cross}.
For better contrast, the front slice of the cross-section is colored in black.
\rme{To aid visualization on the unstructured grid, a smoothing filter is applied.}
We use the modified Bruggeman correlation and the parameter values $\tau=0.005$, $\mu=5$ and $\delta=25$. 
In 3D, a valley-like design is observed.
Taking a cross-section reveals a root-like pattern similar to the 2D counterpart in Figure \ref{tab:pe_eff1d}.

Next, we design a supercapacitor electrode using the parameters $\xi=0.01$, $\tau=0.05$ and $E_{\text{stored}} \ge 0.5$ and the modified Bruggeman correlation  as shown in Figure \ref{fig:supercap_3D}.
In a similar fashion to the two-dimensional designs in Figure \ref{tab:bruggeman_effective_05}, the optimization algorithm creates a network with hierarchical porosity.
Wide channels connect the top boundary to the bottom of the electrode and a series of more narrow channels that spring from them to increase the surface area.
Most notably, a series of holes are carved on the exterior wall.
The cross-section along the XY plane in Figure \ref{fig:supercap_3D_cross_section} resembles the designs in Figure \ref{tab:bruggeman_effective_05}.

\section{Conclusion}\label{sec:conclusion}

\rtwo{In this work we have introduced the use of density-based topology optimization for the design of porous electrodes}. We have demonstrated the generality of this approach by posing and solving optimization problems for two different electrochemical applications: a porous electrode driving a Faradaic reaction and an EDLC/supercapacitor electrode. These serve as model steady and transient systems, respectively. We presented the governing equations for the secondary current distribution of the two systems and identified the key nondimensional groups informing electrode design. The physical parameter inputs to the governing equations were interpolated using a design field representing two different microporous materials. This ensured an inherently multiscale design where the aim of the optimization problem was thus to pattern the two different materials at scales larger than their microporous lengthscale. The approach is general, but we set the porosity of one the microporous materials to 1, effectively modeling it as a pure electrolyte.

For both electrochemical examples, the topology optimization algorithm provides non-trivial optimized electrode designs. In all cases, the optimized design showed improved performance over an undesigned, monolithic single porosity electrode. This was additionally verified for two different Bruggeman correlations for the effective conductivity. For the porous redox electrode, this lead to overpotentials that were up to 84 \% lower, while the supercapacitor electrode showed energy losses that were up to 98 \% lower. All the designs showed that introducing new length scales improved performance, but the resultant designs corresponding to the modified Bruggeman correlation showed the emergence of hierarchical structures, consistent with recent literature \citep{liu2017revitalizing, Zhou_2016,Wu_2019,Cobb_2014,nemani2015design,zhu2016supercapacitors,wang20083d}. Finally, we further demonstrated the utility of these techniques by demonstrating three-dimensional electrode design, thus providing a potential pathway for manufacture and testing of high performance architected electrodes. This work thus provides a new design tool for the computational design of multiscale, architected porous electrodes.

\section*{Funding information}
This work was performed under the auspices of the U.S. Department of Energy by Lawrence Livermore National Laboratory under Contract DE-AC52-07NA27344 and was supported by the LLNL-LDRD program under project numbers 19-ERD-035 and 20-ERD-019. LLNL Release Number LLNL-JRNL-828963.

\bibliography{bibfile}

\begin{thebibliography}{76}
\providecommand{\natexlab}[1]{#1}
\providecommand{\url}[1]{{#1}}
\providecommand{\urlprefix}{URL }
\providecommand{\doi}[1]{\url{https://doi.org/#1}}
\providecommand{\eprint}[2][]{\url{#2}}
 \bibcommenthead

\bibitem[{Ager and Lapkin(2018)}]{Ager_2018}
Ager JW, Lapkin AA (2018) Chemical storage of renewable energy. Science
  360(6390):707--708. \doi{10.1126/science.aat7918}

\bibitem[{Alexandersen and Andreasen(2020)}]{alexandersen_review_2020}
Alexandersen J, Andreasen CS (2020) A review of topology optimisation for
  fluid-based problems. Fluids 5(1):29. \doi{10.3390/fluids5010029}

\bibitem[{Allaire et~al(2002)Allaire, Jouve, and Toader}]{allaire}
Allaire G, Jouve F, Toader AM (2002) A level-set method for shape optimization.
  Comptes Rendus Math{\'e}matique 334(12):1125--1130.
  \doi{10.1016/s1631-073x(02)02412-3}

\bibitem[{Ambrosi et~al(2020)Ambrosi, Shi, and
  Webster}]{ambrosi_3d-printing_2020}
Ambrosi A, Shi RRS, Webster RD (2020) {3D}-printing for electrolytic processes
  and electrochemical flow systems. Journal of Materials Chemistry A
  8(42):21,902--21,929. \doi{10.1039/D0TA07939A}

\bibitem[{Bae et~al(2013)Bae, Erdonmez, Halloran, and Chiang}]{bae_design_2013}
Bae CJ, Erdonmez CK, Halloran JW, et~al (2013) Design of battery electrodes
  with dual-scale porosity to minimize tortuosity and maximize performance.
  Advanced Materials 25(9):1254--1258. \doi{10.1002/adma.201204055}

\bibitem[{Balay et~al(2020)Balay, Abhyankar, Adams, Brown, Brune, Buschelman,
  Dalcin, Dener, Eijkhout, Gropp, Karpeyev, Kaushik, Knepley, May, McInnes,
  Mills, Munson, Rupp, Sanan, Smith, Zampini, Zhang, and
  Zhang}]{petsc-user-ref}
Balay S, Abhyankar S, Adams MF, et~al (2020) {PETS}c users manual. Tech. Rep.
  ANL-95/11 - Revision 3.14, Argonne National Laboratory,
  \urlprefix\url{https://www.mcs.anl.gov/petsc}

\bibitem[{Barton(2020)}]{barton_electrification_2020}
Barton JL (2020) Electrification of the chemical industry. Science
  368(6496):1181--1182. \doi{10.1126/science.abb8061}

\bibitem[{Beck et~al(2021{\natexlab{a}})Beck, Ivanovskaya, Chandrasekaran,
  Forien, Baker, Duoss, and Worsley}]{beck_inertially_2021}
Beck VA, Ivanovskaya AN, Chandrasekaran S, et~al (2021{\natexlab{a}})
  Inertially enhanced mass transport using {3D}-printed porous flow-through
  electrodes with periodic lattice structures. Proceedings of the National
  Academy of Sciences 118(32):e2025562,118. \doi{10.1073/pnas.2025562118}

\bibitem[{Beck et~al(2021{\natexlab{b}})Beck, Wong, Jekel, Tortorelli, Baker,
  Duoss, and Worsley}]{Beck_2021}
Beck VA, Wong JJ, Jekel CF, et~al (2021{\natexlab{b}}) Computational design of
  microarchitected porous electrodes for redox flow batteries. Journal of Power
  Sources 512:230,453. \doi{10.1016/j.jpowsour.2021.230453}

\bibitem[{Behrou et~al(2019)Behrou, Pizzolato, and
  Forner-Cuenca}]{behrou2019topology}
Behrou R, Pizzolato A, Forner-Cuenca A (2019) Topology optimization as a
  powerful tool to design advanced {PEMFCs} flow fields. International Journal
  of Heat and Mass Transfer 135:72--92.
  \doi{10.1016/j.ijheatmasstransfer.2019.01.050}

\bibitem[{Bends{\o}e(1989)}]{bendsoe}
Bends{\o}e MP (1989) Optimal shape design as a material distribution problem.
  Structural optimization 1(4):193--202. \doi{10.1007/bf01650949}

\bibitem[{Brandt(1977)}]{brandt1977multi}
Brandt A (1977) Multi-level adaptive solutions to boundary-value problems.
  Mathematics of Computation 31(138):333--390.
  \doi{10.1090/s0025-5718-1977-0431719-x}

\bibitem[{Bruggeman(1935)}]{bruggeman1935berechnung}
Bruggeman VD (1935) Berechnung verschiedener physikalischer konstanten von
  heterogenen substanzen. i. dielektrizit{\"a}tskonstanten und
  leitf{\"a}higkeiten der mischk{\"o}rper aus isotropen substanzen. Annalen der
  physik 416(7):636--664. \doi{10.1002/andp.19374210205}

\bibitem[{Chen et~al(2018)Chen, Kotyk, and Sheehan}]{Chen_2018}
Chen C, Kotyk JFK, Sheehan SW (2018) Progress toward commercial application of
  electrochemical carbon dioxide reduction. Chem 4(11):2571--2586.
  \doi{10.1016/j.chempr.2018.08.019}

\bibitem[{Chen et~al(2019)Chen, Yaji, Yamasaki, Tsushima, and
  Fujita}]{chen2019computational}
Chen CH, Yaji K, Yamasaki S, et~al (2019) Computational design of flow fields
  for vanadium redox flow batteries via topology optimization. Journal of
  Energy Storage 26:100,990. \doi{10.1016/j.est.2019.100990}

\bibitem[{Chu and Majumdar(2012)}]{Chu_2012}
Chu S, Majumdar A (2012) Opportunities and challenges for a sustainable energy
  future. Nature 488(7411):294--303. \doi{10.1038/nature11475}

\bibitem[{Chu et~al(2016)Chu, Cui, and Liu}]{Chu_2016}
Chu S, Cui Y, Liu N (2016) The path towards sustainable energy. Nature
  Materials 16(1):16--22. \doi{10.1038/nmat4834}

\bibitem[{Cobb and Blanco(2014)}]{Cobb_2014}
Cobb CL, Blanco M (2014) Modeling mass and density distribution effects on the
  performance of co-extruded electrodes for high energy density lithium-ion
  batteries. Journal of Power Sources 249:357--366.
  \doi{10.1016/j.jpowsour.2013.10.084}

\bibitem[{Falgout and Yang(2002)}]{falgout2002hypre}
Falgout RD, Yang UM (2002) hypre: A library of high performance
  preconditioners. In: International Conference on Computational Science,
  Springer, pp 632--641, \doi{10.1007/3-540-47789-6_66}

\bibitem[{Forner-Cuenca and Brushett(2019)}]{forner-cuenca_engineering_2019}
Forner-Cuenca A, Brushett FR (2019) Engineering porous electrodes for
  next-generation redox flow batteries: recent progress and opportunities.
  Current Opinion in Electrochemistry 18:113--122.
  \doi{10.1016/j.coelec.2019.11.002}, publisher: Elsevier Ltd

\bibitem[{Fuller and Harb(2018)}]{fuller2018electrochemical}
Fuller TF, Harb JN (2018) Electrochemical {Engineering}. John Wiley \& Sons,
  Hoboken, NJ USA

\bibitem[{Geuzaine and Remacle(2009)}]{geuzaine2009gmsh}
Geuzaine C, Remacle JF (2009) Gmsh: A {3-D} finite element mesh generator with
  built-in pre-and post-processing facilities. International journal for
  numerical methods in engineering 79(11):1309--1331. \doi{10.1002/nme.2579}

\bibitem[{Golmon et~al(2012)Golmon, Maute, and Dunn}]{golmon2012multiscale}
Golmon S, Maute K, Dunn ML (2012) Multiscale design optimization of lithium ion
  batteries using adjoint sensitivity analysis. International Journal for
  Numerical Methods in Engineering 92(5):475--494. \doi{10.1002/nme.4347}

\bibitem[{Golmon et~al(2014)Golmon, Maute, and Dunn}]{Golmon_2014}
Golmon S, Maute K, Dunn ML (2014) A design optimization methodology for
  {Li}$^+$ batteries. Journal of Power Sources 253:239--250.
  \doi{10.1016/j.jpowsour.2013.12.025}

\bibitem[{Griewank and Walther(2000)}]{revolve}
Griewank A, Walther A (2000) Algorithm 799: Revolve: An implementation of
  checkpointing for the reverse or adjoint mode of computational
  differentiation. ACM Trans Math Softw 26(1):19–45.
  \doi{10.1145/347837.347846}

\bibitem[{Guest et~al(2004)Guest, Pr{\'e}vost, and
  Belytschko}]{guest2004achieving}
Guest JK, Pr{\'e}vost JH, Belytschko T (2004) Achieving minimum length scale in
  topology optimization using nodal design variables and projection functions.
  International journal for numerical methods in engineering 61(2):238--254.
  \doi{10.1002/nme.1064}

\bibitem[{G\"ur(2018)}]{Gur_2018}
G\"ur TM (2018) Review of electrical energy storage technologies, materials and
  systems: challenges and prospects for large-scale grid storage. Energy {\&}
  Environmental Science 11(10):2696--2767. \doi{10.1039/c8ee01419a}

\bibitem[{Haverkort(2019)}]{haverkort_theoretical_2019}
Haverkort J (2019) A theoretical analysis of the optimal electrode thickness
  and porosity. Electrochimica Acta 295:846--860.
  \doi{10.1016/j.electacta.2018.10.065}

\bibitem[{Henson and Yang(2002)}]{henson2002boomeramg}
Henson VE, Yang UM (2002) Boomer{AMG}: A parallel algebraic multigrid solver
  and preconditioner. Applied Numerical Mathematics 41(1):155--177.
  \doi{10.1016/s0168-9274(01)00115-5}

\bibitem[{Iwai et~al(2011)Iwai, Kuroyanagi, Saito, Konno, Yoshida, Yamada, and
  Nishiwaki}]{iwai2011power}
Iwai H, Kuroyanagi A, Saito M, et~al (2011) Power generation enhancement of
  solid oxide fuel cell by cathode--electrolyte interface modification in
  mesoscale assisted by level set-based optimization calculation. Journal of
  Power Sources 196(7):3485--3495. \doi{10.1016/j.jpowsour.2010.12.024}

\bibitem[{Koresh and Soffer(1977)}]{koresh}
Koresh J, Soffer A (1977) Double layer capacitance and charging rate of
  ultramicroporous carbon electrodes. Journal of The Electrochemical Society
  124(9):1379--1385. \doi{10.1149/1.2133657}

\bibitem[{Lamaison et~al(2021)Lamaison, Wakerley, Kracke, Moore, Zhou, Lee,
  Wang, Hubert, Acosta, Gregoire, Duoss, Baker, Beck, Spormann, Fontecave,
  Hahn, and Jaramillo}]{lamaison_designing_2021}
Lamaison S, Wakerley D, Kracke F, et~al (2021) Designing a {Zn}–{Ag} catalyst
  matrix and electrolyzer system for \chem{CO_2} conversion to {CO} and beyond.
  Adv Mater p~11. \doi{10.1002/adma.202103963}

\bibitem[{Lazarov and Sigmund(2011)}]{lazarov2011filters}
Lazarov BS, Sigmund O (2011) Filters in topology optimization based on
  {H}elmholtz-type differential equations. International Journal for Numerical
  Methods in Engineering 86(6):765--781. \doi{10.1002/nme.3072}

\bibitem[{Lin et~al(2022)Lin, Baker, Duoss, and Beck}]{lin2022topology}
Lin TY, Baker SE, Duoss EB, et~al (2022) Topology optimization of 3d flow
  fields for flow batteries. arXiv preprint arXiv:220213032

\bibitem[{Liu et~al(2017)Liu, Zhang, Song, and Li}]{liu2017revitalizing}
Liu T, Zhang F, Song Y, et~al (2017) Revitalizing carbon supercapacitor
  electrodes with hierarchical porous structures. Journal of Materials
  Chemistry A 5(34):17,705--17,733. \doi{10.1039/c7ta05646j}

\bibitem[{Lu et~al(2020)Lu, Bertei, Finegan, Tan, Daemi, Weaving, O'Regan,
  Heenan, Hinds, Kendrick, Brett, and Shearing}]{Lu_2020}
Lu X, Bertei A, Finegan DP, et~al (2020) {3D} microstructure design of
  lithium-ion battery electrodes assisted by {X}-ray nano-computed tomography
  and modelling. Nature Communications 11(1). \doi{10.1038/s41467-020-15811-x}

\bibitem[{Madabattula and Kumar(2020)}]{Madabattula_2020}
Madabattula G, Kumar S (2020) Model and measurement based insights into double
  layer capacitors: Voltage-dependent capacitance and low ionic conductivity in
  pores. Journal of The Electrochemical Society 167(8):080,535.
  \doi{10.1149/1945-7111/ab90aa}

\bibitem[{Mitusch et~al(2019)Mitusch, Funke, and Dokken}]{mitusch2019dolfin}
Mitusch SK, Funke SW, Dokken JS (2019) dolfin-adjoint 2018.1: automated
  adjoints for {FEniCS} and {Firedrake}. Journal of Open Source Software
  4(38):1292. \doi{10.21105/joss.01292}

\bibitem[{Nemani et~al(2015)Nemani, Harris, and Smith}]{nemani2015design}
Nemani VP, Harris SJ, Smith KC (2015) Design of bi-tortuous, anisotropic
  graphite anodes for fast ion-transport in {Li}-ion batteries. Journal of The
  Electrochemical Society 162(8):A1415. \doi{10.1149/ma2016-03/2/848}

\bibitem[{Newman and Thomas-Alyea(2012)}]{newman2012electrochemical}
Newman J, Thomas-Alyea KE (2012) Electrochemical systems. John Wiley \& Sons

\bibitem[{Newman and Tiedemann(1975)}]{newman1975porous}
Newman J, Tiedemann W (1975) Porous-electrode theory with battery applications.
  AIChE Journal 21(1):25--41. \doi{10.1002/aic.690210103}

\bibitem[{O’Brien et~al(2021)O’Brien, Miao, Liu, Xu, Lee, Robb, Huang, Xie,
  Bertens, Gabardo, Edwards, Dinh, Sargent, and Sinton}]{obrien_single_2021}
O’Brien CP, Miao RK, Liu S, et~al (2021) Single pass \chem{CO_2} conversion
  exceeding 85\% in the electrosynthesis of multicarbon products via local
  \chem{CO_2} regeneration. ACS Energy Letters 6(8):2952--2959.
  \doi{10.1021/acsenergylett.1c01122}

\bibitem[{Park et~al(2020)Park, Goodall, and Kim}]{park_perspective_2020}
Park SH, Goodall G, Kim WS (2020) Perspective on {3D}-designed
  micro-supercapacitors. Materials \& Design 193:108,797.
  \doi{10.1016/j.matdes.2020.108797}

\bibitem[{Ramadesigan et~al(2010)Ramadesigan, Methekar, Latinwo, Braatz, and
  Subramanian}]{ramadesigan2010optimal}
Ramadesigan V, Methekar RN, Latinwo F, et~al (2010) Optimal porosity
  distribution for minimized ohmic drop across a porous electrode. Journal of
  The Electrochemical Society 157(12):A1328. \doi{10.1149/1.3495992}

\bibitem[{Rathgeber et~al(2016)Rathgeber, Ham, Mitchell, Lange, Luporini,
  McRae, Bercea, Markall, and Kelly}]{rathgeber2016firedrake}
Rathgeber F, Ham DA, Mitchell L, et~al (2016) Firedrake: automating the finite
  element method by composing abstractions. ACM Transactions on Mathematical
  Software (TOMS) 43(3):1--27. \doi{10.1145/2998441}

\bibitem[{Roy et~al(2022)Roy, Salazar~de Troya, and
  Beck}]{thomas_roy_2022_6366849}
Roy T, Salazar~de Troya MA, Beck VA (2022) {LLNL}/{TOPE}: Topology optimization
  for porous electrodes. \doi{10.5281/zenodo.6366849}

\bibitem[{Ruge and St{\"u}ben(1987)}]{ruge1987algebraic}
Ruge JW, St{\"u}ben K (1987) Algebraic multigrid. In: Multigrid methods, vol 3
  of Frontiers in Applied Mathematics. SIAM, Philadelphia, chap~4, p 73--130,
  \doi{10.1137/1.9781611971057.ch4}

\bibitem[{Sawant et~al(2021)Sawant, Yim, Henry, Miller, and
  McKone}]{sawant_harnessing_2021}
Sawant TV, Yim CS, Henry TJ, et~al (2021) Harnessing interfacial electron
  transfer in redox flow batteries. Joule 5(2):360--378.
  \doi{10.1016/j.joule.2020.11.022}

\bibitem[{Schiffer and Manthiram(2017)}]{schiffer_electrification_2017}
Schiffer ZJ, Manthiram K (2017) Electrification and decarbonization of the
  chemical industry. Joule 1(1):10--14. \doi{10.1016/j.joule.2017.07.008}

\bibitem[{Sethian and Wiegmann(2000)}]{SETHIAN2000489}
Sethian J, Wiegmann A (2000) Structural boundary design via level set and
  immersed interface methods. Journal of Computational Physics 163(2):489 --
  528. \doi{10.1006/jcph.2000.6581}

\bibitem[{Shatskiy et~al(2019)Shatskiy, Lundberg, and
  K\"ark\"as}]{shatskiy_organic_2019}
Shatskiy A, Lundberg H, K\"ark\"as MD (2019) Organic electrosynthesis:
  Applications in complex molecule synthesis. ChemElectroChem 6(16):4067--4092.
  \doi{10.1002/celc.201900435}, publisher: John Wiley \& Sons, Ltd

\bibitem[{Song et~al(2013)Song, Diaz, Benard, and Nicholas}]{song20132d}
Song X, Diaz A, Benard A, et~al (2013) A {2D} model for shape optimization of
  solid oxide fuel cell cathodes. Structural and Multidisciplinary Optimization
  47(3):453--464. \doi{10.1007/s00158-012-0837-x}

\bibitem[{Stankiewicz and Nigar(2020)}]{stankiewicz_beyond_2020}
Stankiewicz AI, Nigar H (2020) Beyond electrolysis: old challenges and new
  concepts of electricity-driven chemical reactors. Reaction Chemistry \&
  Engineering 5(6):1005--1016. \doi{10.1039/D0RE00116C}

\bibitem[{St{\"o}ckl et~al(2021)St{\"o}ckl, Schill, and
  Zerrahn}]{stockl_optimal_2021}
St{\"o}ckl F, Schill WP, Zerrahn A (2021) Optimal supply chains and power
  sector benefits of green hydrogen. Scientific Reports 11(1):14,191.
  \doi{10.1038/s41598-021-92511-6}

\bibitem[{Svanberg(1987)}]{svanberg1987method}
Svanberg K (1987) The method of moving asymptotes -- a new method for
  structural optimization. International journal for numerical methods in
  engineering 24(2):359--373. \doi{10.1002/nme.1620240207}

\bibitem[{Thorat et~al(2009)Thorat, Stephenson, Zacharias, Zaghib, Harb, and
  Wheeler}]{thorat}
Thorat IV, Stephenson DE, Zacharias NA, et~al (2009) Quantifying tortuosity in
  porous {Li}-ion battery materials. Journal of Power Sources 188(2):592--600.
  \doi{10.1016/j.jpowsour.2008.12.032}

\bibitem[{Tjaden et~al(2016)Tjaden, Cooper, Brett, Kramer, and
  Shearing}]{TJADEN201644}
Tjaden B, Cooper SJ, Brett DJ, et~al (2016) On the origin and application of
  the {{Bruggeman}} correlation for analysing transport phenomena in
  electrochemical systems. Current Opinion in Chemical Engineering 12:44--51.
  \doi{10.1016/j.coche.2016.02.006}

\bibitem[{Tjaden et~al(2018)Tjaden, Brett, and Shearing}]{tjaden2018tortuosity}
Tjaden B, Brett DJ, Shearing PR (2018) Tortuosity in electrochemical devices: a
  review of calculation approaches. International Materials Reviews
  63(2):47--67. \doi{10.1080/09506608.2016.1249995}

\bibitem[{Salazar~de Troya(2021)}]{miguel_salazar_de_troya_2021_4456055}
Salazar~de Troya MA (2021) {LLNL/pyMMAopt: Method of Moving Asymptotes for
  Firedrake}. \doi{10.5281/zenodo.5524961}

\bibitem[{Salazar~de Troya and Tortorelli(2020)}]{Salazar_de_Troya_2020}
Salazar~de Troya MA, Tortorelli DA (2020) Three-dimensional adaptive mesh
  refinement in stress-constrained topology optimization. Structural and
  Multidisciplinary Optimization 62(5):2467--2479.
  \doi{10.1007/s00158-020-02618-z}

\bibitem[{Salazar~de Troya et~al(2021)Salazar~de Troya, Oxberry, Petra, and
  Tortorelli}]{salzardetroya2021}
Salazar~de Troya MA, Oxberry GM, Petra CG, et~al (2021) Another source of mesh
  dependence in topology optimization. arXiv preprint arXiv:210612098

\bibitem[{Wang et~al(2008)Wang, Li, Liu, Lu, and Cheng}]{wang20083d}
Wang DW, Li F, Liu M, et~al (2008) {3D} aperiodic hierarchical porous graphitic
  carbon material for high-rate electrochemical capacitive energy storage.
  Angewandte Chemie International Edition 47(2):373--376.
  \doi{10.1002/ange.200702721}

\bibitem[{Wang et~al(2011)Wang, Lazarov, and Sigmund}]{wang2011projection}
Wang F, Lazarov BS, Sigmund O (2011) On projection methods, convergence and
  robust formulations in topology optimization. Structural and
  Multidisciplinary Optimization 43(6):767--784.
  \doi{10.1007/s00158-010-0602-y}

\bibitem[{Wang et~al(2003)Wang, Wang, and Guo}]{wang2003level}
Wang MY, Wang X, Guo D (2003) A level set method for structural topology
  optimization. Computer methods in applied mechanics and engineering
  192(1-2):227--246. \doi{10.1016/s0045-7825(02)00559-5}

\bibitem[{Wathen(2015)}]{wathen2015preconditioning}
Wathen AJ (2015) Preconditioning. Acta Numerica 24:329--376.
  \doi{10.1017/s0962492915000021}

\bibitem[{Weber et~al(2011)Weber, Mench, Meyers, Ross, Gostick, and
  Liu}]{weber_redox_2011}
Weber AZ, Mench MM, Meyers JP, et~al (2011) Redox flow batteries: a review.
  Journal of Applied Electrochemistry 41(10):1137--1164.
  \doi{10.1007/s10800-011-0348-2}

\bibitem[{Wu et~al(2019)Wu, Lv, Lin, Zhang, Liu, and Zhou}]{Wu_2019}
Wu Q, Lv Y, Lin L, et~al (2019) An improved thin-film electrode for vanadium
  redox flow batteries enabled by a dual layered structure. Journal of Power
  Sources 410-411:152--161. \doi{10.1016/j.jpowsour.2018.11.020}

\bibitem[{Xue et~al(2015)Xue, Du, Martins, and Shyy}]{xue2016lithium}
Xue N, Du W, Martins JR, et~al (2015) Lithium-ion batteries: Thermomechanics,
  performance, and design optimization. In: Handbook of Clean Energy Systems,
  John Wiley \& Sons Ltd., vol. 6. Wiley Online Library, p 2849--2864,
  \doi{10.1002/9781118991978.hces225}

\bibitem[{Yaji et~al(2018)Yaji, Yamasaki, Tsushima, Suzuki, and
  Fujita}]{yaji2018topology}
Yaji K, Yamasaki S, Tsushima S, et~al (2018) Topology optimization for the
  design of flow fields in a redox flow battery. Structural and
  multidisciplinary optimization 57(2):535--546.
  \doi{10.1007/s00158-017-1763-8}

\bibitem[{Yan et~al(2017)Yan, Kawamata, and Baran}]{yan_synthetic_2017}
Yan M, Kawamata Y, Baran PS (2017) Synthetic organic electrochemical methods
  since 2000: On the verge of a renaissance. Chemical Reviews
  117(21):13,230--13,319. \doi{10.1021/acs.chemrev.7b00397}

\bibitem[{Zadin et~al(2013)Zadin, Brandell, Kasem{\"a}gi, Lellep, and
  Aabloo}]{zadin2013designing}
Zadin V, Brandell D, Kasem{\"a}gi H, et~al (2013) Designing the
  {3D}-microbattery geometry using the level-set method. Journal of power
  sources 244:417--428. \doi{10.1016/j.jpowsour.2012.12.004}

\bibitem[{Zhang et~al(2022)Zhang, Constantinescu, and Smith}]{zhang2022petsc}
Zhang H, Constantinescu EM, Smith BF (2022) {PETSc TSAdjoint}: a discrete
  adjoint {ODE} solver for first-order and second-order sensitivity analysis.
  SIAM Journal on Scientific Computing 44(1):C1--C24

\bibitem[{Zhang and Ran(2021)}]{zhang_design_2021}
Zhang T, Ran F (2021) Design strategies of {3D} carbon‐based electrodes for
  charge/ion transport in lithium ion battery and sodium ion battery. Advanced
  Functional Materials 31(17):2010,041. \doi{10.1002/adfm.202010041}

\bibitem[{Zhang et~al(2021)Zhang, Hui, King, Wang, Ju, Wu, Takeuchi,
  Marschilok, West, Takeuchi, and Yu}]{zhang_tunable_2021}
Zhang X, Hui Z, King S, et~al (2021) Tunable porous electrode architectures for
  enhanced {Li}-ion storage kinetics in thick electrodes. Nano Letters
  21(13):5896--5904. \doi{10.1021/acs.nanolett.1c02142}

\bibitem[{Zhou et~al(2016)Zhou, Zeng, Zhu, Wei, and Zhao}]{Zhou_2016}
Zhou X, Zeng Y, Zhu X, et~al (2016) A high-performance dual-scale porous
  electrode for vanadium redox flow batteries. Journal of Power Sources
  325:329--336. \doi{10.1016/j.jpowsour.2016.06.048}

\bibitem[{Zhu et~al(2016)Zhu, Liu, Qian, Han, Duoss, Kuntz, Spadaccini,
  Worsley, and Li}]{zhu2016supercapacitors}
Zhu C, Liu T, Qian F, et~al (2016) Supercapacitors based on three-dimensional
  hierarchical graphene aerogels with periodic macropores. Nano letters
  16(6):3448--3456. \doi{10.1021/acs.nanolett.5b04965}

\end{thebibliography}

\appendix
\section{Variational formulation}
\label{sec:variational}
Here we provide the variational formulation of the system \eqref{eq:porous_electrode_problem}.
It is necessary for the finite element method, which is used in
our numerical experiments.

Let the bilinear forms
\begin{equation}
    \begin{aligned}
        a_1(\Phi_1, p_1) & = \int_\Omega \sigma \nabla \Phi_1 \cdot \nabla p_1 \diff V , \\
        a_2(\Phi_2, p_2) & = \int_\Omega \kappa \nabla \Phi_2 \cdot \nabla p_2 \diff V,
    \end{aligned}
\end{equation}
the nonlinear mapping
\begin{equation}
    b(\Phi_1, \Phi_2; p) =  \int_\Omega a i_n(\Phi_1, \Phi_2) p \diff V,
\end{equation}
and functional
\begin{equation}
    l(p) = \int_{\Gamma_2} g p \diff s.
\end{equation}
Let the function spaces
\begin{equation}
    \begin{aligned}
        V   & = \left\{ \Phi \in H^1 (\Omega) \right\},                                         \\
        V_1 & = \left\{ \Phi \in H^1 (\Omega) \;\vert\; \Phi = 0 \text{ on } \Gamma_1 \right\}.
    \end{aligned}
\end{equation}
First, in the case where we have a Neumann boundary condition on $\Gamma_2$, the
variational formulation for \eqref{eq:porous_electrode_problem} is given by: Find $\Phi_1 \in V_1$,
$\Phi_2\in V$ such that
\begin{multline}
    a_1(\Phi_1, p_1) + a_2(\Phi_2, p_2) \\+ b(\Phi_1,\Phi_2; p_1) - b(\Phi_1, \Phi_2; p_2) = l(p_2),
\end{multline}
for all $p_1\in V_1$, $p_2\in V$
Next, in the case where we have a Dirichlet boundary condition on $\Gamma_2$, we require the additional function spaces
\begin{equation}
    \begin{aligned}
        V_D & = \left\{ \Phi \in H^1 (\Omega) \;\vert\; \Phi = g \text{ on } \Gamma_2 \right\}, \\
        V_2 & = \left\{ \Phi \in H^1 (\Omega) \;\vert\; \Phi = 0 \text{ on } \Gamma_2 \right\}.
    \end{aligned}
\end{equation}
The variational formulation for \eqref{eq:porous_electrode_problem} is given by: Find $\Phi_1 \in V_1$,
$\Phi_2\in V_D$ such that
\begin{multline}
    a_1(\Phi_1, p_1) + a_2(\Phi_2, p_2) \\+ b(\Phi_1,\Phi_2; p_1) - b(\Phi_1, \Phi_2; p_2) = 0,
\end{multline}
for all $p_1\in V_1$, $p_2\in V_2$.

\section{Two-point flux approximation for the PDE filter}\label{sec:DG0}
\rtwo{The PDE filter \eqref{eq:filter} is solved using a two-point flux approximation (TPFA) finite volume method that preserves the \emph{minimum principle}: if $\gamma(\mathbf{x})\geq 0 $ for all $\mathbf{x}\in \Omega$, then $\hat\gamma$ attains its minimum on $\partial\Omega$ and $\hat\gamma\geq 0$. This principle is important to comply because negatives values of $\hat\gamma$ result in non-positive matrices and affect the iterative solver.
Using Lagrange finite elements to discretize \eqref{eq:filter} does not ensure satisfaction of the minimum principle.
Indeed, for a sufficiently small $r / h$ where $h$ is the mesh element size, the reaction term $\hat\gamma$ in \eqref{eq:filter} dominates the diffusion term.
As such, $\hat\gamma$ effectively becomes the $L^2$-orthogonal projection of $\gamma$ onto $H^1$, which necessitates oscillations to minimize the $L^2$-distance \citep{Salazar_de_Troya_2020}.}

Given a partition on $\Omega$, let $\Gamma_\mathrm{int}$ denote the union of all
interior facets. We define the jump at a facet by $[v] = v^+ - v^-$, where $v^+$
and $v^-$ are the limit values of $v$ on either side of the facet. Let
$\mathbf{c}$ be the piecewise constant function of cell-centered coordinates.
Let $\mathbb P_\mathrm{DG}^0$ be the space of piecewise constant functions on
our partition of $\Omega$. The variational problem is given by: Find $\gfp\in
    \mathbb P_\mathrm{DG}^0$ such that
\begin{equation}
    \int_{\Gamma_\mathrm{int}} r^2[v]\frac{[\gfp]}{\|[\mathbf c]\|} \diff s +
    \int_{\Omega} \gfp v \diff V = \int_{\Omega} \gamma v \diff V ,
\end{equation}
for all $v\in\mathbb P_\mathrm{DG}^0$.

It is known that in order for TPFA to converge, the jump of the cell centers
must be orthogonal to the facet between the cells. This is not in general
satisfied by unstructured meshes.
\rtwo{We employ the Frontal-Delaunay algorithm in Gmsh \citep{geuzaine2009gmsh} that ensures most of the mesh elements are equilateral triangles/tetrahedra and therefore, their centroids are connected with lines orthogonal to the facets.} One could also instead use a mixed formulation
where piecewise constant elements are used for $\gfp$ and lowest-order
Raviart-Thomas elements are used for the flux $\bm{\psi} = - \nabla \gfp$.

\section{PETSc solver options}\label{sec:solveroptions}
For the redox electrode, the \texttt{PETSc} nonlinear solver options are:
\begin{lstlisting}
    "snes_type": "newtonls",
    "snes_linesearch_type": "l2",
    "snes_rtol": 1e-4,
\end{lstlisting}
The preconditioner \eqref{eq:precon} is specially suited for block
preconditioning using \texttt{Firedrake}'s solver interface with \texttt{PETSc}. The solver
options are given by:
\begin{lstlisting}
    "mat_type": "aij",
    "ksp_rtol": 1e-4,
    "ksp_type": "cg",
    "pc_type": "fieldsplit",
    "pc_fieldsplit_type": "symmetric_multiplicative",
    "fieldsplit_1_ksp_type": "preonly",
    "fieldsplit_1_pc_type": "hypre",
    "fieldsplit_0_ksp_type": "preonly",
    "fieldsplit_0_pc_type": "hypre",
\end{lstlisting}
We use the following \texttt{PETSc} solver options for the PDE filter:
\begin{lstlisting}
    "mat_type": "aij",
    "ksp_rtol": 1e-6,
    "ksp_type": "cg",
    "pc_type": "hypre",
\end{lstlisting}

\end{document}